\documentclass{article}

\usepackage{arxiv}

\usepackage[utf8]{inputenc} 
\usepackage[T1]{fontenc}    
\usepackage{hyperref}       
\usepackage{url}            
\usepackage{booktabs}       
\usepackage{amsfonts}       
\usepackage{nicefrac}       
\usepackage{microtype}      
\usepackage{lipsum,enumerate,bm,mathrsfs,yfonts}
\usepackage{graphicx,amsmath,multicol,multirow}
\usepackage{url}
\usepackage{amsthm}
\usepackage[style=apa, backend=biber, natbib=true]{biblatex}
\usepackage{adjustbox,xr,comment}

\usepackage{siunitx}
\usepackage{rotating}   
\usepackage{booktabs}
\usepackage{import}
\usepackage{bm,amsmath,multicol,multirow,amssymb,comment}
\usepackage{placeins}
\usepackage{graphicx}
\usepackage{xcolor} 
\usepackage[table]{xcolor}
\usepackage{caption,subcaption}
\definecolor{lightgray}{gray}{0.9}
\usepackage{longtable,setspace}
\usepackage{xr}

\usepackage{amsmath}
\usepackage{amssymb}
\usepackage{amsthm,tikz}
\usepackage{colortbl,caption,xr}
\usepackage{adjustbox}  
\usepackage{graphicx}
\usepackage{lscape}
\usepackage{rotating}
\usepackage{graphicx}
\usepackage{float}
\usepackage{url}
\usepackage{dcolumn}
\usepackage{multirow}
\usepackage{verbatim,booktabs, array,arydshln}
\usepackage{color}
\usepackage{setspace}
\usepackage{bbm}
\usepackage{booktabs}
\usepackage{hhline}
\usepackage{bm}
\usepackage{dsfont,import}
\usepackage{enumerate}
\usepackage{threeparttable}
\usepackage{array}
\usepackage{caption}
\usepackage{tikz}
\usetikzlibrary{shapes.geometric, arrows, shadows}
\usepackage{graphicx}
\usetikzlibrary{shapes.geometric, arrows, shadows,positioning,arrows.meta,calc}
\usepackage{graphicx}
\usepackage{booktabs,multicol,multirow,tabularx,array}
\usepackage{longtable}
\usepackage{siunitx,array}
\usepackage{caption}
\usepackage{graphicx}

\usepackage{siunitx}

\newtheorem{theorem}{Theorem}[section]   
\newtheorem{lemma}[theorem]{Lemma}       

\theoremstyle{definition}

\theoremstyle{remark}
\newtheorem*{remark}{Remark}

\theoremstyle{assumption}

\theoremstyle{proposition}

\tikzstyle{state} = [ellipse, draw, fill=gray!20, text centered, text width=1.5cm, minimum height=1.2cm, drop shadow]
\tikzstyle{arrow} = [thick,->,>=stealth]

\title{Sample size determination for win statistics in cluster-randomized trials}

\author{
 Xi Fang \\
    Data Science Institute \\
    Medical College of Wisconsin \\
    Milwaukee, WI, USA\\
\And
Zhiqiang Cao \\
School of Artificial Intelligence\\
Shenzhen Technology University\\
Guangdong, China
\And
 Fan Li \\
Department of Biostatistics \\
Yale School of Public Health \\
New Haven, CT, USA\\
  \texttt{fan.f.li@yale.edu} \\
}

\begin{document}
\maketitle
\begin{abstract}
Composite endpoints are increasingly used in clinical trials to capture treatment effects across multiple or hierarchically ordered outcomes. Although inference procedures based on win statistics---such as the win ratio, win odds, and win difference (net benefit)--have gained traction in individually randomized trials, their methodological development for cluster-randomized trials remains limited. In particular, there is no formal framework for power and sample size determination when using win statistics with composite time-to-event outcomes. We develop a unified framework for power and sample size calculation for win statistics under cluster randomization. Analytical variance expressions are derived for a class of win statistics, yielding closed-form variance expressions and power procedures that avoid computationally intensive simulations. The variance expressions explicitly characterize the roles of the rank intracluster correlation coefficient, cluster size, tie probability, and outcome prioritization for study planning purposes. Importantly, our variances nest existing formulas for univariate outcomes as special cases while extending them to complex, hierarchically ordered composite endpoints. Simulation studies confirm accurate finite-sample performance, and we supply a case study to illustrate the use of our proposed method.
\end{abstract}


\section{Introduction}
\label{sec:intro}

Cluster-randomized trials (CRTs), in which entire groups of individuals are randomized to different conditions, are widely used in clinical research. A primary complication of cluster randomization is that individual observations are correlated within clusters. This correlation, often quantified by the intracluster correlation coefficient (ICC), inflates the variance of treatment effect estimators and hence the required sample size, compared to individual randomization \citep{murray1998design}. Due to the positive ICC, CRTs require a larger total sample size than individually randomized trials to maintain the same statistical power, and the amount of sample size inflation depends on the design effect that increases linearly in both the ICC and the cluster size \citep{turner2017review}. Closed-form sample size formulas have been previously developed to address continuous, binary, count, and censored time-to-event outcomes; see, for example, \citet{rutterford2015methods} on a synthesis of sample size formulas in CRTs. 
Despite these advances, existing sample size methods for CRTs are restricted to support the primary analysis of a single endpoint. Modern clinical trials have seen a growing interest in incorporating multiple prioritized outcomes, both to better assess the overall net benefit and to enhance statistical efficiency without requiring multiplicity adjustment \citep{mao2021statistical_review}. Such composite outcomes capture the overall treatment effect when more than one specific endpoint is of interest, but introduce new challenges for the design and analysis in the context of CRTs.


When analyzing composite outcomes, traditional time-to-first-event analyses, while straightforward to implement, may overweight less critical outcomes and obscure clinically meaningful treatment differences. To address this limitation, alternative summary measures that incorporate clinical prioritization have been proposed. Among these, the win ratio, introduced by \citet{pocock2012win}, has emerged as a popular summary measure. The win ratio is based on pairwise comparisons between individuals in the treatment and control groups, following a pre-specified hierarchy of clinically relevant outcomes. For each pair, a ``win'' is assigned to the treatment arm if the treated participant has a more favorable outcome on the highest-priority component; if outcomes are tied, the comparison proceeds sequentially to lower-priority components. The treatment effect is then summarized as the ratio of total wins to total losses across all admissible comparisons. There is a burgeoning literature studying methods targeting win ratio along with their variants for applications to individually randomized trials (IRT). For example, \citet{luo2015alternative} proposed a closed-form variance estimator and hypothesis testing procedure for unmatched comparisons. The asymptotic properties for win statistics have been developed based on the theory of U-statistic \citep{bebu2016large}. These results have only recently been expanded to CRTs. Specifically, \citet{zhang2021inference} extended the win ratio estimator to prioritized outcomes under the semi-competing risks structure, accounting for within-cluster correlations. With non-prioritized outcomes, \citet{smith2024rank} proposed a cluster-adjusted global win probability by summarizing each individual's outcomes into a win fraction, which was further analyzed with a working linear mixed model to address within-cluster correlations.

Despite the advancements in analytical methods, methods for sample size determination when using win statistics are relatively sparse, with a few recent exceptions. \citet{yu2022sample} proposed a simple and interpretable formula based on the \citet{finkelstein1999combining} test, requiring design parameters such as the expected effect size, tie probability, and win/loss ratio. \citet{mao2022sample} extended the study design method by explicitly deriving the asymptotic variance of the win ratio estimator using the theory of U-statistics, and considered reparameterizations of the effect size in the presence of ordinal and survival endpoints. \citet{barnhart2025sample} further developed unified analytic formulas for sample size and power calculations across multiple win statistics, providing practical expressions that replace simulation-based approaches for trials with hierarchical endpoints. Nevertheless, these existing approaches are not directly applicable to cluster-correlated data, as they are not developed to account for ICC nor cluster size variation that are essential ingredients of sample size formulas for CRTs.


To fill in this methodological gap, we derive a set of new sample size formulas tailored for CRTs with win statistics-based inference, addressing the within-cluster correlations and variable cluster sizes. Specifically, we first propose testing procedures by modeling the rank statistics based on pairwise comparison that accounts for within-cluster dependence. Our development covers three tests that utilize different win statistics: win ratio, win odds \citep{brunner2021win}, and win difference (net benefit) \citep{buyse2010generalized}. This provides a complementary approach for win ratio based test in \citet{zhang2021inference}. Based on our testing procedure, we then derive the explicit design effect under cluster randomization. Interestingly, the derived design effect exhibits a familiar form compared to the usual design effect previously developed for single outcomes, but now depends on the rank-based analog of the ICC (rank ICC) \citep{tu2023rank} and the coefficient of variation in cluster sizes. In fact, a broader implication of this design effect is that it is applicable not only to prioritized composite outcomes but also to conventional, single outcomes when they are analyzed via win statistics. Finally, we discuss considerations for implementing our sample size methods, and develop a free R shiny app at \url{https://fangxx.shinyapps.io/WR_CRT/} that can facilitate the design of CRTs in the presence of prioritized composite outcomes. To the best of our knowledge, this work represents the first analytic sample size methodology tailored for win statistical analysis of correlated single and composite outcomes under cluster randomization.


\raggedbottom
\section{Notation, setup and introduction to win statistics}
\label{sec:win_stat}

Let \(M\) denote the total number of clusters in a parallel CRT, where \(qM\) clusters are randomized to intervention and the remaining \((1-q)M\) clusters to the usual care with $q\in (0,1)$. For each cluster \(i \in \{1, \dots, M\}\), let \(N_i\) be the number of individuals within cluster \(i\), and \(A_i \in \{0, 1\}\) denote the binary cluster-level treatment (\(A_i = 1\) indicates intervention and \(A_i = 0\) control), then $q=E(A_i)$. For each individual \(j \in \{1, \dots, N_i\}\) within cluster \(i\), we write the observed outcome vector \(\bm{Y}_{ij} = \{Y_{ij1}, \dots, Y_{ijV}\}\), where \(Y_{ijv} \in \mathcal{Y}^V \) corresponds to the \(v\)-th endpoint ($V\geq 1$). These components are assumed to be ordered by clinical importance, representing, for example, time to mortality, hospitalization, and symptom progression, among others. We note that this is a general setup and includes both the composite outcomes setting (when \(V \geq 2\)), and conventional single-outcome settings (when \(V = 1\)). We assume between-cluster independence such that \(\{\bm{Y}_{i1}, \dots, \bm{Y}_{iN_i}\} \perp \{\bm{Y}_{k1}, \dots, \bm{Y}_{kN_k}\}\) for all \(i \ne k\), while allowing for correlation between endpoints within each individual, and across individuals within the same cluster. We also define \(n_1 = \sum_{i=1}^{M} A_i N_i\) and \(n_0 = \sum_{i=1}^{M} (1 - A_i) N_i\) as the sample sizes under the treatment and control groups, respectively; the total sample size is $n=n_1+n_0$.

For testing the existence of treatment effect, we consider all pairwise comparisons between individuals from different treatment arms. For any pair of individuals---individual $j$ in cluster $i$ and individual $l$ in cluster $k$---abbreviated as $(i,j)$ and $(k,l)$ with $A_i \ne A_k$, we define a win operator `$\succ$' to indicate the favourability of outcomes under a pre-specified clinical ranking rule that sequentially evaluates the components of the outcome vector. Mathematically, $\bm Y_{ij}\succ \bm Y_{kl}$ indicates that individual $(i,j)$ had a more favorable outcome than individual $(k,l)$, $\bm Y_{ij}\prec \bm Y_{kl}$ denotes the opposite, and $\bm Y_{ij}=\bm Y_{kl}$ indicates a tie under the comparison rule. For example, in a prioritized composite endpoint that ranks mortality above hospitalization, $\bm Y_{ij}\succ \bm Y_{kl}$ indicates that individual $(i,j)$ survived but individual $(k,l)$ did not, or that both survived but individual $(i,j)$ experienced later hospitalization. A win is recorded when $\bm Y_{ij}\succ \bm Y_{kl}$; a loss is recorded when $\bm Y_{ij}\prec \bm Y_{kl}$; and a tie $\bm{Y}_{ij} = \bm{Y}_{kl}$ is recorded if the rule does not identify a winner \citep{pocock2012win}. Here $\bm{Y}_{ij} = \bm{Y}_{kl}$ denotes equality under the prespecified comparison rule, not necessarily equality of the full raw outcome vector. For ordinal outcomes, a tie occurs when two individuals fall in the same ordered category. For continuous outcomes, exact ties are absent under a continuous distribution, but ties may arise from measurement rounding, coarsening, or a protocol-specified clinically meaningful indifference margin. For right-censored survival or prioritized composite outcomes, a tie denotes an unresolved or non-discriminating comparison, including equal observed event times, administrative censoring before a discriminating event, or censoring that prevents the endpoint from determining a winner. Several population summary measures are useful in quantifying treatment benefits. These include $\pi_{\text{win}}=\mathbb P(\bm Y_{ij}\succ\bm Y_{kl}\mid A_i=1,A_k=0)$, the probability that an intervention individual has a more favorable outcome than a control individual; $\pi_{\text{loss}}=\mathbb P(\bm Y_{ij}\prec\bm Y_{kl}\mid A_i=1,A_k=0)$, the probability of the opposite comparison; and $\pi_{\text{tie}}=\mathbb P(\bm Y_{ij}=\bm Y_{kl}\mid A_i=1,A_k=0)$, the probability of a tie. They satisfy $\pi_{\text{win}}+\pi_{\text{loss}}+\pi_{\text{tie}}=1$. In this work, we allow cluster sizes to vary randomly across clusters but follow the standard convention in the CRT study-planning literature that cluster size is noninformative; that is, the cluster size distribution is unrelated to the outcome distribution or the magnitude of the treatment effect. Accordingly, $\pi_{\mathrm{win}}$, $\pi_{\mathrm{loss}}$, and $\pi_{\mathrm{tie}}$ are defined marginally over individuals, clusters, and the random cluster size distribution in the target trial population. Under the noninformative cluster size assumption, sampling an individual from each treatment arm or first sampling a cluster with equal probability and then sampling an individual within that cluster yields the same marginal win distribution \citep{wang2026mixed}. Thus, the resulting win estimands may equivalently be interpreted as individual-pair-average or cluster-pair-average estimands in the setting considered here. Unequal cluster sizes nevertheless remain relevant to the sampling variance and design effect and are therefore explicitly accommodated in our sample size development. When cluster size is informative of the outcome distribution or treatment effect, however, individual-pair-average and cluster-pair-average win estimands generally differ, as discussed recently by \citet{lee2026s} and \citet{chen2026optimal}. Extending the proposed sample size methodology to distinguish and accommodate these alternative estimands under informative cluster size is an important direction for future research.


Three population-level estimands are defined based on the above quantities. The win difference or net benefit \citep{buyse2010generalized} is given by $W_D = \pi_{\text{win}} - \pi_{\text{loss}}$, representing the net probability that treatment leads to more favorable outcomes than control. The win ratio is defined as $W_R = \frac{\pi_{\text{win}}}{\pi_{\text{loss}}}$, and measures the relative probability of treatment wins to losses \citep{pocock2012win}. The win odds \citep{brunner2021win} is a modification of win ratio by incorporating comparisons of ties, and takes the form of $W_O = \frac{\pi_{\text{win}} + 0.5 \pi_{\text{tie}}}{\pi_{\text{loss}} + 0.5 \pi_{\text{tie}}}$ (assigning half probability of ties to win and loss scores). These estimands generalize traditional effect measures to accommodate hierarchical composite endpoints. To estimate these quantities, we count the total number of pairwise wins, losses and ties as 
$W = \sum_{i=1}^{M} \sum_{j=1}^{N_i} \sum_{k=1}^{M} \sum_{l=1}^{N_k} A_i (1-A_k) \mathbb{I}(\bm{Y_{ij}} \succ \bm{Y_{kl}})$, $L = \sum_{i=1}^{M} \sum_{j=1}^{N_i} \sum_{k=1}^{M} \sum_{l=1}^{N_k}A_i (1-A_k)  \mathbb{I}(\bm{Y_{ij}} \prec \bm{Y_{kl}})$ and \( T = \sum_{i=1}^{M} \sum_{j=1}^{N_i} \sum_{k=1}^{M} \sum_{l=1}^{N_k} A_i (1-A_k)  \mathbb{I}( \bm{Y_{ij}} = \bm{Y_{kl}}) \), respectively, where \( \mathbb{I}(\cdot) \) is the indicator function. Then, consistent estimators of these win estimands are given by
\[
\widehat{W}_D = \frac{W - L}{n_1 n_0}, \quad 
\widehat{W}_R = \frac{W}{L}, \quad 
\widehat{W}_O = \frac{W + 0.5 T}{L + 0.5 T},
\]
provided \( L > 0 \). For the ratio-type estimators, $\widehat{W}_R$ and $\widehat{W}_O$, their relationship with the difference-type estimator $\widehat{W}_D$ is also conveniently expressed on the log scale:
\begin{equation} \label{eq:relation}
      \log(\widehat W_R) \;=\; 
  2\,\mathrm{atanh}\!\left(
    \frac{\widehat W_D}{\,1 - \widehat{\pi}_{\text{tie}}\,}
  \right), 
\qquad
\log(\widehat W_O) \;=\; 2\,\mathrm{atanh}(\widehat W_D),  
\end{equation}
where $\mathrm{atanh}(x) = \tfrac{1}{2}\log\!\left(\tfrac{1+x}{1-x}\right)$ for $-1 < x < 1$ is the inverse hyperbolic tangent function and \(\widehat{\pi}_{\text{tie}} = T/(n_0n_1)\) is consistent estimator of \(\pi_{\text{tie}}\).

\section{Hypothesis testing and statistical power in CRTs} \label{sec:test}

To derive the large-sample distribution of the win statistics for inference, we let \(s_{ij,kl}= s(\bm{Y}_{ij}, \bm{Y}_{kl}) = \mathbb{I}(\bm Y_{ij}\succ \bm Y_{kl})-\mathbb{I}(\bm Y_{ij}\prec \bm Y_{kl})\) denote the antisymmetric pairwise comparison kernel and \(\phi_{ij}=\sum_{k=1}^{M}\sum_{l=1}^{N_k} s_{ij,kl}\) be the cumulative individual-level score. For each cluster \(i\), we define the centered cluster score \(S_i=\sum_{j=1}^{N_i}\phi_{ij}\), which satisfies \(\sum_{i=1}^M S_i=0\). Specifically, we denote \(\mu_a=\mathbb{E}(S_i\mid A_i=a)\) and \(\sigma_a^2=\operatorname{Var}(S_i\mid A_i=a)\) as the arm-specific mean and variance of the cluster score, and let \(\overline{N} = \mathbb{E}(N_i)\) be the mean cluster size. To derive the following central limit theorem, the arm-specific cluster scores are assumed to have finite, nondegenerate second moments and to satisfy a Lindeberg condition (detailed assumptions are provided in Web Appendix \ref{supp:sec:theorem1}).  These are conditions on \(S_i\), so arbitrary within-cluster dependence is still allowed. The large-sample property below is taken with the number of clusters $M \rightarrow \infty$. Clusters are assumed to be independent of each other, and treatment is randomized at the cluster level with allocation probability $q$ bounded away from 0 and 1. Cluster sizes are allowed to be unequal and assumed to be random draws from an unknown population distribution. We assume that their first two moments exist, for example $M^{-1}\sum_{i=1}^{M} N_i \xrightarrow{p} \overline{N}$ and $M^{-1}\sum_{i=1}^{M} N_i^2  \xrightarrow{p} E(N_i^2)$, and that no cluster is asymptotically dominant such that $\max_i N_i / \sum_{i=1}^{M} N_i  \xrightarrow{p} 0$. These expressions permit us to write \(W_D = M^{-1}\overline{N}^{-2}(\mu_1-\mu_0) \) with its estimator \(\widehat{W}_D = (n_1n_0)^{-1}\sum_{i=1}^{M}A_iS_i\). The estimators \(\log(\widehat{W}_R)\) and \(\log(\widehat{W}_O)\) can then be obtained using the relationship in \eqref{eq:relation}. By considering $\widehat{W}_D$ as a two-sample U-statistic with kernel \(s_{ij,kl}\) and leveraging the Hoeffding-Hájek decomposition \citep{hajek1968asymptotic}, we first provide the asymptotic distribution of $\widehat{W}_D$ along with the variance expressions for the associated win statistics.
\begin{theorem}\label{theorem:u-var}
Under cluster randomization and the cluster-score regularity conditions described above, we have the following central limit theorem
\[
M^{3/2}\left(\widehat{W}_D-W_D\right)\ \xrightarrow{d}\ \mathcal N\!\left(0,\ \overline{N}^{-4}\left\{\frac{\sigma_1^2}{q}+\frac{\sigma_0^2}{1-q}\right\}\right).
\]
Consequently the asymptotic variances for $\widehat{W}_D,\log(\widehat{W}_R),\log(\widehat{W}_O)$ are 
\[
\sigma_D^2
= M^{-3}\overline{N}^{-4}\left(\frac{\sigma_1^2}{q}+\frac{\sigma_0^2}{1-q}\right),\quad
\sigma_R^2
= \left\{\frac{2\,\{1/(1-\pi_{\text{tie}})\}}{\,1-\left(\{1/(1-\pi_{\text{tie}})\}W_D\right)^2}\right\}^{\!2}\sigma_D^2,\quad
\sigma_O^2
= \frac{4\sigma_D^2}{\left(1-W_D^2\right)^{\!2}}.
\]
\end{theorem}
According to Theorem \ref{theorem:u-var}, the variance estimator $\widehat{\sigma}^2_\star$ for $\star\in\{D,R,O\}$ is obtained by replacing population quantities with their finite-sample analogues: $\overline{N}$ by $M^{-1}\sum_{i=1}^{M}N_i$, $\sigma_a^2$ by
\[
\widehat{\sigma}_a^2
= \bigl\{q^{a}(1-q)^{1-a}M -1\bigr\}^{-1}\sum_{i=1}^{M} \left\{ A_i^{a}(1-A_i)^{1-a}\,(S_i-\overline{S}_a)^2 \right\},
\]
where \(\overline{S}_a = \bigl\{q^{a}(1-q)^{1-a}M\bigr\}^{-1} \sum_{i=1}^{M} \left\{ A_i^a(1-A_i)^{(1-a)} S_i\right\} \), and $\pi_{\text{tie}}$ by $\widehat{\pi}_{\text{tie}}$. A detailed proof is provided in Web Appendix \ref{supp:sec:theorem1}. 
Theorem \ref{theorem:u-var} thus provides a principled basis for deriving testing procedures for $\Delta\in\{W_D,\log(W_R),\log(W_O)\}$. Consider the Wald-type test based on the studentized statistic \(Z_\star=\widehat{\Delta}/\widehat{\sigma}_\star\), where $\widehat{\sigma}_\star$ consistently estimates $\sigma_\star$. Under $H_0:\Delta=0$, we have $Z_\star \overset{d}{\to}\mathcal{N}(0,1)$. A closely related Wald test for $\log(W_R)$ has also been proposed by \citet{zhang2021inference}, based on two clustered U-statistics. In their framework, inference is built on a pairwise kernel that compares outcomes across treatment arms, which yields an estimator of \(\log (W_R)\) identical to the expression in \eqref{eq:relation}. But our asymptotic variance in Theorem \ref{theorem:u-var} is derived from a different kernel formulation. We show in Web Appendix \ref{supp:sec:u-stat} that our variance is asymptotically equivalent to that proposed by \citet{zhang2021inference}, but our tests demonstrate better finite-sample properties throughout (see Section \ref{sec:simu} for numerical comparisons). 

Assuming $\Delta$ is the effect size for $W_D$, $\log(W_R)$, or $\log(W_O)$, with a two-sided level-$\alpha$ test of $H_0:\Delta=0$ versus $H_1:\Delta\neq 0$, the power of our Wald test is
\begin{eqnarray}\label{eq:power_normal}
\mathrm{Power}
\approx \Phi\!\left({|\Delta|}/{\sigma_\star}-z_{1-\alpha/2}\right),
\end{eqnarray}
where $\Phi(\cdot)$ is the standard normal CDF and $z_q$ its $q$th quantile. Inverting the normal approximation gives the required sample size, which is the smallest \(M\) such that \( (z_{1-\alpha/2} + z_{1-\beta})^2\sigma_{\star}^2/\Delta^2 \leq 1 \). Considering that $M$ may be modest in some CRTs, one can also apply a small-sample adjustment by considering the $t$ reference distribution, leading to 
\begin{eqnarray}\label{eq:power_t}
\mathrm{Power}
  \approx \Phi_t\!\left({|\Delta|}/{\sigma_\star}-t_{M-2,\,1-\alpha/2}\right),
\end{eqnarray}
where 
$t_{M-2,\,q}$ is the $q$th quantile of the $t$ distribution with $M-2$ degrees of freedom, and $\Phi_t(\cdot)$ is the standard $t$ CDF. Inverting this gives the required sample size, which is the smallest $M$ such that \((t_{M-2,\,1-\alpha/2}+t_{M-2,\,1-\beta}\bigr)^{2}\,\sigma_\star^{2}/{\Delta^{2}}\;\le\;1\). Below, we provide explicit forms of $\sigma_\star^{2}$ for the cases $V=1$ and $V\ge 2$ to facilitate sample size determination.

\subsection{Single non-survival endpoints} \label{sec:single}

We first consider a single non-survival endpoint (continuous, ordinal, count), so that $V=1$ and $Y_{ij}$ is scalar. The pairwise sign kernel $s_{ij,kl}$ leads to a net score $\phi_{ij}$ that depends linearly on the mid-rank $R_{ij}$ of $Y_{ij}$, namely $\phi_{ij} = 2R_{ij} - (1+n)$ with  $R_{ij}=\sum_{(k,l)\neq(i,j)} \mathbb{I}\{{Y}_{ij} \succ {Y}_{kl}\}+ \frac{1}{2}\sum_{(k,l)\neq(i,j)} \mathbb{I}\{{Y}_{ij} = {Y}_{kl}\}+1$ (see Web Appendix \ref{supp:sec:theorem2}). This relationship allows the variance $\sigma_\star^2$ for $\star \in \{D,R,O\}$ to be expressed in rank statistics. Since the rank $R_{ij}$ is closely related to the empirical CDF through $\widehat{\mathcal F}^{*}(Y_{ij}) = R_{ij}/n$, we characterize the within-cluster dependence using the rank intraclass correlation (rank ICC), through the concept of mid-distribution score $\mathcal F^{*}(y) = \{\mathcal F(y) + \mathcal F(y^-)\}/2$. Because the score $\phi_{ij}$ is a function of the mid-rank, the relevant within-cluster dependence measure should be based on individuals' ranking positions in the pooled outcome distribution. The rank ICC quantifies this dependence by correlating the mid-distribution scores of two individuals from the same cluster. In contrast, a conventional Pearson ICC correlates the original outcome random variables and can critically depend on the scale of the outcome. The rank ICC therefore matches the information used by win statistics, which only depends on pairwise ordering and ties rather than on the original outcome variable. Formally, the rank ICC is given by
\begin{align} \label{eq:rank_icc}
    \rho = \mathrm{corr}\left\{\mathcal F^{*}(Y_{ij}), \mathcal F^{*}(Y_{ij'})\right\},  
\end{align}
where $\mathcal F$ is the marginal CDF of the clustered outcome and $\mathcal F(y^-)={\rm lim}_{t\uparrow y}\mathcal F(t)$ \citep{tu2023rank}. For a continuously distributed Gaussian outcome, let \(\rho_{\mathrm{P}}=\operatorname{corr}(Y_{ij},Y_{ij'})\) denote the Pearson ICC for two distinct individuals \(j\neq j'\) in the same cluster. The rank ICC in \eqref{eq:rank_icc} then satisfies \(\rho=(6/\pi)\arcsin(\rho_{\mathrm{P}}/2)\) \citep{pearson1907further}, whereas the Kendall's \(\tau\) satisfies \(\tau=(2/\pi)\arcsin(\rho_{\mathrm{P}})\) \citep{romdhani2014kendall}. Thus, \(\rho\approx 3\tau/2\) under weak association. For a binary outcome coded as \(0\) and \(1\), \(\mathcal F^{*}(Y)\) is a positive affine transformation of \(Y\), and therefore the rank ICC equals the Pearson ICC of the observed binary responses. This equality applies on the observed binary scale and can not be interpreted as equality with a latent Gaussian or tetrachoric correlation. Furthermore, the rank ICC and the intraclass odds ratio are alternative summaries of the same exchangeable \(2\times2\) joint distribution \citep{locatelli2016assessing}. For ordinal and count outcomes, the rank ICC can be calculated from the joint category probabilities and the corresponding mid-distribution scores. A Gaussian copula model can be used to specify those joint probabilities, but with discrete margins the resulting rank ICC depends on both the latent copula correlation and the marginal distributions. Therefore, the continuous Gaussian arcsine relationship does not generally carry over to discrete outcomes \citep{inouye2017review}.

In summary, the rank ICC provides a common rank scale measure of within-cluster dependence across ordered endpoint types and is invariant to strictly monotone transformations. We notice that the rank ICC $\rho$ in \eqref{eq:rank_icc} is defined through the pooled mid-distribution score $\mathcal F^{*}$ and is therefore a single dependence parameter for the entire trial population. This construction differs from certain model-based ICC measures in the existing literature, such as those defined from linear mixed models where the true ICCs can differ by treatment arms \citep{eldridge2009intra, candel2015sample}. Because win statistics in CRTs compare individuals exclusively across arms on the pooled rank scale, and clusters are nested within arms, no between-arm rank correlation parameter naturally arises, and the arm-specific dependence measure do not directly apply here. This is a special feature of the rank-based dependence measure. However, when within-cluster dependence is nevertheless expected to differ materially by arm (for example, when the intervention leads to between-arm heterogeneity), the arm-specific cluster-score variances $\sigma_a^2$ in Theorem \ref{theorem:u-var} will be affected and reflect this heterogeneity even though the rank ICC is defined globally as a single quantity. 
Finally, due to the connections between win statistics and ranks, the large-sample variances of win statistics are written in closed form in terms of the rank ICC $\rho$, the tie probability $\pi_{\text{tie}}$, and standard CRT design quantities, as we detail below. 

\begin{theorem}
\label{theorem:var_single}
Under regularity conditions in Web Appendix \ref{supp:sec:theorem2}, the asymptotic variances of \(\widehat{W}_D, \log(\widehat{W}_R), \log(\widehat{W}_O)\) for $V=1$ are $\sigma_\star^2=\nu_\star^2+o_p(M^{-1})$, $\star\in\{D,R,O\}$, where
\begin{align*}
\nu_{D}^2
&= \frac{1-\pi_{\text{tie}}^2}{3\,M\,\overline N}
\left(\frac{1}{q}+\frac{1}{1-q}\right)
\Big[1+\rho\{(1+\mathrm{CV}^2)\,\overline N-1\}\Big]
\;-\;
\frac{W_D^2}{M},\\[4pt]
\nu_R^2
&=\left\{\frac{2\{1/(1 - \pi_{\text{tie}})\}}{1-(\{1/(1 - \pi_{\text{tie}})\} W_D)^2}\right\}^{\!2}\,\nu_D^2,
\qquad
\nu_O^2
=\frac{4\nu_D^2}{\left(1-W_D^2\right)^{\!2}}\,,
\end{align*}
where $\mathrm{CV}=\sqrt{\operatorname{Var}(N_i)}/\overline N$ is the coefficient of variation for the cluster size distribution. 
\end{theorem}
Several remarks immediately follow from Theorem \ref{theorem:var_single}. First, the second subtraction term in $\nu_D^2$ reflects an explicit shift under $H_1$. Under contiguous alternatives with $W_D=O_p(M^{-1/2})$, it becomes $o_p(M^{-1})$ and is ignorable with a sufficiently large number of clusters; hence the leading variance term is $\Delta$-free to first order and often used for convenient sample size calculation. For finite $M$ and fixed noncontiguous alternatives, however, the subtraction term is $O_p(M^{-1})$ and non-negligible, especially with moderate effect sizes. Second, in the leading term of $\nu_{D}^2$, the rank ICC and the cluster size heterogeneity define the familiar variance inflation factor under cluster randomization \citep{rutterford2015methods, eldridge2006sample, vanbreukelen2007relative},
$$\mathrm{VIF}=1+\rho\{(1+\mathrm{CV}^2)\,\overline N-1\}.$$
Therefore, this leading term can be viewed as the individual-randomization variance scaled by $\mathrm{VIF}$. To see this, if $\rho=0$ (or $\overline N=1$ and $\text{CV}=0$), then $\mathrm{VIF}=1$ and the leading variance term of $\nu_D^2$ recovers the variance expression derived in \citet{yu2022sample} under individual randomization and $H_0$: \(\Delta = 0\) or contiguous alternative (Remark \ref{supp:rmk:irt} in Web Appendix \ref{supp:sec:theorem2}). This CV-based expression can be considered as a second moment approximation. It assumes that the cluster size distribution affects the leading variance through $\overline{N}$ and CV, rather than through a systematic association between cluster size and the rank-score distribution. 
The approximation is expected to perform well when the cluster size distribution is not extremely skewed and no small number of clusters contributes a large fraction of participants. It can be less reliable with highly right skewed cluster sizes, very small clusters, or a few dominant clusters, because two cluster size distributions with the same $\overline{N}$ and CV can then have different distributions. Third, Theorem \ref{theorem:var_single} clarifies the impact of different design parameters. For example, the multiplicative term $(1-\pi_{\text{tie}})$ quantifies information loss due to ties, and each $\nu_\star^2$ is minimized at $q=1/2$, indicating that balanced randomization yields optimal efficiency (Web Appendix \ref{supp:sec:optimalq}). Fourth, for continuous outcomes with small $\pi_{\text{tie}}$, $\nu_D^2$ is asymptotically equivalent to the variance of the clustered Mann--Whitney $U$ statistic under exchangeable within-cluster rank dependence and appropriate assumptions on the ICC under the null; see details in Web Appendix \ref{supp:sec:mann_withney} \citep{rosner1999use}, recovering classical rank-based sample size estimation as a special case. For binary outcomes, this connection is exact at the estimand level, as formalized in the following Remark.

\begin{remark}\label{remk:binary}
For binary outcomes, if $Y_{ij} \in \{0,1\}$, with 1 denoting the more favorable outcome and $p_a = P(Y_{ij} = 1 \mid A_i = a)$, then $\pi_{\text{win}} = p_1(1-p_0)$, $\pi_{\text{loss}} = p_0(1-p_1)$, and $\pi_{\text{tie}} = p_1 p_0 + (1-p_1)(1-p_0)$. Hence $W_D = p_1 - p_0$, the usual risk difference, and $W_R = p_1(1-p_0)/\{p_0(1-p_1)\}$, the usual odds ratio, and their estimators coincide with the sample risk difference and sample odds ratio estimators, respectively. The asymptotic variance $\sigma^2_{\star}$ for $\star \in \{D, R\}$ is asymptotically equivalent, under the null, to the traditional variance expressions for the risk difference and log odds ratio previously derived for CRTs \citep{cornfield1978symposium, donner1981randomization}. The proposed U-statistic formulation therefore includes the classical CRT sample size formulas for binary outcomes as special cases.
\end{remark}
The detailed derivations are provided in Web Appendix \ref{supp:sec:binary}. Building on these special cases, we next extend the variance expressions to composite survival endpoints in Section \ref{sec:composite}.

\begin{sidewaystable}[htbp]
\centering
\setlength{\tabcolsep}{5pt}
\renewcommand{\arraystretch}{1.15}
\caption{Smallest sample size (\(M\)) required for detecting target effect \(\Delta\) at two-sided level \(\alpha\) and power \(1-\beta\) for single non-survival and composite prioritized survival endpoints, based on Theorems \ref{theorem:var_single} and \ref{theorem:var_comp}. For one-sided tests, replace \(z_{1-\alpha/2}\) by \(z_{1-\alpha}\). For \(t\) distribution, replace \(z_{1-\alpha/2}\) by \(t_{M-2,1-\alpha/2}\). All expressions retain finite-effect (non-contiguous) terms.}
\label{tab:sample_size}

\resizebox{\linewidth}{!}{%
\begin{tabular}{ccc}
\toprule
\multicolumn{3}{c}{\textbf{Single non-survival endpoint}}\\
\addlinespace[4pt]
\textbf{\(\Delta\)} 
& \textbf{Asymptotic variance \(\nu_\star^2\)}
& \textbf{Required number of clusters \(M\)} \\ 
\midrule
\(W_D\) 
& \(\displaystyle
\frac{1-\pi_{\text{tie}}^2}{3\,M\,\overline N}
\left(\frac{1}{q}+\frac{1}{1-q}\right)
\mathrm{VIF}
\;-\;\frac{\Delta^2}{M}
\) 
& \(\displaystyle
\frac{(z_{1-\alpha/2}+z_{1-\beta})^2}{\Delta^2}
\left[
\frac{1-\pi_{\text{tie}}^2}{3\,\overline N}
\left(\frac{1}{q}+\frac{1}{1-q}\right)
\mathrm{VIF}
-\Delta^2
\right]
\) \\[14pt]

\(\log(W_R)\) 
& \(\displaystyle
\left[\frac{2\,\{1/(1 - \pi_{\text{tie}})\}}{1-\tanh^2(\Delta/2)}\right]^{\!2}
\left\{
\frac{1-\pi_{\text{tie}}^2}{3\,M\,\overline N}
\left(\frac{1}{q}+\frac{1}{1-q}\right)
\mathrm{VIF}
\;-\;\frac{\big[(1-\pi_{\text{tie}})\tanh(\Delta/2)\big]^2}{M}
\right\}
\) 
& \(\displaystyle
\frac{(z_{1-\alpha/2}+z_{1-\beta})^2}{\Delta^2}\;
\left[\frac{2\,\{1/(1 - \pi_{\text{tie}})\}}{1-\tanh^2(\Delta/2)}\right]^{\!2}
\left[
\frac{1-\pi_{\text{tie}}^2}{3\,\overline N}
\left(\frac{1}{q}+\frac{1}{1-q}\right)
\mathrm{VIF}
-\frac{\tanh^2(\Delta/2)}{\{1/(1 - \pi_{\text{tie}})\}^{2}}
\right]
\) \\[14pt]

\(\log(W_O)\)
& \(\displaystyle
\left[\frac{2}{1-\tanh^2(\Delta/2)}\right]^{\!2}
\left\{
\frac{1-\pi_{\text{tie}}^2}{3\,M\,\overline N}
\left(\frac{1}{q}+\frac{1}{1-q}\right)
\mathrm{VIF}
\;-\;\frac{\tanh^2(\Delta/2)}{M}
\right\}
\) 
& \(\displaystyle
\frac{(z_{1-\alpha/2}+z_{1-\beta})^2}{\Delta^2}\;
\left[\frac{2}{1-\tanh^2(\Delta/2)}\right]^{\!2}
\left[
\frac{1-\pi_{\text{tie}}^2}{3\,\overline N}
\left(\frac{1}{q}+\frac{1}{1-q}\right)
\mathrm{VIF}
-\tanh^2(\Delta/2)
\right]
\) \\[10pt]

\midrule
\multicolumn{3}{c}{\textbf{Composite endpoint}}\\
\addlinespace[4pt]
\textbf{\(\Delta\)} 
& \textbf{Asymptotic variance \(\widetilde{\nu}_\star^2\)}
& \textbf{Required number of clusters \(M\)} \\ 
\midrule

\(W_D\) 
& \(\displaystyle
\frac{1}{M^3\,\overline{N}^4}
\left( \frac{1}{q} + \frac{1}{1-q} \right)
\mathrm{VIF}^{*}\,
\mathcal{V}(M)
 - \frac{\Delta^2}{M}
\)
& \(\displaystyle
M_{\min} \;=\; \min\!\left\{ M :
\widetilde{\nu}_D^2(M) \;\le\; \frac{\Delta^2}{(z_{1-\alpha/2}+z_{1-\beta})^2} \right\}
\) \\[16pt]

\(\log(W_R)\)
& \(\displaystyle
\left[\frac{2\,\{1/(1 - \pi_{\text{tie}})\}}{1-\tanh^2(\Delta/2)}\right]^{\!2}
\Bigg[
\frac{1}{M^3\,\overline{N}^4}
\left( \frac{1}{q} + \frac{1}{1-q} \right)
\mathrm{VIF}^{*}\,
\mathcal{V}(M)
 - \frac{\big[(1-\pi_{\text{tie}})\tanh(\Delta/2)\big]^2}{M}
\Bigg]
\)
& \(\displaystyle
M_{\min} \;=\; \min\!\left\{ M :
\widetilde{\nu}_R^2(M) \;\le\; \frac{\Delta^2}{(z_{1-\alpha/2}+z_{1-\beta})^2} \right\}
\) \\[16pt]

\(\log(W_O)\)
& \(\displaystyle
\left[\frac{2}{1-\tanh^2(\Delta/2)}\right]^{\!2}
\Bigg[
\frac{1}{M^3\,\overline{N}^4}
\left( \frac{1}{q} + \frac{1}{1-q} \right)
\mathrm{VIF}^{*}\,
\mathcal{V}(M)
 - \frac{\tanh^2(\Delta/2)}{M}
\Bigg]
\)
& \(\displaystyle
M_{\min} \;=\; \min\!\left\{ M:
\widetilde{\nu}_O^2(M) \;\le\; \frac{\Delta^2}{(z_{1-\alpha/2}+z_{1-\beta})^2} \right\}
\) \\[10pt]

\bottomrule
\end{tabular}
}

\vspace{4pt}
\begin{minipage}{0.98\linewidth}
\footnotesize
\(\mathrm{VIF} := 1+\rho\{(1+\mathrm{CV}^2)\,\overline N-1\}\).
For the composite endpoint: 
\(P = 3p_W + \frac{5}{4}p_T\), 
\(Q = p_{WW} + p_{WT} + \frac{1}{4}p_{TT}\), 
\(\mathrm{VIF}^{*} = 1 + \{ (1+\mathrm{CV}^2)\overline{N} - 1 \}\rho^{*}\), and  
\(\mathcal{V}(M)= \left\{ 4\!\left[ 1+ (M\overline{N} - 1)P + (M\overline{N} - 1)(M\overline{N}-2)Q \right] - (M\overline{N}+1)^2 \right\},
\)
\end{minipage}
\end{sidewaystable}

\subsection{Composite survival endpoints} \label{sec:composite}
\begin{figure}[htbp]
    \centering
    \includegraphics[width=1\linewidth]{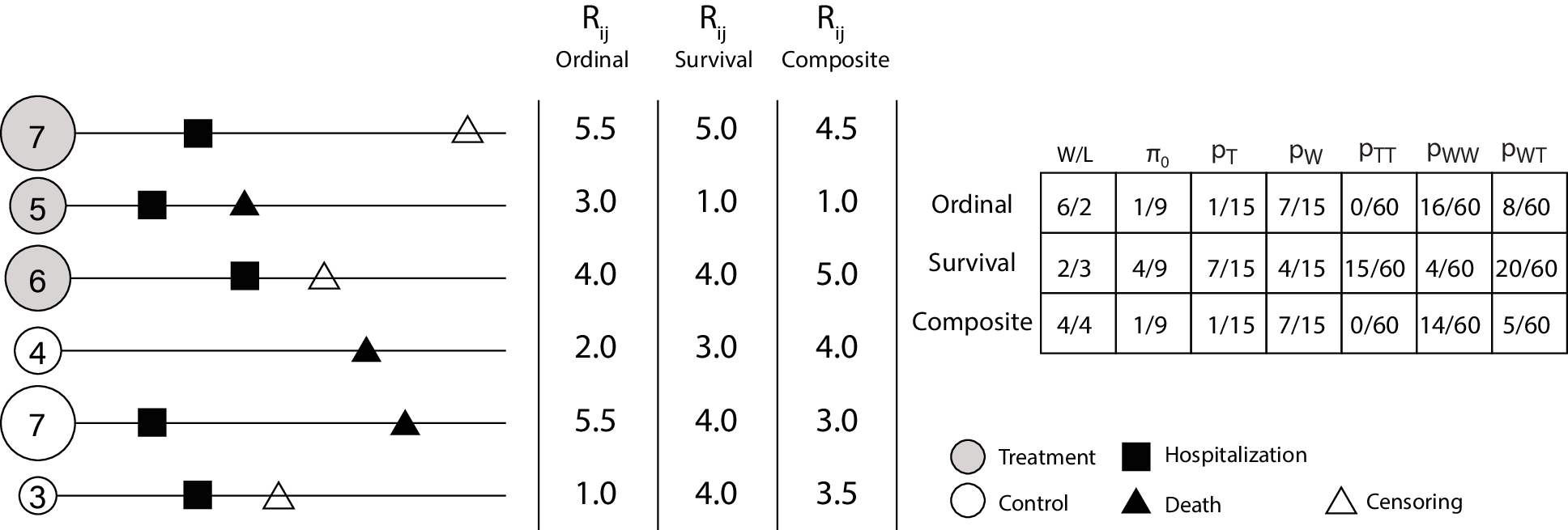}
    \caption{Illustration of endpoint-specific pairwise rules, mid-ranks, and pooled probabilities. From the top to the bottom, each row is one individual with composite outcome $\bm{Y}_i$, $i=1,\dots,6$. The number inside the circle is the ordinal outcome. The horizontal line is the time scale, symbols on the line represent time-to-event components. The mid-rank \(R_{ij}\) is computed from pairwise counts over all \(6\) individuals. For survival and the composite endpoint, pairwise comparison follows Gehan/FS rules.}
    \label{fig:toy_ex}
\end{figure}

In Section \ref{sec:single} with non-censored endpoints, the sign kernel \(s_{ij,kl}\) induces a complete ranking (a total reorder with mid-ranks for ties): every pair of individuals is comparable, and we can decide \(s_{ij,kl} \in \{-1,0,1\}\). Even with a single survival endpoint, pairwise rules produce a partial ranking under right censoring \citep{gehan1965generalized}. A strict win is declared only when the earlier observed time in a pair is an event, and the pair is unsolved if the earlier time is censored. In this case, \(s_{ij,kl}\) is concluded as a tie. Strict wins remain transitive, but the fraction of comparable pairs now depends on the joint outcome-censoring law and produce partial ranking. For the composite endpoint, we consider the comparison by a mapping score \(\mathcal{W}: \mathcal{Y}^V \rightarrow \mathbb{R}\), and thus  \(s_{ij,kl} =  \mathbb{I} \{ \mathcal{W}(\bm{Y}_{ij}) \succ \mathcal{W}(\bm{Y}_{kl}) \} - \mathbb{I}\{ \mathcal{W}(\bm{Y}_{ij}) \prec \mathcal{W}(\bm{Y}_{kl}) \} \) for \(V\geq 2\). If all endpoint components are fully observed and compared by a fixed hierarchy, following the lexicographic rule, a single fixed \(\mathcal{W}\) can produce a complete, transitive ranking. Under censoring, however, the hierarchy may reach an endpoint level where observations are not fully available for a given pair, which creates a partial ranking and leads to non-transitive patterns across different pairs. Figure \ref{fig:toy_ex} provides a concrete illustration; in that example, the pairwise comparisons are not transitive since \(\bm{Y}_{4} \succ \bm{Y}_{3}\), \(\bm{Y}_{3} \succ \bm{Y}_1\), but \(\bm{Y}_1 \succ \bm{Y}_4\). This non-transitivity is not a pathology of the win statistic itself, but a structural consequence of pair-specific censoring in the hierarchy, where the component that resolves one comparison may be censored in another, so the effective component of comparison varies across pairs.
Despite this caveat, 
for our testing procedures, we can still express $\phi_{{ij}} = 2R_{ij}^{*} - (N+1)$, where \(R_{ij}^{*} = \sum_{(k,l)\neq(i,j)} \mathbb{I}\!\left\{\mathcal{W}(\mathbf{Y}_{ij}) \succ \mathcal{W}(\mathbf{Y}_{kl})\right\}
\;+\; \tfrac{1}{2} \sum_{(k,l)\neq(i,j)} \mathbb{I}\!\left\{\mathcal{W}(\mathbf{Y}_{ij}) = \mathcal{W}(\mathbf{Y}_{kl})\right\}
\;+\; 1 \)
is the generalized definition of rank $R_{ij}$. In the special case $V=1$ and $\mathcal{W}$ is the identity function, $R_{ij}^{*}=R_{ij}$, which aligns with the regular mid-rank definition for a single non-survival endpoint. For composite endpoints with $V\geq 2$, the direct relationship between $R_{ij}^{*}$ and the mid-CDF no longer applies, and the rank-based variance form in Theorem \ref{theorem:var_single} no longer holds. Intuitively, when the comparison rule depends on both individuals' censoring realizations, the induced ordering is no longer generated by a single scalar, hence \(R_{ij}^{*}\) is not a monotone transform of one-dimensional marginal distribution. Instead, the final variance expression depends on second moments of comparisons sharing the same focal outcome, through the terms \(\sigma_1^2\), and \(\sigma_0^2\) in Theorem \ref{theorem:u-var}, which involve the second moment \(\mathbb{E}(S_i^2\mid A_i=a)\). Unlike $V=1$ where this structure can collapse to the marginal tie rate, these second moments can not be determined by the marginal tie rate or win/loss rate with $V\geq 2$. 

Therefore, we introduce agnostic probabilities that capture how wins and ties co-occur across comparisons sharing the same focal individual. Specifically, let
$p_W = \mathbb{P}(\bm{Y}_{ij} \succ \bm{Y}_{kl})$, $p_T = \mathbb{P}(\bm{Y}_{ij} = \bm{Y}_{kl})$, and define triplet probabilities
\begin{align*}
p_{WW} &= \mathbb{P}(\bm{Y}_{ij} \succ \bm{Y}_{kl},\, \bm{Y}_{ij} \succ \bm{Y}_{mn}),\quad
p_{WT} = \mathbb{P}(\bm{Y}_{ij} \succ \bm{Y}_{kl},\, \bm{Y}_{ij} = \bm{Y}_{mn}), \\
p_{TT} &= \mathbb{P}(\bm{Y}_{ij} = \bm{Y}_{kl},\, \bm{Y}_{ij} = \bm{Y}_{mn}),   \quad (i,j)\neq(k,l)\neq (m,n).
\end{align*}
These pairwise and triplet probabilities characterize the joint behavior of two comparisons that share the same focal outcome and are required for the variance derivation. To define the generalized rank ICC, let \(s(\bm{y},\bm{y}')=\mathbb{I}\{\bm{y}\succ\bm{y}'\}-\mathbb{I}\{\bm{y}\prec\bm{y}'\}\in\{-1,0,1\}\) denote the signed pairwise comparison kernel, where \(s(\bm{y},\bm{y}')=1\), \(-1\), or \(0\) indicates that \(\bm{y}\) is a win, loss, or tie, respectively, relative to \(\bm{y}'\) under the prespecified comparison rule. Let \(\bm{Y}'\) be an independent draw from the pooled participant-level marginal distribution of \(\bm{Y}\), independent of cluster \(i\), and define \(\mathcal{F}^{**}(\bm{Y}_{ij})=\mathbb{P}\{s(\bm{Y}_{ij},\bm{Y}')=1\mid\bm{Y}_{ij}\}+\frac{1}{2}\mathbb{P}\{s(\bm{Y}_{ij},\bm{Y}')=0\mid\bm{Y}_{ij}\}\). Thus, \(\mathcal{F}^{**}(\bm{Y}_{ij})\in[0,1]\) is the conditional probability of superiority score of \(\bm{Y}_{ij}\) against a random comparator, with a tie receiving one-half credit. Equivalently, \(2\mathcal{F}^{**}(\bm{Y}_{ij})-1=\mathbb{E}\{s(\bm{Y}_{ij},\bm{Y}')\mid\bm{Y}_{ij}\}\), and therefore \(2\{\mathcal{F}^{**}(\bm{Y}_{ij})-\mathbb{E}[\mathcal{F}^{**}(\bm{Y}_{ij})]\}\) is the centered first order H\'ajek projection of the signed comparison kernel. For two distinct individuals \(j\neq j'\) in cluster \(i\), define the generalized rank ICC as \(\rho^{*}=\operatorname{corr}\{\mathcal F^{**}(\bm{Y}_{ij}),\mathcal F^{**}(\bm{Y}_{ij'})\}\), provided that \(\operatorname{Var}\{\mathcal{F}^{**}(\bm{Y}_{ij})\}>0\). For a single fully observed scalar endpoint, with larger values taken to be more favorable, \(\mathcal{F}^{**}(Y)=\mathbb{P}(Y'<Y\mid Y)+\frac{1}{2}\mathbb{P}(Y'=Y\mid Y)=\mathcal{F}^{*}(Y)\), as in \eqref{eq:rank_icc}. Hence, \(\rho^{*}=\operatorname{corr}\{\mathcal{F}^{*}(Y_{ij}),\mathcal{F}^{*}(Y_{ij'})\}=\rho\) degenerates to the rank ICC defined in \eqref{eq:rank_icc}. For censored or composite endpoints, the same definition applies through the prespecified comparison rule, with unresolved comparisons treated as ties. Thus, \(\rho^{*}\) measures the within-cluster correlation between conditional probability of superiority scores under the endpoint specific comparison rule. Under these definitions, we obtain the following generalized variance expressions for win statistics.
\begin{theorem} \label{theorem:var_comp}
Under regularity conditions in Web Appendix \ref{supp:sec:theorem3}, the asymptotic variances of \(\widehat{W}_D, \log(\widehat{W}_R), \log(\widehat{W}_O)\) for an arbitrary integer $V\geq 2$ are $\sigma_\star^2=\widetilde\nu_\star^2+o_p(M^{-1})$, $\star\in\{D,R,O\}$, where
\allowdisplaybreaks
\begin{align*}
     \widetilde\nu_D^2 &=  \frac{1}{M^3\overline{N}^4}  \left( \frac{1}{q} + \frac{1}{1-q} \right) \mathrm{VIF}^{*} \\
     & \times \left\{ 4\left[ 1+ (M\overline{N} - 1)P + (M\overline{N} - 1)(M\overline{N}-2)Q \right] - (M\overline{N}+1)^2  \right\} -   \frac{W_D^2}{M},\\
\widetilde \nu_R^2
&=\left\{\frac{2\{1/(1 - \pi_{\text{tie}})\}}{1-(\{1/(1 - \pi_{\text{tie}})\} W_D)^2}\right\}^{\!2}\,\widetilde{\nu}_D^2,
\qquad
\widetilde \nu_O^2
=\left\{\frac{2}{1-W_D^2}\right\}^{\!2}\,\widetilde{\nu}_D^2,
\end{align*}
where \(P = 3p_W + \frac{5}{4}p_T\), \(Q = p_{WW} + p_{WT} + \frac{1}{4}p_{TT}\), and  \(\mathrm{VIF}^{*} = 1 + \left\{ \left( 1 + \ \text{CV}^2 \right)\overline{N} -1  \right\} \rho^{*}  \).
\end{theorem}

Theorem \ref{theorem:var_comp} extends Theorem \ref{theorem:var_single} by showing that, for composite endpoints, the variance depends on the pooled comparison structure summarized by \(P\) and \(Q\). The quantity \(P\) is determined by the marginal win and tie probabilities, whereas \(Q\) captures second moment information about comparisons sharing the same focal outcome that cannot generally be recovered from those marginal probabilities alone. Under the independent comparator construction used above, \(Q=\mathbb{E}\{\mathcal{F}^{**}(\bm{Y})^{2}\}\) and \(\mathbb{E}\{\mathcal{F}^{**}(\bm{Y})\}=p_W+\frac{1}{2}p_T\). Thus, \(Q-\{p_W+\frac{1}{2}p_T\}^{2}=\operatorname{Var}\{\mathcal{F}^{**}(\bm{Y})\}\geq 0\), with equality if and only if \(\mathcal{F}^{**}(\bm{Y})\) is almost surely constant. Thus, \(Q\) measures heterogeneity across outcome profiles in their conditional probability of superiority scores, where greater heterogeneity increases the sampling variance. Similar in Theorem \ref{theorem:var_single}, when \(\rho^{*}=0\) , the variance expressions reduce to the variance under individual randomization and extend the methods in \citet{yu2022sample} to cluster randomization designs (Web Appendix \ref{supp:sec:theorem3_irt}).


\begin{remark}\label{remk:equiv}
\emph{Theorem \ref{theorem:var_comp} includes Theorem \ref{theorem:var_single} as a special case when the composite reduces to a regular single-endpoint structure. That is, the expressions in Theorem \ref{theorem:var_comp} are asymptotically equivalent to the form in Theorem \ref{theorem:var_single} if and only if: (a) the pairwise comparison rule induces a total preorder on $\mathbf{Y}_{ij}$ that partitions individuals into $G$ disjoint tie groups indexed by $g=1,\dots,G$ with sizes $t_g$, and ties are transitive, namely, if $\mathbf{Y}_{ij}=\mathbf{Y}_{kl}$ and $\mathbf{Y}_{kl}=\mathbf{Y}_{mn}$, then $\mathbf{Y}_{ij}=\mathbf{Y}_{mn}$. (b) Individuals are exchangeable with respect to the joint distribution of $\{\mathbf{Y}_{ij},i=1,\dots,M, j=1,\dots, N_i\}$, so the distributions of pairwise and triplet comparisons are invariant under permutations.}
\end{remark}

The proofs are provided in Web Appendix \ref{supp:sec:theorem2_equiv_3}. In Theorem \ref{theorem:var_comp}, non-transitivity and dependence between comparisons sharing the same focal outcome enter the variance through $Q$, and this is precisely where composite endpoints depart from standard single-endpoint rank behavior. For sample size determination, the primitives $(p_W,p_T,p_{WW},p_{WT},p_{TT})$ can be specified from a generative model (see Section \ref{sec:practice}). Practically, composite endpoints tend to increase required sample size through two sources: unresolved comparison rates (raising \(P\)), and greater heterogeneity in individual win probabilities (raising Q), with both sources often amplified by heavy censoring. The terms $(P,Q)$ are computed from the pooled distribution, which shifts under \(\Delta \ne 0 \) and becomes \(\Delta\)-dependent. Based on the variance expressions we derived, the power and sample size can then be generated based on \eqref{eq:power_normal} and \eqref{eq:power_t}. For reference, Table \ref{tab:sample_size} summarizes the sample size for single non-survival and composite survival endpoints.

\section{Implementation considerations} \label{sec:practice}
To operationalize our sample size approach, we recommend the following workflow. First, specify the clinical win rule and the target effect scale, $W_D$, $\log(W_R)$, or $\log(W_O)$. Second, specify the target estimand described in Section \ref{sec:win_stat} and the anticipated cluster size characteristics, summarized by $\overline{N}$ and CV, or by the planned cluster sizes when these are known. Third, estimate or elicit the effect size $\Delta$, tie probability $\pi_{\text{tie}}$, and rank ICC $\rho$ or $\rho^{*}$. When pilot or historical CRT data are available, these quantities should be computed using the same win rule intended for the new trial. When only individually randomized data are available, the effect size and tie probability can be estimated from cross-arm pairwise comparisons, but the rank ICC and cluster size distribution must be borrowed from comparable CRTs, registries, or expert elicitation (otherwise sensitivity analysis across a range of values is recommended). For prioritized composite survival endpoints, the additional quantities $(p_W, p_T, p_{WW}, p_{WT}, p_{TT})$ can be obtained either from a pilot CRT or from a generative event time and censoring model. Finally, because several inputs are uncertain at the design stage, the sample size calculation should be repeated over plausible ranges of $\Delta$, $\pi_{\text{tie}}$, $\rho$ or $\rho^{*}$, $\overline{N}$, and CV. Underestimating $\rho$, $\rho^{*}$, $\pi_{\text{tie}}$, or the extent of censoring-induced unresolved comparisons can lead to underpowered trials, whereas using plausible upper-range values for these parameters generally yields a larger, more conservative sample size. 

To supply these design inputs, we describe two strategies to apply Theorems \ref{theorem:var_single} and \ref{theorem:var_comp}. The first approach uses direct specification, where key quantities such as the rank ICC, the effect size, and the tie probability are directly specified. For composite survival endpoints, additional quantities including the marginal win and tie probabilities \(p_W\) and \(p_T\), as well as triple comparison probabilities \(p_{WW}\), \(p_{WT}\), and \(p_{TT}\), must also be supplied. These values may be informed by historical studies, pilot data, or expert clinical judgment. Direct specification of $\pi_{\text{tie}}$ deserves particular care. For time-to-event endpoints, the tie probability is not a free-standing quantity but is jointly induced by the event rates, the censoring distribution, the follow-up duration, and the correlation among component endpoints \citep{mou2025correlation}. Consequently, lower-than-anticipated event rates or heavier censoring inflate $\pi_{\text{tie}}$ and dilute the number of decisive pairwise comparisons, which in turn increases the required number of clusters. Rather than positing $\pi_{\text{tie}}$ as a single value, one can generate $\pi_{\text{tie}}$ and the triplet probabilities from these interpretable primitives, either through the refined formulas of \citet{mou2025correlation} or through the generative model described below, and then repeat the calculation over plausible event-rate and censoring scenarios; practical guidance on specifying win-measure design inputs under individual randomization is also available elsewhere \citep{yu2022sample, barnhart2025sample}. Once specified, they are substituted into the analytic formulas to obtain the variance, and subsequently, the required number of clusters for a desired power. 

A second approach is to derive the relevant quantities from a generative model that captures the joint distribution of prioritized events and censoring (but doing so without explicit simulation-based empirical power calculation). For illustration, we consider a canonical, semi-competing risks setting in which each individual may experience both a terminal event (e.g., death) and a non-terminal event (e.g., hospitalization). Let \(T^D_{ij}\) and \(T^H_{ij}\) denote the death and hospitalization times for individual \((i,j)\) in cluster \(i\). To model within-cluster dependence, we specify cause-specific hazard functions with a shared gamma frailty \citep{meng2023simulating}
\begin{equation} \label{eq:hazard_model}
\lambda^H_{ij} = \gamma_i \lambda^H_0 \exp(-\eta_H A_i), \quad
\lambda^D_{ij} = \gamma_i \lambda^D_0 \exp(-\eta_D A_i),
\end{equation}
where \(\lambda^H_0\) and \(\lambda^D_0\) are baseline hazards, \(\eta_H\) and \(\eta_D\) denote log hazard ratios for hospitalization and death for treatment  \(A_i \in \{0,1\} \) in cluster \(i\). The joint distribution of \((T^H_{ij}, T^D_{ij})\) can be specified through a Gumbel-Hougaard copula:
\begin{equation} \label{eq:copula}
    P(T^H_{ij} > t^H,\; T^D_{ij} > t^D \mid A_i, \gamma) =
\exp\left\{ -\left[ (\lambda^H_{ij} t^H)^\phi + (\lambda^D_{ij} t^D)^\phi \right]^{1/\phi} \right\}, 
\end{equation}
where \(\phi \geq 1\) is the copula dependence parameter. A larger value of \(\phi\) corresponds to stronger positive dependence between the different types of event times, which in turn affects the likelihood of discordant comparisons. We specify the hazard function for censoring as \(\lambda_{ij}^C(A_i)=\lambda_0^C e^{-\eta_C A_i}\), which is independent of \((T_{ij}^H,T_{ij}^D)\) conditional on treatment. Pairwise comparisons under prioritized outcomes are conducted according to the hierarchical win rule for \(\bm{Y_{ij}} = \{T_{ij}^H, T_{ij}^D, T_{ij}^C  \} \), and \(T^C_{ij}\) denotes censoring times as
\begin{align} \label{win_rule}
\mathbb{I}\{\mathcal{W}(\bm{Y}_{ij}),\mathcal{W}(\bm{Y}_{kl})\} = &\; \mathbb{I}\left\{T^D_{kl} < \min(T^D_{ij}, T^C_{ij}, T^C_{kl}) \right\} \nonumber \\
&+ \mathbb{I}\left\{\min(T^D_{ij}, T^D_{kl}) > \min(T^C_{ij}, T^C_{kl}),\; T^H_{kl} < \min(T^H_{ij}, T^C_{ij}, T^C_{kl}) \right\},
\end{align}
Individual \((i,j)\) is considered to win over \((k,l)\) if they experience death later; if neither individual dies prior to censoring, the one who is hospitalized later is declared the winner. Using this rule, the population win ratio comparing intervention versus control is given as:
\[
W_R
=\frac{\displaystyle \int_{0}^{\infty}\!\!\int_{0}^{\infty}\!\!\left\{O_{ij,kl}\!\left(1,0\right)+H_{ij,kl}\!\left(1,0\right)\right\}
f_\gamma\!\left(\gamma_{i}\right)f_\gamma\!\left(\gamma_{k}\right)\,d\gamma_{i}\,d\gamma_{k}}
{\displaystyle \int_{0}^{\infty}\!\!\int_{0}^{\infty}\!\!\left\{O_{ij,kl}\!\left(0,1\right)+H_{ij,kl}\!\left(0,1\right)\right\}
f_\gamma\!\left(\gamma_{i}\right)f_\gamma\!\left(\gamma_{k}\right)\,d\gamma_{i}\,d\gamma_{k}},
\]
where \(f_\gamma(.)\) is the probability density function of frailty \(\gamma_i\), and
\begin{align*}
O_{ij,kl}\!\left(a_i,a_k\right)
&=\frac{\lambda_{kl}^D\!\left(a_k\right)}
{\lambda_{ij}^D\!\left(a_i\right)+\lambda_{kl}^D\!\left(a_k\right)+\lambda_{ij}^C\!\left(a_i\right)+\lambda_{kl}^C\!\left(a_k\right)}\\
H_{ij,kl}\!\left(a_i,a_k\right)
&=\int_{0}^{\infty}\!\!\int_{0}^{t_2}
h_{kl}^H\!\left(t_1,t_2\mid a_k\right)
S_{ij}^{HD}\!\left(t_1,t_2\mid a_i\right)S_{kl}^{HD}\!\left(t_1,t_2\mid a_k\right)
S_{ij}^C\!\left(t_2\mid a_i\right)S_{kl}^C\!\left(t_2\mid a_k\right)\,dt_1\,dt_2,
\end{align*}
and furthermore, 
\begin{align*}
S_{ij}^{HD}\!\left(t_1,t_2\mid a_i\right)
&=\mathbb{P}\!\left(T_{ij}^H>t_1,\,T_{ij}^D>t_2 \mid A_i=a_i\right)
=\exp\!\left(-\left[\left\{\lambda_{ij}^H(a_i)t_1\right\}^{\phi}+\left\{\lambda_{ij}^D(a_i)t_2\right\}^{\phi}\right]^{1/\phi}\right),\\
h_{ij}^H\!\left(t_1,t_2\mid a_i\right)
&=\mathbb{P}\!\left(T_{ij}^H = t_1\mid T_{ij}^H>t_1,\,T_{ij}^D>t_2,\,A_i=a_i\right)=\lambda_{ij}^H(a_i)\!\left\{\lambda_{ij}^H(a_i)t_1\right\}^{\phi-1}
\!\left[\left\{\lambda_{ij}^H(a_i)t_1\right\}^{\phi}+\left\{\lambda_{ij}^D(a_i)t_2\right\}^{\phi}\right]^{1/\phi-1},\\
S_{ij}^C\!\left(t\mid a_i\right)&=\exp\!\left\{-\lambda_{ij}^C(a_i)t\right\},
\end{align*}
Under the model in \eqref{eq:hazard_model}--\eqref{eq:copula}, the cross-arm probability that an intervention individual loses to a control individual can be decomposed into two mutually exclusive pathways: an ``early death'' component \(O_{ij,kl}\), corresponding to the event that the control dies first before either subject is censored, and a ``deferred-to-hospitalization'' component \(H_{ij,kl}\), corresponding to the case where neither death is observed before censoring and the hierarchy is resolved at hospitalization. The win and tie probabilities (i.e., \(p_W\) and \(p_T\)), as well as their triplet pairwise comparison can be estimated via Monte Carlo integration under the above generative model. The details are provided in the Web Appendix \ref{supp:sec:generating_composite}

Each model component plays a distinct role in shaping the comparison structure. 
The frailty \(\gamma\) governs within-cluster dependence and directly contributes to the generalized rank ICC \(\rho^{*}\). If \(\gamma \sim \text{Gamma}(\text{shape}=\nu,\ \text{rate}=\nu)\), then the between-individual Kendall's \(\tau\) for the same event type is given by \(\tau = 1/(2\nu+1)\) \citep{balan2020tutorial}. The copula parameter \(\phi\) modulates individual-level dependence between the two event types, influencing both the frequency of ties and the joint comparison probabilities \(p_{WW}, p_{WT}, \text{ and } p_{TT}\). 
It has a direct relationship with Kendall's as: \(\tau = 1-1/\phi\) \citep{hougaard2000analysis}. The marginal hazard parameters \(\lambda^H_0, \lambda^D_0\) and treatment effects \(\eta_H, \eta_D\) shape the overall event incidence and the expected effect size \(\Delta\). Importantly, the censoring distribution substantially influences \(p_T\), \(p_{TT}\), and  \(\pi_{\text{tie}}\), as ties often arise when both individuals are censored before either event occurs. Estimation of \(\rho^{*}\) under composite endpoints is complicated by the non-transitive nature of the comparison rule and the presence of censoring. In practice, one may approximate \(\rho^{*}\) by computing the empirical intracluster correlation of individual-level score functions under simulated data (details are provided in Web Appendix \ref{supp:sec:generating_composite}). 

While we illustrate this approach using a Gumbel--Hougaard model for semi-competing risks, the same framework extends naturally to more complex generative models, including those incorporating recurrent events and multi-state transitions. In such settings, the key steps, such as specifying the event-time distribution, defining a priority-based comparison rule, estimating the induced comparison probabilities, and evaluating the corresponding variance, remain applicable with straightforward modifications. Because the variance inflation factor is linear in the rank ICC and scales with the average cluster size, the accuracy of the specified $\rho$ or $\rho^{*}$ is often the single most consequential planning input. Unlike conventional Pearson ICCs, however, empirical estimates of rank ICCs under clinically relevant win rules are rarely reported. We therefore encourage investigators to report estimated rank ICCs from completed CRTs under the intended win rule, as illustrated in Section \ref{sec:real_data} and implemented in the output of the accompanying R package, so that an empirical basis accumulates for future trial planning, in the same spirit as the CONSORT extension recommendation to report ICC estimates in CRTs \citep{campbell2012consort}.

\section{Simulation Study} 
\label{sec:simu}

We conducted simulation studies to validate the proposed sample-size and power formulas for CRTs with win statistics, and, in the meantime, to examine the finite-sample performance of the proposed testing procedures. The primary objective was to determine whether the empirical power observed in finite samples agrees with the power predicted by Theorems \ref{theorem:var_single} and \ref{theorem:var_comp}, under both equal and unequal cluster sizes. To complement this, we also evaluated empirical type I error under the null hypothesis, with attention to small numbers of clusters. We considered two outcome scenarios: (i) a scalar ordinal endpoint and (ii) prioritized composite survival endpoints. Under each scenario, clusters were randomized with a \(1:1\) ratio to treatment or control, and hypothesis tests were conducted at the nominal two-sided significance level \(\alpha=0.05\). We focused primarily on \(\Delta=\log(W_R)\) here. Results for the alternative win measures, \(\Delta=W_D\) and \(\Delta=\log(W_O)\), are reported in Web Appendix \ref{supp:sec:additional_simulation}. We evaluated the power trajectory over a grid of \(M\), which allows us to compare the predicted and empirical power curves across underpowered, moderately powered, and well-powered designs. For power evaluation, we used \(M\in\{30,40,50,75,100\}\). For type I error evaluation, we used \(M\in\{12,20,30,50,75,100\}\), so that calibration could also be assessed when the number of clusters is very small. We considered four cluster size designs: fixed \(N_i=30\), \(N_i\sim\mathrm{Uniform}\{10,\ldots,50\}\), fixed \(N_i=50\), and \(N_i\sim\mathrm{Uniform}\{10,\ldots,90\}\). These correspond approximately to \((\overline N,\mathrm{CV})=(30,0)\), \((30,0.39)\), \((50,0)\), and \((50,0.47)\), respectively. Each empirical power or type I error estimate was based on \(2{,}000\) Monte Carlo replicates per configuration.

For each simulated dataset, we computed the three win measures \(\Delta\in\{W_D,\log(W_R),\log(W_O)\}\), together with the replication-specific cluster-score standard error from Theorem \ref{theorem:u-var}. We evaluated both the Wald \(z\)-test in \eqref{eq:power_normal} and the Wald \(t\)-test in \eqref{eq:power_t}, where the latter used \(M-2\) degrees of freedom. The \(z\)-test represents the large-sample normal approximation, whereas the \(t\)-test is intended as a finite-sample reference distribution when the number of independent clusters is small. For \(B\) Monte Carlo replications, the average estimated standard error (ASE) is the average of the \(B\) replication-specific standard-error estimates from Theorem \ref{theorem:u-var}. In contrast, the design-based standard error (DSE) is the standard error implied by the closed-form design-stage variance formula and is calculated once for each design configuration. Specifically, the DSE is \(\sqrt{\nu_\star^2}\) from Theorem \ref{theorem:var_single} for scalar ordinal outcomes and \(\sqrt{\widetilde{\nu}_\star^2}\) from Theorem \ref{theorem:var_comp} for composite survival outcomes, after substituting the true design inputs from the corresponding data-generating mechanism. Thus, the ASE summarizes the finite-sample behavior of the standard-error estimator used for inference, whereas the DSE is the design-stage quantity used to calculate formula predicted power. The required inputs include the effect size \(\Delta\), the tie probability \(\pi_{\mathrm{tie}}\), the rank ICC \(\rho\) for scalar ordinal outcomes, and the generalized rank ICC \(\rho^*\), together with the pairwise and triplet probabilities \((p_W,p_T,p_{WW},p_{WT},p_{TT})\), for composite survival outcomes. For composite survival endpoints, these design inputs were calculated from large Monte Carlo samples following the Gehan-type comparison rule. The power prediction error is defined as empirical power minus formula-predicted power and is reported in \%.

\subsection{Scalar ordinal outcomes} 
\label{simu:ordinal}

For the scalar ordinal endpoint, we generated \(Y_{ij}\in\{1,\ldots,6\}\) from a proportional-odds mixed model with a cluster-level random intercept,
\[
\Pr(Y_{ij}\le k\mid A_i,b_i)=\operatorname{logit}^{-1}(\theta_k-\beta A_i-b_i),
\qquad k=1,\ldots,5,
\]
where \(A_i\) is the cluster-level treatment indicator and \(b_i\sim \mathcal N(0,\sigma_b^2)\) induces within-cluster dependence. The cutpoints \(\{\theta_k\}\) were calibrated so that the marginal control-arm distribution matched the prespecified probability vector \(\pi_0=\{0.217,\,0.093,\,0.173,\,0.241,\,0.036,\,0.241\}\). The latent-scale ICC was set to \(0.05\) or \(0.10\), with \(\sigma_b^2\) chosen to satisfy \(\mathrm{ICC}=\frac{\sigma_b^2}{\sigma_b^2+\pi^2/3}\). For power simulations, we used \(\beta\in\{\log(1.5),\log(2.0)\}\), corresponding to moderate and larger treatment effects. For type I error simulations, we set \(\beta=0\), so that \(\Delta=0\) for \(W_D\), \(\log(W_R)\), and \(\log(W_O)\). The true values of the win measures and all auxiliary quantities entering Theorem \ref{theorem:var_single} were computed from the corresponding ordinal data-generating model and then held fixed for the design-stage power calculation. The computational details for the ordinal truth calculation and the empirical estimation of the rank ICC are provided in Web Appendix \ref{supp:sec:ordinal_est}.

\subsection{Composite survival outcomes}
\label{simu:composite}

We next considered prioritized composite survival endpoints in CRTs under the same cluster size generating mechanism as in Section \ref{simu:ordinal}. The composite endpoint consisted of death as the first-priority component and hospitalization as the second-priority component. Within-cluster dependence in event times was induced by a shared multiplicative frailty \(\gamma_i\sim\mathrm{Gamma}(7.5,7.5)\), corresponding to Kendall’s \(\tau=0.0625\). Conditional on \((A_i,\gamma_i)\), the cause-specific hazards for hospitalization and death followed the semi-competing risks model in \eqref{eq:hazard_model} with constant baselines \(\lambda_0^H=0.10\) and \(\lambda_0^D=0.08\), and the joint law of \((T^H_{ij},T^D_{ij})\) was coupled via a Gumbel--Hougaard copula as in \eqref{eq:copula}. Independent right censoring times were generated as \(T^C_{ij}\sim\mathrm{Exp}(\lambda^C_{ij})\), with \(\lambda^C_{ij}=\lambda_0^C\exp(-\eta_C A_i)\), \(\lambda_0^C=0.03\), and \(\eta_C=0.15\). For power simulations, we varied the treatment effects on death and hospitalization, \(\eta_D\in\{0.3,0.5\}\) and \(\eta_H\in\{0.3,0.5\}\), and the within-individual association parameter \(\phi\in\{1,3\}\), corresponding to Kendall's \(\tau\in\{0,2/3\}\). Defining the marginal censoring rate as \(\Pr(T^C_{ij}<T^D_{ij})\), these settings lead to approximately \(29.2\%\) censoring in the control arm and \(32.3\%\) or \(36.7\%\) censoring in the treatment arm for \(\eta_D=0.3\) or \(0.5\), respectively. Under equal allocation, the corresponding overall censoring rates were approximately \(30.8\%\) and \(33.0\%\).  For type I error, we included the null \(\eta_H=\eta_D=\eta_C=0\), which implies \(\Delta=0\) under each win measure.

Each individual could contribute a hospitalization record if hospitalization occurred before death or censoring, and also contributed a terminal or censoring record at the minimum of death and censoring times. We formed prioritized win comparisons following Gehan's rule \citep{gehan1965generalized}: the pair was first compared on death, and the comparison descended to hospitalization only when death did not resolve the pair before censoring. We then evaluated \(\Delta\in\{W_D,\log(W_R),\log(W_O)\}\), estimating empirical power and type I error using the proposed test and comparing empirical power to the analytic prediction computed from the true design inputs. Under these simulation settings, we also benchmarked our \(\log(W_R)\) testing procedure against the Wald-type method of \citet{zhang2021inference} following clustered U-statistics, implemented through the \texttt{cWR} package. This comparison was restricted to type I error for \(\log(W_R)\) only.

\subsection{Simulation results} \label{sec:sim:results}

\begin{figure}[!htbp]
\centering
\includegraphics[width=0.98\linewidth]{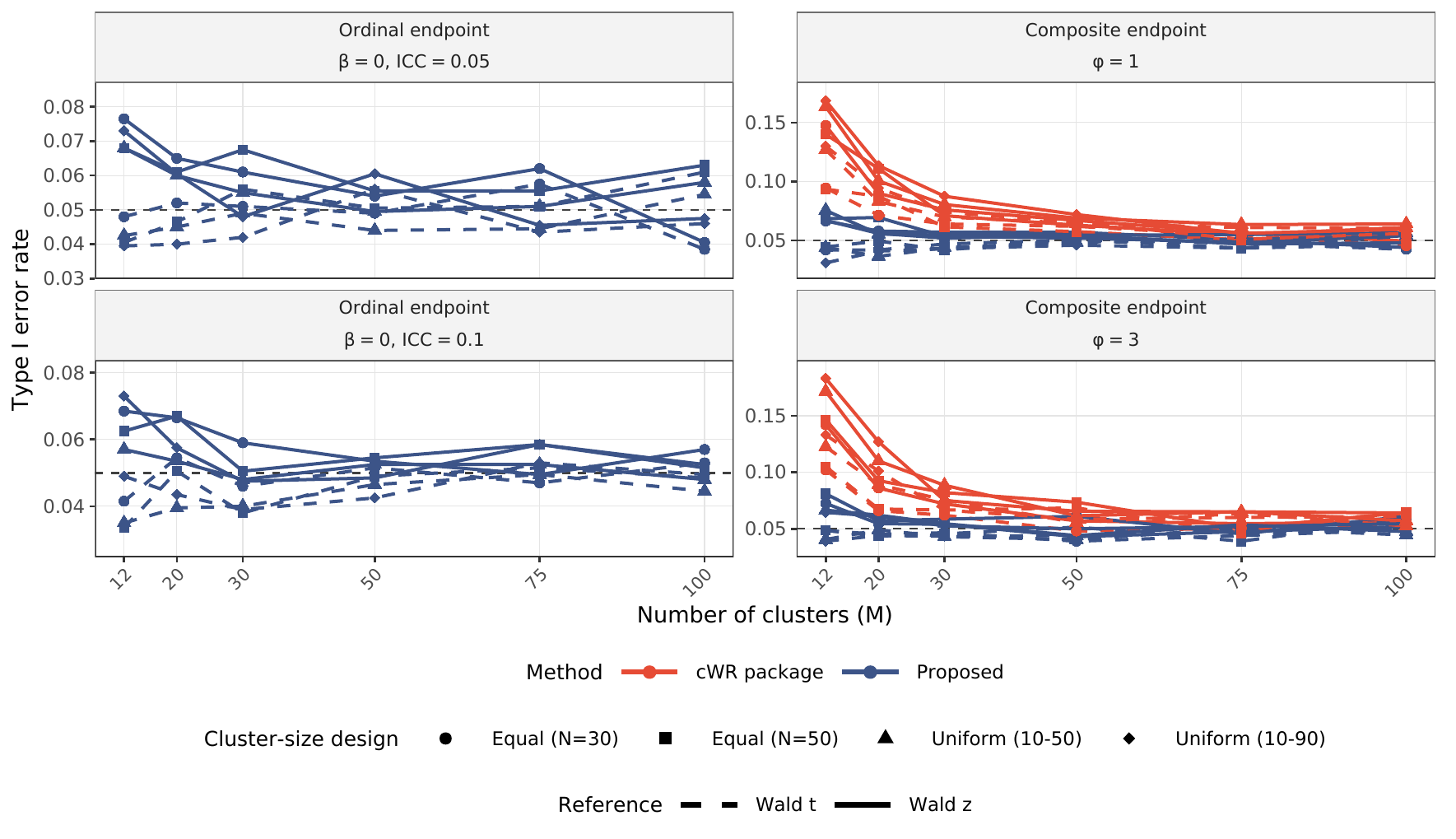}
\caption{Empirical type I error for \(\Delta=\log(W_R)\). Results are shown for the proposed Wald \(z\)- and Wald \(t\)-tests. For composite survival endpoints, the \texttt{cWR} package is included as an additional comparator. The horizontal dashed line marks the nominal 5\% level.}
\label{fig:sim_typeI_logWR}
\end{figure}

We first examine type I error, because valid power calculation is only meaningful when the corresponding test is appropriately sized. Figure \ref{fig:sim_typeI_logWR} summarizes empirical type I error for \(\Delta=\log(W_R)\). For the ordinal endpoint, the proposed Wald \(z\)-test has empirical type I error ranging from \(4.05\%\) to \(7.65\%\). The largest inflation occurs at \(M=12\), where the median type I error is approximately \(6.82\%\). In contrast, the Wald \(t\)-test ranges from \(3.35\%\) to \(6.10\%\), and its median type I error at \(M=12\) is approximately \(4.10\%\). Thus, for ordinal outcomes, the \(t\)-reference distribution removes most of the small-\(M\) liberal behavior of the normal approximation, although it can be slightly conservative in the smallest sample size settings. By \(M=75\) or \(100\), the two reference distributions become similar and both remain close to the nominal \(5\%\) level.

For composite survival endpoints, the same pattern is observed but the small-\(M\) sensitivity is slightly more pronounced. For the proposed procedure, the Wald \(z\)-test ranges from \(4.30\%\) to \(8.10\%\). At \(M=12\), its median type I error is approximately \(6.78\%\). The Wald \(t\)-test ranges from \(3.10\%\) to \(5.85\%\), and has median type I error approximately \(4.30\%\) at \(M=12\). The type I error inflation is primarily a small number of clusters phenomenon rather than a failure of the cluster score variance estimator itself. The comparison between ordinal and composite endpoints also suggests that censoring and partial rankings add some finite-sample variability, but do not fundamentally change the pattern of the simulation results.
For the composite endpoint, Figure \ref{fig:sim_typeI_logWR} also includes the \texttt{cWR} comparator. The \texttt{cWR} test is substantially more anti-conservative when \(M\) is small. For example, at \(M=12\), its median type I error is approximately \(15.55\%\) under the normal reference distribution and \(11.38\%\) under the \(t\)-reference distribution, compared with approximately \(6.78\%\) and \(4.30\%\) for the proposed method. Even at \(M=20\), the median type I error for \texttt{cWR} remains about \(10.52\%\) with the normal reference and \(8.52\%\) with the \(t\)-reference. This inflation decreases as \(M\) increases. When \(M=100\), the median type I error are approximately \(6.02\%\) and \(5.62\%\), respectively. This convergence is consistent with the asymptotic equivalence discussed in Web Appendix \ref{supp:sec:u-stat}. However, for the finite numbers of clusters commonly encountered in CRTs, the proposed variance estimator provides better type I error calibration. This comparison highlights that the main advantage of the proposed testing procedure is improved finite-sample stability under cluster randomization. The type I error results for \(W_D\) and \(\log(W_O)\) are reported in Web Appendix \ref{supp:sec:typeI_additional}. These results are consistent with the \(\log(W_R)\) findings. Across all three estimands, the proposed Wald \(z\)-test is mildly liberal when \(M\) is very small, while the Wald \(t\)-test improves calibration.

\begin{figure}[ht!]
\centering
\includegraphics[width=0.98\linewidth]{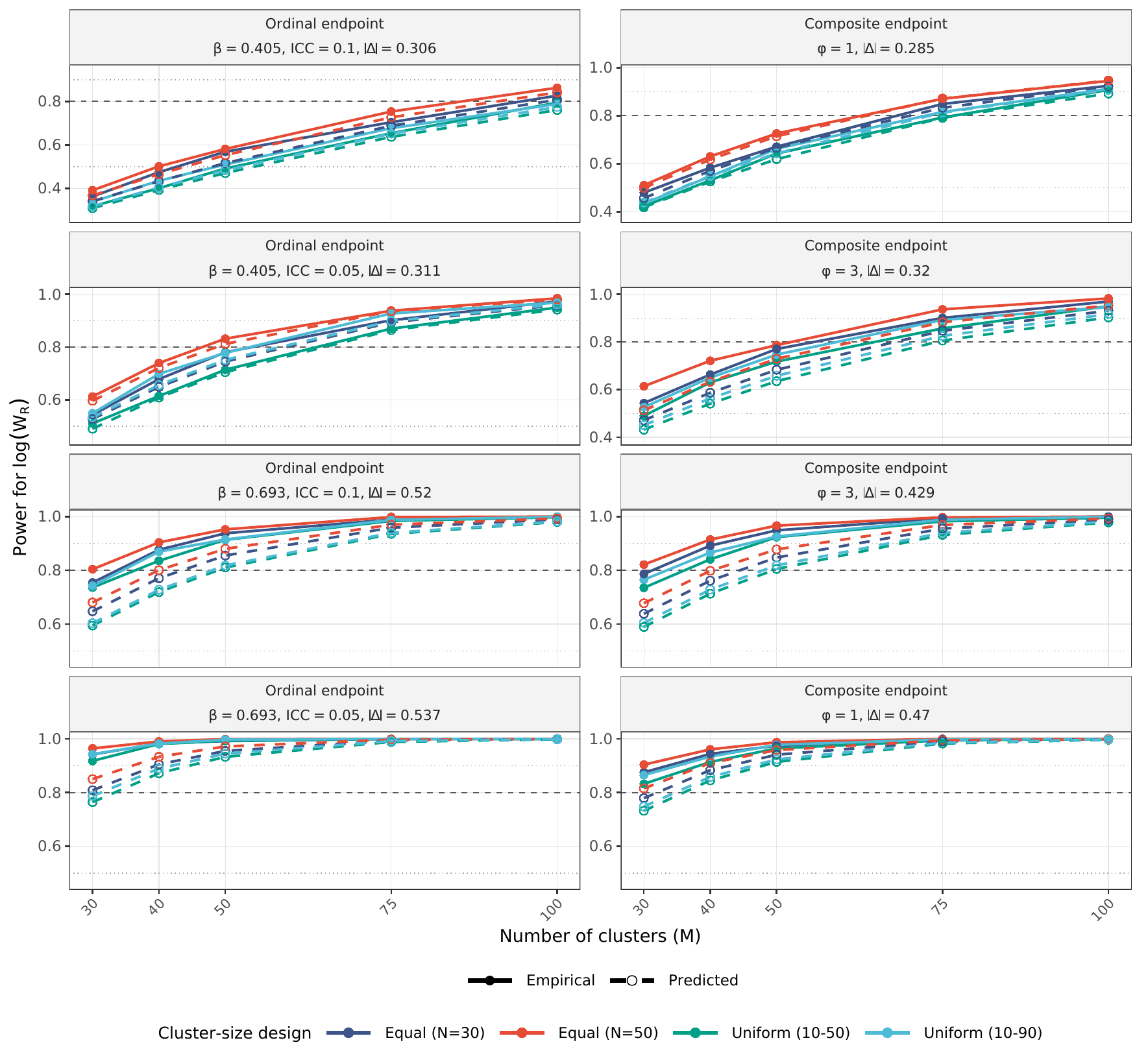}
\caption{Power curves for \(\Delta=\log(W_R)\). Solid lines denote empirical power, and dashed lines denote predicted power based on the from Theorem \ref{theorem:var_single} for ordinal endpoints or Theorem \ref{theorem:var_comp} for composite endpoints. The horizontal dashed line marks \(80\%\) power. Rows correspond to selected alternatives spanning weaker to stronger effects, and columns distinguish ordinal and composite endpoint settings.}
\label{fig:sim_power_logWR}
\end{figure}

We next examine the power for \(\Delta=\log(W_R)\). Figure \ref{fig:sim_power_logWR} summarizes empirical and predicted power curves. For the ordinal endpoint, empirical power increases monotonically with the number of clusters. Across the displayed ordinal settings, the empirical Wald \(z\)-test power at \(M=30\) ranges from \(31.8\%\) to \(96.4\%\). By \(M=100\), the range increases to \(78.8\%\) to nearly \(100\%\). The predicted power follows the same ordering, ranging from \(30.9\%\) to \(85.0\%\) at \(M=30\) and from \(76.0\%\) to nearly \(100\%\) at \(M=100\). Thus, the formula correctly captures the shape of the power curve and the relative difficulty of the different design settings. The largest absolute power prediction errors occur in smaller \(M\) settings and for stronger alternatives, where the empirical curve rises somewhat faster than the predicted curve.

The power prediction error for ordinal outcome is generally positive, meaning that the predicted power is mildly conservative. For the Wald \(z\)-test, empirical power minus predicted power ranges from \(0.02\) to \(15.57\) percentage points. The median discrepancy decreases from \(6.55 \%\) at \(M=30\), to \(4.98\) at \(M=40\), \(3.61\%\) at \(M=50\), \(1.67\%\) at \(M=75\), and \(0.89\%\) at \(M=100\). The Wald \(t\)-test shows the same pattern, with median discrepancy decreasing from \(6.11\%\) at \(M=30\) to \(1.05\%\) at \(M=100\). Therefore, the proposed formula is most conservative when the cluster count is small and becomes increasingly accurate as the cluster-level asymptotic approximation improves.

For composite survival endpoints, the same pattern is observed, but the power prediction error is somewhat larger. Empirical Wald \(z\)-test power ranges from \(41.4\%\) to \(90.4\%\) at \(M=30\), and increases to \(88.0\%\) to nearly \(100\%\) at \(M=100\). The predicted power ranges from \(36.9\%\) to \(81.6\%\) at \(M=30\), and from \(84.2\%\) to \(99.9\%\) at \(M=100\). The median power prediction error is \(9.03\%\) at \(M=30\), \(7.34\%\) at \(M=40\), \(5.44\%\) at \(M=50\), \(2.95\%\) at \(M=75\), and \(1.03\%\) at \(M=100\). The Wald \(t\)-test again shows a parallel pattern, with median discrepancy \(9.28\%\), \(7.60\%\), \(6.04\%\), \(3.35\%\), and \(1.16\%\) \% over the same sequence of \(M\). This shows that the formula remains directionally accurate for composite endpoints, but is more conservative in smaller designs than in the scalar ordinal setting.

This mild conservativeness pattern is also reflected in the standard-error diagnostics in Web Appendix \ref{supp:sec:se_diagnostics}. The ASE from Theorem \ref{theorem:u-var} closely tracks the Monte Carlo standard deviation (MCSD), with median ASE/MCSD ratios of \(1.03\) for ordinal \(\log(W_R)\) and \(1.02\) for composite \(\log(W_R)\). In contrast, the DSE from Theorem \ref{theorem:var_single} for the ordinal endpoint or Theorem \ref{theorem:var_comp} for the composite endpoint is larger, with a median DSE/MCSD ratio of approximately \(1.11\) in both endpoint settings. Equivalently, the DSE is about \(7.6\%\) larger than the ASE from Theorem \ref{theorem:u-var} for ordinal \(\log(W_R)\) and about \(8.3\%\) larger for composite \(\log(W_R)\). Because predicted power is calculated using the DSE, this modest overestimation of the standard error directly lowers the predicted power. The conservativeness is more visible for composite endpoints because censoring creates partial rankings and unresolved comparisons and because the composite variance depends on triplet probabilities and the generalized rank ICC, which enter through low-dimensional design-stage summaries. These summaries provide a reasonable study planning approximation but slightly overstate finite sample uncertainty in some settings. The level of conservativeness becomes negligible as \(M\) increases.

Cluster size heterogeneity also contributes to the conservative pattern. In both endpoint settings, the unequal cluster size designs have lower power than their equal size counterparts with the same mean, especially at smaller \(M\). This is exactly the role of the \(\mathrm{CV}\) term in the variance inflation factor. However, the empirical variance inflation from cluster size heterogeneity is not always identical to the leading order CV approximation in finite samples. When \(M\) is small, the realized distribution of cluster sizes can differ from its assumed, study planning distribution, and a few larger clusters can contribute more pairwise comparisons. Because the design formula uses the population \(\overline N\) and \(\mathrm{CV}\) rather than conditioning on each realized cluster size vector, the predicted power tends to be slightly conservative under unequal cluster sizes. As \(M\) increases, the realized cluster size distribution stabilizes around its population distribution, and the predicted and empirical powers become closer. The corresponding power results for \(W_D\) and \(\log(W_O)\) are provided in Web Appendix \ref{supp:sec:additional_simulation}. Those results lead to the same conclusions as Figure \ref{fig:sim_power_logWR}.

Finally, because \(\rho^{*}\) and \(\pi_{\mathrm{tie}}\) may be difficult to specify prospectively, we assessed the sensitivity of the required number of clusters to these two design inputs in Web Appendix \ref{supp:sec:tie_sensitivity}. Because the derived variance expressions increase monotonically with the generalized rank ICC and the tie probability, the required number of clusters is expected to increase as \(\rho^{*}\) and \(\pi_{\mathrm{tie}}\) increase. To evaluate this pattern, we repeated the sample size calculation over a grid of planning values \(\rho^{*}\in\{0.03,0.06,0.09,0.12\}\), holding the data-generating process for the fatal and hospitalization events fixed while varying the censoring rate to induce a range of \(\pi_{\mathrm{tie}}\) values. Figure \ref{supp:fig:tie_sensitivity} displays the resulting relationship, where each curve traces the required number of clusters \(M\) for 80\% power over the censoring-induced \(\pi_{\mathrm{tie}}\) at a given planning value of \(\rho^{*}\). Two monotone trends are evident. Along each curve, \(M\) increases steadily with \(\pi_{\mathrm{tie}}\). For example, at \(\rho^{*}=0.09\), the required number of clusters rose from 74 to 120 for \(\phi=1\) and from 76 to 156 for \(\phi=3\). Across curves, larger planning values of \(\rho^{*}\) uniformly shift the required \(M\) upward at every tie probability. Because \(\log(W_R)\) remained stable between 0.282 and 0.295 across all censoring scenarios, the increase in \(M\) is attributable to the loss of decisive comparisons rather than to any change in the treatment effect. Across these sensitivity scenarios, the close agreement between the formula-predicted power and the empirical power continued to hold, indicating that the proposed formulas remain accurate over the plausible ranges of \(\rho^{*}\) and \(\pi_{\mathrm{tie}}\) considered.

\section{An illustrative data example} \label{sec:real_data}

We illustrate the proposed sample size and power method using the Strategies to Reduce Injuries and Develop Confidence in Elders (STRIDE) trial, a pragmatic cluster-randomized trial evaluating a multifactorial intervention to prevent fall-related injuries among older adults at risk of falling \citep{bhasin2020randomized}. The trial enrolled 5{,}451 community-dwelling adults aged $\ge 70$ years from $M=86$ primary care practices (10 health systems), randomized $1\!:\!1$ to intervention ($43$ practices; $2{,}802$ individuals) or control ($43$ practices; $2{,}649$ individuals); the primary outcome was time to first serious fall-related injury over up to 44 months. 
For this analysis, we retained each patient’s time to death, time to first fall-related injury, and censoring, and compared pairs using a prioritized win rule in \eqref{win_rule} ranking death above injury: pairs were compared first on death, then on fall-related injury. We report all three prioritized win measures and their estimated finite-sample standard errors from Theorem \ref{theorem:u-var}: the estimated win difference (net benefit) is $\widehat W_D=0.04$ with SE $=0.014$, the estimated log win ratio is $\widehat\Delta=\log\widehat W_R=0.13$ with SE $=0.044$, and the estimated log win odds is $\log\widehat W_O=0.08$ with SE $=0.028$, where the \(\log(W_O)\) is slightly more efficient than \(\log(W_R)\). Given the relatively large number of clusters in STRIDE ($M=86$), we used the Wald $z$-test, yielding a two-sided $p$-value of 0.004. To benchmark future designs, we used the pooled probabilities to enter the power formulas: $\widehat p_W=31.4\%$, $\widehat p_T=37.2\%$, $\widehat p_{WW}=12.1\%$, $\widehat p_{WT}=13.1\%$, $\widehat p_{TT}=21.8\%$, with tie rate $\widehat\pi_{\text{tie}}=37.1\%$, rank ICC $\widehat\rho^{*}=0.003$, and average cluster size $\overline N=63.4$ (CV$=0.517$). Following Theorem \ref{theorem:var_comp}, the predicted powers (Wald $z$-test) are 82.9\% for $W_D$, 82.7\% for $\log(W_R)$, and 82.8\% for $\log(W_O)$, suggesting that the realized STRIDE configuration provided adequate operating characteristics for the observed effect and dependence structure. Although post hoc power assessment is not generally recommended for design planning, these quantities offer an internally coherent benchmark for exploring alternative scenarios. Using the STRIDE estimates as baseline inputs, we next examine how the required number of clusters $M$ varies with design factors, computing the $M$ needed to achieve 80\% power for the Wald $z$-test of $\log(W_R)$ across grids of $(\overline N,\mathrm{CV})$ and combinations of $\Delta=\log(W_R)$ and $\rho^{*}$.

\begin{figure}[ht!]
\centering
\begin{subfigure}[b]{0.48\linewidth}
  \includegraphics[width=\linewidth]{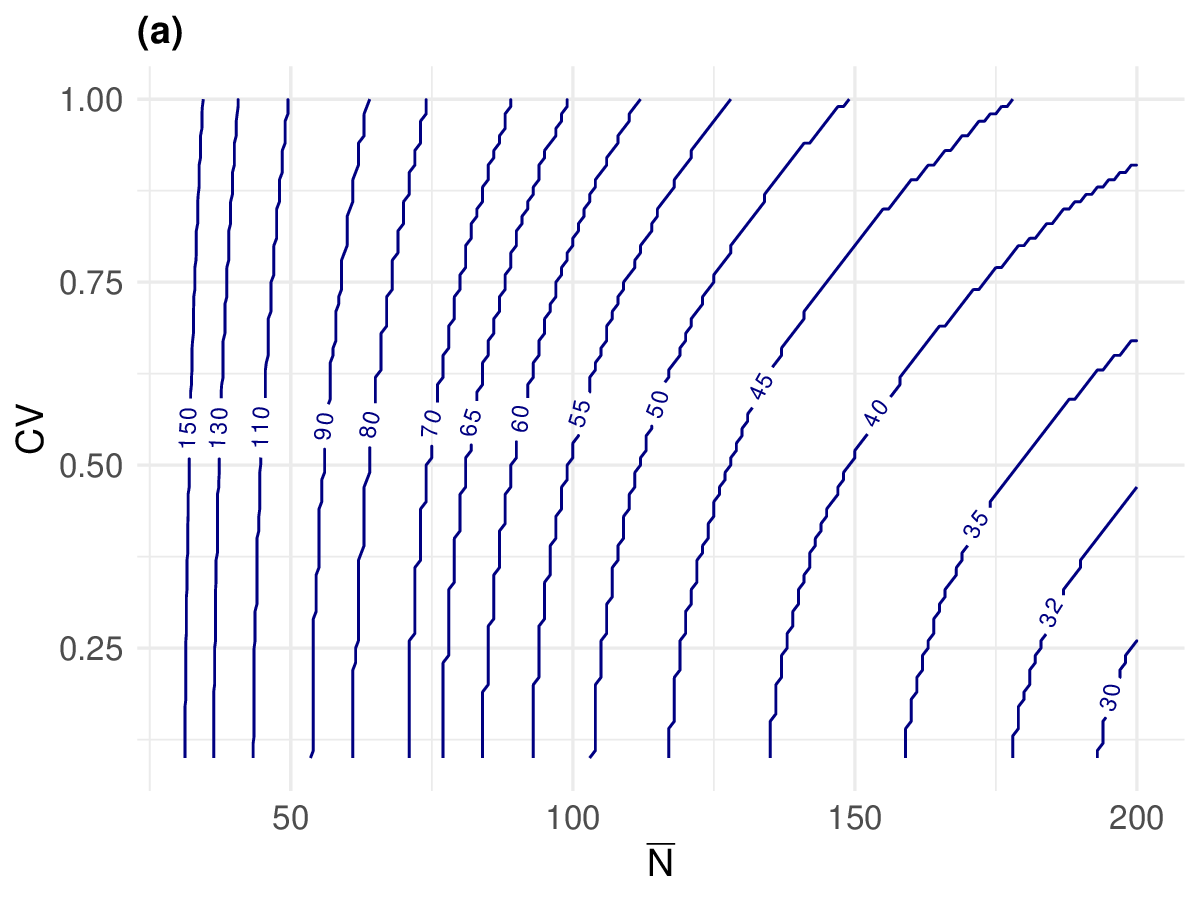}
  \caption{}
  \label{fig:stride_a}
\end{subfigure}
\begin{subfigure}[b]{0.48\linewidth}
  \includegraphics[width=\linewidth]{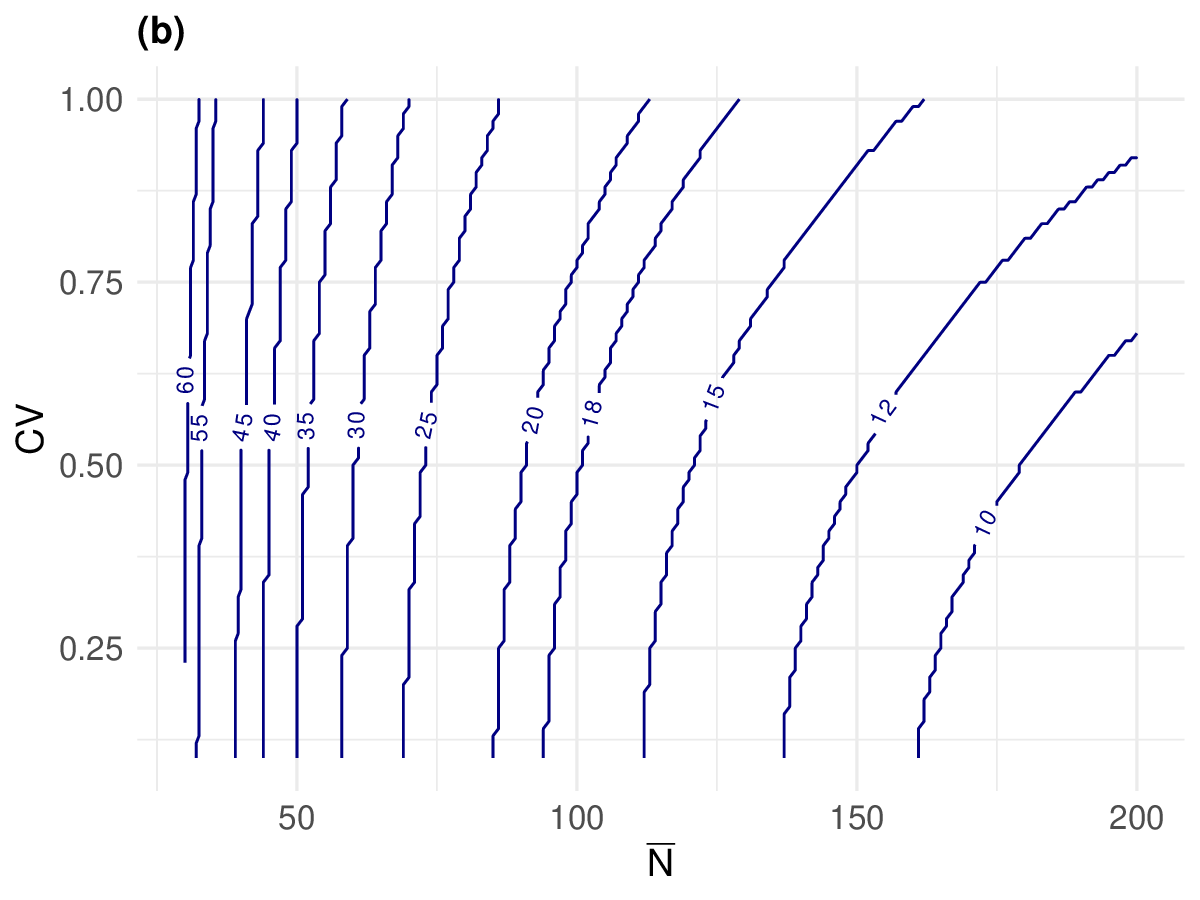}
  \caption{}
  \label{fig:stride_b}
\end{subfigure}

\begin{subfigure}[b]{0.48\linewidth}
  \includegraphics[width=\linewidth]{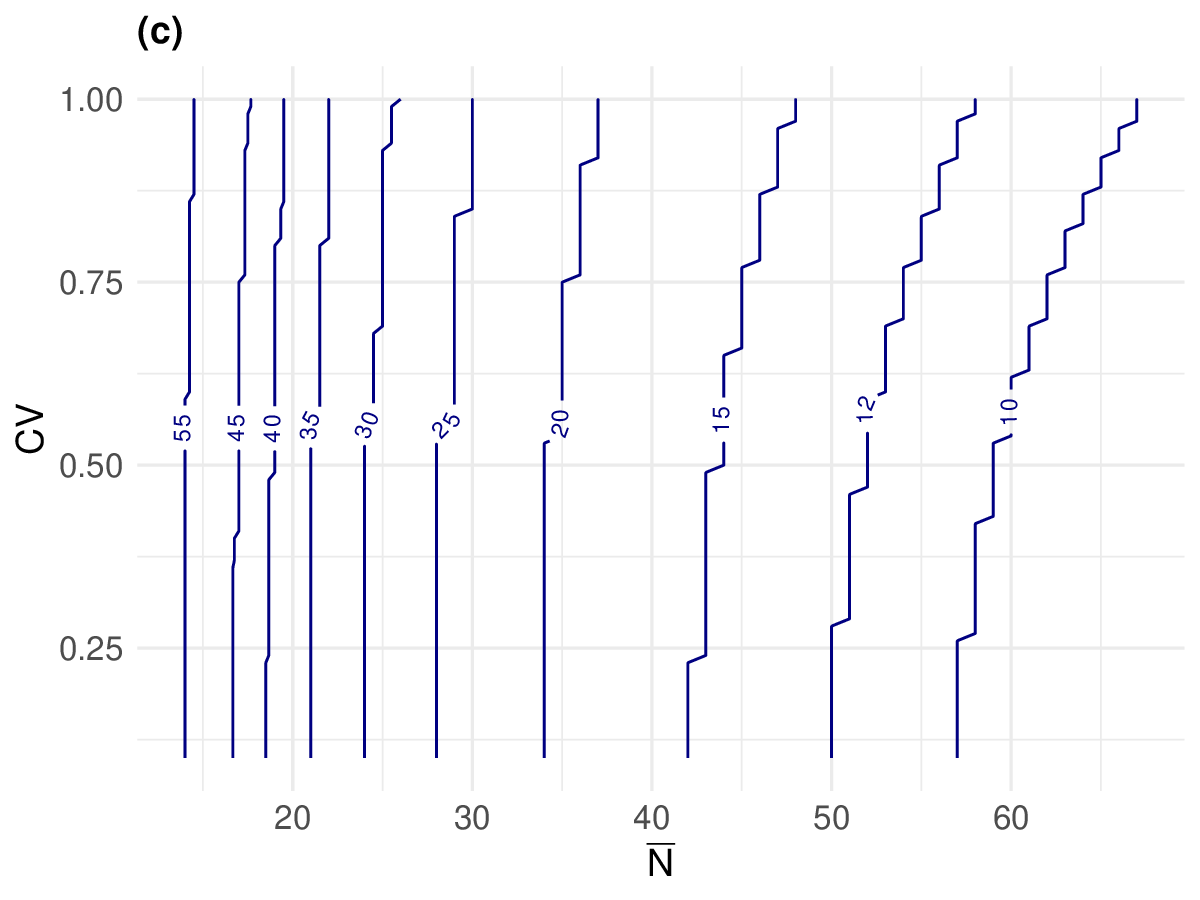}
  \caption{}
  \label{fig:stride_c}
\end{subfigure}
\begin{subfigure}[b]{0.48\linewidth}
  \includegraphics[width=\linewidth]{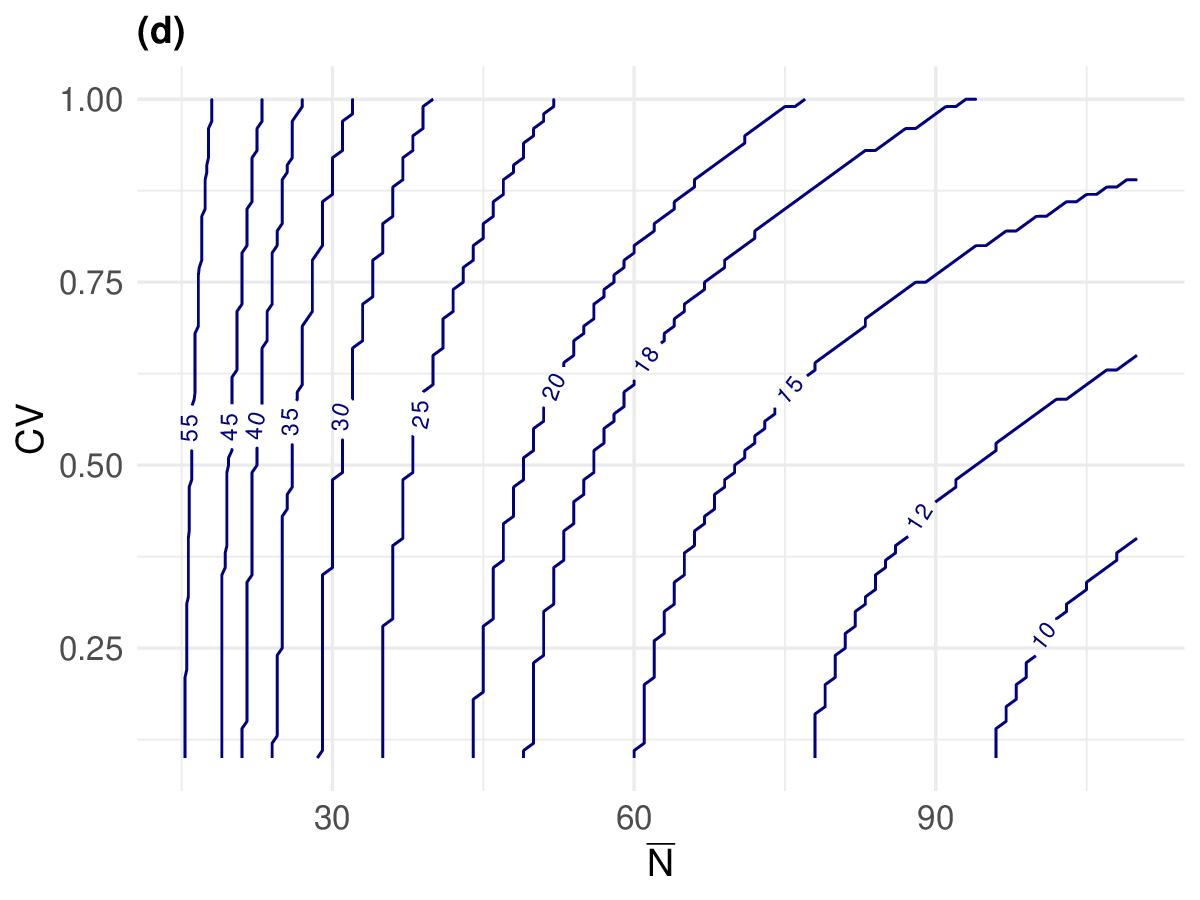}
  \caption{}
  \label{fig:stride_d}
\end{subfigure}
\caption{Contour plots of the required number of clusters \(M\) to achieve approximately 80\% power for the win measure \(\Delta = \log(W_R)\) using Wald \(z\)-test. Results are shown as functions of the CV in cluster size and the mean cluster size \(\overline{N}\). Each panel represents a different combination of the rank ICC \(\rho^{*}\) and true effect size: (a) \(\rho^{*} = 0.003,\ \Delta = 0.127\); (b) \(\rho^{*} = 0.003,\ \Delta = 0.200\); (c) \(\rho^{*} = 0.003,\ \Delta = 0.300\); (d) \(\rho^{*} = 0.010,\ \Delta = 0.300\).}
\label{fig:stride}
\end{figure}

Figure \ref{fig:stride} presents contour plots of the required number of clusters \(M\) to achieve a fixed power of \(80\%\) for the Wald \(z\)-test of \(\log(W_R)\), evaluated over the design axes \((\overline{N},\mathrm{CV})\) and across combinations of true effect \(\Delta=\log(W_R)\) and rank-based ICC \(\rho^{*}\). The contours represent level sets of the implicit sample-size surface \(M(\overline{N},\mathrm{CV})\). Panels (a)-(c) keep \(\rho^{*}=0.003\) fixed while increasing \(\Delta\in\{0.127,0.200,0.300\}\). As \(\Delta\) increases, the entire surface shifts downward, and the contours move leftward; that is, for any given \((\overline{N},\mathrm{CV})\), fewer clusters suffice. This is consistent with the usual noncentrality structure, whereby the detectable signal grows with \(\Delta\), reducing the \(M\) needed to reach the same power. The rate of leftward shift is not uniform. At larger \(\overline{N}\) the contours become more widely spaced, with diminishing marginal gains from further increases in \(\overline{N}\). Conversely, at larger \(\mathrm{CV}\) the contours pack more tightly (steeper surface), showing higher sensitivity of \(M\) to \(\mathrm{CV}\) when the cluster size heterogeneity is already substantial. Panel (d) fixes \(\Delta=0.3\) and increases the ICC to \(\rho^{*}=0.01\). Two qualitative changes are apparent. First, all contours move upward, showing a global inflation in \(M\) attributable to stronger within-cluster correlation. Second, the contours ``fan out'' with increasing \(\mathrm{CV}\): the spacing between successive \(M\)-levels widens as \(\mathrm{CV}\) grows, revealing an interaction between \(\rho^{*}\) and \(\mathrm{CV}\) whereby heterogeneity in cluster sizes amplifies the effective correlation penalty as presented in Theorem \ref{theorem:var_comp}.  Across all panels, we observe two general monotonicity patterns; that is, \(M\) decreases with \(\overline{N}\) and increases with \(\mathrm{CV}\).This sensitivity analysis is intended to illustrate how the formula should be used during planning. In practice, the same calculation can be repeated over plausible values of \(\pi_{\mathrm{tie}}\) or, for composite survival endpoints, over event-rate and censoring scenarios that induce different \(\pi_{\mathrm{tie}}\), \((P,Q)\), and \(\rho^*\).

\section{Discussion}
\label{s:discuss}

We developed a unified sample size methodology for CRT with either single endpoints or prioritized composite endpoints analyzed via win statistics, thereby addressing a notable gap in the study design literature. In contrast to existing sample size approaches for the win ratio that assume independent and identically distributed observations, our framework explicitly incorporates correlation induced by cluster randomization through a rank-based intracluster correlation coefficient (rank ICC) and accounts for unequal cluster sizes via a generalized design effect depending jointly on the rank ICC and the coefficient of variation in cluster sizes. The derivation leads to new variance expressions of pairwise comparison-based rank statistics under intracluster dependence, yielding analytic expressions that obviate the need for simulation-based design. Our framework is applicable to both prioritized composite endpoints and single-component outcomes such as survival or ordinal data, providing a unified approach to sample size determination in the clustered settings. A practical contribution is the implementation of the method in a Shiny application, which accepts design inputs (anticipated effect size, tie probability, rank ICC, number and sizes of clusters) or parameters from generative survival models described in Section \ref{sec:practice}, and returns power or sample size estimates in a computationally efficient manner. Beyond quantifying the variance inflation due to clustering, our formulation draws attention to the influence of non-transitivity in hierarchical pairwise comparisons \citep{lee2025note}, which can increase variability and affect interpretability. Our developed R Shiny app supports visual exploration of these properties to promote transparent and reproducible trial planning. 

We point out several key considerations for applying the proposed formulas. First, the estimand interpretation and the closed-form variance expressions in Theorems \ref{theorem:var_single} and \ref{theorem:var_comp} are developed under the noninformative cluster size assumption. Under this assumption, unequal cluster sizes are randomly varying, and their leading contribution to the variance enters through \(\mathbb{E}(N_i^2)=\overline{N}^{\,2}(1+\mathrm{CV}^2)\), which produces the CV-based variance inflation factor. What is excluded is a systematic association of \(N_i\) with the outcome distribution, the rank-score distribution, or the magnitude of the treatment effect. The proposed VIF therefore accounts for unequal cluster sizes under noninformative cluster size, but it is not an adjustment for informative cluster size. If cluster size is informative \citep{seaman2014review}, the win estimands constructions are more complicated and we refer to the recent work of \citet{lee2026s} and \citet{chen2026optimal} for analytic considerations. Sample size development for the informative setting requires a separate derivation and will be pursued in future work.

Second, even under noninformative cluster size, the tests and power calculations are justified by large-\(M\) approximations, but their empirical performance may be affected when the number of clusters is small. Two features of a CRT design deserve particular attention. The first is a small number of clusters. The second is substantial imbalance in cluster sizes. When a small number of clusters contribute a disproportionately large share of the enrolled participants (or equivalently of \(\sum_{i=1}^{M}N_i^2\)), the effective number of independent cluster-level contributions is smaller than the nominal number of clusters would suggest, making the large-sample approximation less accurate. This issue is most pronounced when the cluster size distribution is highly right-skewed. By contrast, the presence of many small clusters alone is generally not problematic. In our simulations, the Wald test using a \(t\) reference distribution with \(M-2\) degrees of freedom was slightly better calibrated than the corresponding \(z\)-test for smaller designs. Accordingly, for CRTs with relatively few clusters or substantial cluster size imbalance, we recommend reporting the \(t\)-based sample size calculation. 

Third, as with any analytic sample size procedure, the operating characteristics of the proposed formulas depend on the accuracy of the design inputs. Because the derived variance expressions increase monotonically with the rank ICC and the tie probability, the direction of the impact of input misspecification is predictable. Underestimating these quantities may lead to an insufficient sample size and an underpowered trial, whereas using plausible upper-range values provides a conservative sample size. As discussed in Section \ref{sec:practice}, sensitivity analyses over a range of plausible rank ICC values are therefore recommended. For composite time-to-event endpoints, this process is more naturally carried out at the level of a generative outcome model rather than by varying the tie and triplet probabilities independently. Specifically, the tie probability and triplet probabilities are induced jointly by the event rates, censoring distribution, follow-up duration, and dependence structure among endpoint components \citep{mou2025correlation}. Consequently, lower-than-anticipated event rates or heavier censoring generally increase the proportion of unresolved comparisons and alter the triplet probabilities simultaneously, thereby increasing the required sample size. We therefore recommend sensitivity analyses based on plausible ranges of the underlying event and censoring mechanisms, allowing the induced tie and triplet probabilities to vary jointly according to the assumed data-generating model.


\section*{Supplementary Materials}
Web Appendix referenced in Section \ref{sec:test} - \ref{sec:practice}, and additional simulation results are available with this paper at the \emph{Scandinavian Journal of Statistics} website on Wiley Online
Library.\vspace*{-8pt}

\section*{Acknowledgements}
F. Li is supported by the United States National Institutes of Health (NIH), National Heart, Lung, and Blood Institute (NHLBI, grant number 1R01HL178513). All statements in this report, including its findings and conclusions, are solely those of the authors and do not necessarily represent the views of the NIH. The STRIDE study was funded primarily by the Patient Centered Outcomes Research Institute (PCORI\textsuperscript{\textregistered}), with additional support from the National Institute on Aging (NIA) at NIH. Z. Cao contributed to this work on a voluntary collaborative basis, and his contribution was not supported by any grant or external funding source.

\section*{Data Availability Statement}
An R package implementing our method is available at \url{https://cran.r-project.org/web/packages/WinsCRT}. The STRIDE data can be obtained via the National Institute on Aging (NIA) Aging Research Biobank at\\ \url{https://agingresearchbiobank.nia.nih.gov/studies/stride/details}.

\clearpage
\newpage
\printbibliography

\newpage
\appendix

\setcounter{section}{0}
\renewcommand{\thesection}{A.\arabic{section}}
\renewcommand{\thesubsection}{\thesection.\arabic{subsection}}
\renewcommand{\thesubsubsection}{\thesubsection.\arabic{subsubsection}}

\section{Proof of Theorem 1} \label{supp:sec:theorem1}
\allowdisplaybreaks
\raggedbottom
In this section, we first prove Theorem 1.
\begin{proof}
Note that $\sum_{i=1}^{M} \sum_{j=1}^{N_i} \sum_{k=1}^{M} \sum_{l=1}^{N_k} s_{ij,kl}=\sum_{k=1}^{M} \sum_{l=1}^{N_k} \sum_{i=1}^{M} \sum_{j=1}^{N_i} s_{kl,ij}$, \(s_{ij,kl} = -s_{kl,ij}\) and \(s_{ij,ij} = 0 \), we can show that
\begin{align*}
    \sum_{i=1}^{M} S_i &= \sum_{i=1}^{M} \sum_{j=1}^{N_i} \phi_{ij}  = \sum_{i=1}^{M} \sum_{j=1}^{N_i} \sum_{k=1}^{M} \sum_{l=1}^{N_k} s_{ij,kl}  \\
   & = \frac{1}{2}\left[\sum_{i=1}^{M} \sum_{j=1}^{N_i} \sum_{k=1}^{M} \sum_{l=1}^{N_k} s_{ij,kl}
   +\sum_{k=1}^{M} \sum_{l=1}^{N_k} \sum_{i=1}^{M} \sum_{j=1}^{N_i} s_{kl,ij}\right]\\
   &=\frac{1}2\sum_{i=1}^{M} \sum_{j=1}^{N_i} \sum_{k=1}^{M} \sum_{l=1}^{N_k} (s_{ij,kl} + s_{kl,ij}) = 0.
\end{align*}

For the estimator \(\widehat{W}_D\), we can rewrite it as
\[\widehat{W}_D = \frac{qM(1-q)M}{M n_1n_0} (\overline{S}_1 - \overline{S}_0), \]
where \(\overline{S}_1 = \frac{1}{qM}\sum_{i=1}^{M} A_iS_i\), \(\overline{S}_0 = \frac{1}{(1-q)M} \sum_{i=1}^{M} (1-A_i)S_i\), and \(qM\overline{S}_1 + (1-q)M\overline{S}_0 = 0\). We have the arm-specific mean \(\mu_a = \mathbb{E}(S_i\mid A_i=a)\), and arm-specific variance \(\sigma_a^2 = \operatorname{Var}(S_i \mid A_i= a)\). Using the Hoeffding-Hájek projection of the U-statistic \citep{hajek1968asymptotic}, we have
\[\overline{S}_1 - \overline{S}_0 - (\mu_1 - \mu_0) = \frac{1}{M} \sum_{i-1}^{M} \psi_i + o_p(M^{-1/2}), \]
with \(\psi_i = \frac{A_i}{q} (S_i - \mu_1)  - \frac{1-A_i}{1-q} (S_i - \mu_0) \). It is straightforward to show that \(\mathbb{E}(\psi_i) = 0\), and  \(\operatorname{Var}(\psi_i)  = \frac{\sigma_1^2}{q} + \frac{\sigma_0^2}{1-q} \). Then under cluster randomization, for \(\widehat{W}_D\), we have
\begin{align*}
    \widehat{W}_D & = \frac{qM(1-q)M}{M n_1n_0} \left[(\mu_1-\mu_0) + M^{-1} \sum_{i=1}^{M} \psi_i + o_p(M^{-1/2})   \right] \\
    & = \frac{\mu_1-\mu_0}{M\overline{N}^2}  + \frac{1}{M^2\overline{N}^2}  \sum_{i=1}^{M} \psi_i + o_p(M^{-3/2}).
\end{align*}
Thus, \(M^{3/2} \left( \widehat{W}_D - W_D \right) \stackrel{p}{\rightarrow}\mathcal{N} \left( 0, \overline{N}^{-4} \left[\frac{\sigma_1^2}{q} + \frac{\sigma_0^2}{1-q}  \right]  \right)    \).
The asymptotic variance of \(\widehat{W}_D\) is 
\begin{align}\label{wd_asymp}
 \sigma_{D}^{2}=\operatorname{Var}[\widehat{W}_D] =  \frac{1}{M^3} \cdot \frac{1}{\overline{N}^4} \left[\frac{\sigma_1^2}{q} + \frac{\sigma_0^2}{1-q}  \right] + o_p(M^{-3}).    \end{align}
And a consistent variance estimator of  $\sigma_{D}^{2}$ is given by
\begin{align*}
\widehat{\sigma}_{D}^2 &= \left(\frac{Mq(1-q)}{n_1n_0} \right)^2 \left( \frac{\widehat{\sigma}_1^2}{qM} + \frac{\widehat{\sigma}_0^2}{(1-q)M} \right)\\
& = \left(\frac{1}{M\overline{N}}\right)^2  \left( \frac{\widehat{\sigma}_1^2}{qM} + \frac{\widehat{\sigma}_0^2}{(1-q)M} \right) ,
\end{align*}
where \(\widehat{\sigma}_a^2 = 1/(q^a(1-q)^{1-a}M-1)\sum_{i=1}^{M} A_i^{a}(1-A_i)^{1-a} (S_i - \overline{S}_a)^2 \), and \\ $\overline{S}_a= \{q^a(1-q)^{1-a}M\}^{-1}\sum_{i=1}^{M} A_i^{a}(1-A_i)^{1-a}S_i$ for $a \in \{0,1\}$.
By the Delta method, we can show that
\[
\sigma_R^2
= \left\{\frac{2\,\{1/(1-\pi_{\text{tie}})\}}{\,1-\left(\{1/(1-\pi_{\text{tie}})\}W_D\right)^2}\right\}^{\!2}\sigma_D^2,\quad
\sigma_O^2
= \left\{\frac{2}{1-W_D^2}\right\}^{\!2}\sigma_D^2.
\]
\end{proof}

\subsection{Connection with permutation variance based on clustered \citet{finkelstein1999combining} test} \label{supp:sec:permutation}
The estimator \(\widehat{W}_D\) can be viewed as an extension of the FS permutation test (originally proposed in \citet{finkelstein1999combining} for independent data), in which the cluster-level scores \(S_i\) are mutually independent under the null and serve as the test statistics in the cluster-level FS framework. Since \(\sum_{i=1}^{M} S_i = 0\) and \(s_{ij,kl} = -\,s_{kl,ij}\), the pairwise win–loss structure ensures antisymmetry across all cluster pairs.

\begin{eqnarray*}
\sum_{i=1}^{M} \sum_{j=1}^{N_i} \sum_{k=1}^{M} \sum_{l=1}^{N_k} A_i (1-A_i) s_{ij,kl}&=&\sum_{k=1}^{M} \sum_{l=1}^{N_k}\sum_{i=1}^{M} \sum_{j=1}^{N_i}  A_k (1-A_k) s_{kl,ij}\\
&=&\sum_{i=1}^{M} \sum_{j=1}^{N_i} \sum_{k=1}^{M} \sum_{l=1}^{N_k} A_k (1-A_k) [-s_{ij,kl}], 
\end{eqnarray*}
for the net benefit, we have
\begin{align*}
    W - L & = \sum_{i=1}^{M} \sum_{j=1}^{N_i} \sum_{k=1}^{M} \sum_{l=1}^{N_k} A_i (1-A_i) s_{ij,kl} \\
    & = \sum_{i=1}^{M} \sum_{k=1}^{M} A_i (1-A_i) \left( \sum_{j=1}^{N_i} \sum_{l=1}^{N_k} s_{ij,kl}  \right) \\
    & = \frac{1}{2} \sum_{i=1}^{M} \sum_{k=1}^{M} [A_i(1-A_i) - A_k(1-A_k)] \left( \sum_{j=1}^{N_i} \sum_{l=1}^{N_k} s_{ij,kl}  \right) \\
    & = \frac{1}{2} \sum_{i=1}^{M} \sum_{k=1}^{M} (A_i - A_k) \left( \sum_{j=1}^{N_i} \sum_{l=1}^{N_k} s_{ij,kl}  \right) \\
    & = \frac{1}{2} \left[ \sum_{i=1}^{M} A_i \left( \sum_{j=1}^{N_i}\sum_{k=1}^{M} \sum_{l=1}^{N_k} s_{ij,kl}  \right) - \sum_{k=1}^{M} A_k \left( \sum_{l=1}^{N_k}\sum_{i=1}^{M} \sum_{j=1}^{N_i} s_{ij,kl}  \right)    \right] \\
    & = \frac{1}{2}\left[ \sum_{i=1}^{M} A_i S_i - \sum_{k=1}^{M} A_k (-S_k)\right] \\
    & = \sum_{i=1}^{M} A_i S_i.
\end{align*}
By the definition of  \(W_D\), we have
\[\widehat{W}_D = \frac{\sum_{i=1}^{M} A_i S_i }{n_1n_0}. \]

Next, we derive the permutation variance under the null hypothesis. For permutation variance, we treat the outcomes as fixed and consider randomness arising solely from the treatment assignments under the null (no treatment effect, \(\Delta=0\)). The outcomes and consequently all pairwise scores \(s_{ij,kl}\) are invariant to the treatment assignment \(A_i\). Hence, under complete randomization, we have
\[
\mathbb{E}(A_i)=q,\qquad \operatorname{Var}(A_i)=q(1-q),\qquad
\operatorname{Cov}(A_i,A_k)=-\frac{q(1-q)}{M-1}\ (i\neq k).
\]
Therefore, we can show that the permutation variance is
\begin{align*}
\operatorname{Var}\!\left(\sum_{i=1}^{M} A_i S_i \,\Big|\, S_1,\ldots,S_M\right)
&= \sum_{i=1}^{M} S_i^2 \operatorname{Var}(A_i)+2\!\sum_{i<k} S_i S_k \operatorname{Cov}(A_i,A_k) \\
&= q(1-q)\sum_{i=1}^{M} S_i^2-\frac{2q(1-q)}{M-1}\sum_{i<k} S_i S_k \\
&= q(1-q)\sum_{i=1}^{M} S_i^2-\frac{2q(1-q)}{M-1}\cdot\frac{(\sum_{i=1}^{M} S_i)^2-\sum_{i=1}^M S_i^2}{2} \\
&= q(1-q)\sum_{i=1}^{M} S_i^2+\frac{q(1-q)}{M-1}\sum_{i=1}^{M} S_i^2
= q(1-q)\,\frac{M}{M-1}\sum_{i=1}^{M} S_i^2,
\end{align*}
Conditioning on the observed number of cross-arm pairs $n_1n_0$,
\begin{align*}
\operatorname{Var}\!\bigl(\widehat W_D\,\big|\,S_1,\ldots,S_M,n_1,n_0\bigr)
=\frac{q(1-q)}{(n_1n_0)^2}\,\frac{M}{M-1}\,\sum_{i=1}^{M} S_i^2.
\end{align*}
Under the null $\Delta=0$ (i.e., $W_D=0$) and $\sigma_1^2=\sigma_0^2=\sigma^2$, the asymptotic variance in Theorem \ref{theorem:u-var} becomes
\[
\varsigma_D^2 \;=\; \frac{\sigma^2}{M^{3}\,\overline{N}^{4}\,q(1-q)},\quad
\varsigma_R^2 \;=\; \frac{4\,\varsigma_D^2}{\left(1-\pi_{\text{tie}}\right)^{2}},\quad
\varsigma_O^2 \;=\; 4\,\varsigma_D^2.
\]
By replacing the  consistent estimator of \(\widehat{\sigma}_1^2 = \sigma_0^2 = \widehat{\sigma}^2=1/(M-1) \sum_{i=1}^{M}S_i^2\), the estimator of \(\varsigma_D^2\) becomes
\[\widehat{\varsigma}_D^2 = \left( \frac{1}{M\overline{N}^2} \right)^2 \frac{1}{q(1-q)M} \frac{\sum_{i=1}^{M}S_i^2}{M-1} = \frac{\sum_{i=1}^{M} S_i^2}{q(1-q)M^3(M-1)\overline{N}^2}, \]
which is asymptotically  equivalent to the permutation variance \(\operatorname{Var}\!\bigl(\widehat W_D\,\big|\,S_1,\ldots,S_M,n_1,n_0\bigr)\) since \(n_1 \xrightarrow{p} qM\overline N\), \(n_0 \xrightarrow{p} (1-q)M\overline N\). By the Delta method, the permutation variance for \(\log (W_R)\) and \(\log (W_D)\) are also the same as the estimators of \(\varsigma_R^2\) and \(\varsigma_O^2\).

\subsection{Binary outcomes} \label{supp:sec:binary}
For binary outcomes \(Y_{ij} \in \{0,1\}\), the distribution is degenerate, and then \(s_{ij,kl}=s(Y_{ij},Y_{kl}) = Y_{ij} - Y_{kl} = 1\) if and only if \(Y_{ij}=1\), and \(Y_{kl}=0\). Suppose \(p_1 = \mathbb{P}(Y_{ij}=1 \mid A_i = 1)\), \(p_0 = \mathbb{P}(Y_{ij}=1 \mid A_i = 0)\), and \(\rho_a = \operatorname{Corr}(Y_{ij},Y_{ij}'\mid A_i=a)\), then \(\pi_{\text{win}} = \mathbb{P}(Y_{ij}=1, Y_{kl}=0, A_i=1, A_k=0 ) = p_1(1-p_0) \), \(\pi_{\text{loss}} = \mathbb{P}(Y_{ij}=0, Y_{kl}=1, A_i=1, A_k=0 ) = p_0(1-p_1)\), and \(\pi_{\text{tie}} = \mathbb{P}(Y_{ij} = Y_{kl} ) = p_1p_0 + (1-p_1)(1-p_0) \). Following this definition, we can show that
\begin{align*}
    & W_D = \pi_{\text{win}} - \pi_{\text{loss}} = p_1(1-p_0) - (1-p_1)p_0 = p_1- p_0, \\
    & W_R = \frac{\pi_{\text{win}}}{\pi_{\text{loss}}} = \frac{p_1(1-p_0)}{(1-p_1)p_0} = \frac{p_1/(1-p_1)}{p_0(1-p_0)},\\
    & W_O = \frac{\pi_{\text{win}} + 0.5\pi_{\text{tie}} }{\pi_{\text{loss}}+ 0.5\pi_{\text{tie}}} = \frac{0.5 + 0.5(p_1-p_0)}{0.5 - 0.5(p_1-p_0)} = \frac{1 + (p_1-p_0)}{1 - (p_1-p_0)}.
\end{align*}
Thus, with a binary endpoint and cross-arm comparisons, the win difference (net benefit) equals to the risk difference, and the win ratio equals to the odds ratio. Based on the kernel \(s_{ij,kl}\), we can show that 
\[S_i = \sum_{j=1}^{N_i} \sum_{k=1}^{M} \sum_{l=1}^{N_k} (Y_{ij} - Y_{kl}) = n\sum_{j=1}^{N_i}Y_{ij} - N_i \sum_{k=1}^{M} \sum_{l=1}^{N_k} Y_{kl}. \]
Under then null, the variance of  \(\sigma_a^2\) can be further expressed as 
\begin{align*}
    \sigma_a^2  =& \mathbb{E} \left[(n-N_i)^2 \operatorname{Var}\left(\sum_{j=1}^{N_i}Y_{ij} \mid A_i=a, N_i  \right) + N_i^2 \sum_{k\neq i} \operatorname{Var}\left(\sum_{l=1}^{N_k} Y_{kl} \mid A_k, N_k  \right)  \mid N_i \right]\\
     =& \mathbb{E}\left\{(n-N_i)^2 \mid N_i \right\} \cdot \mathbb{E} \left[ \operatorname{Var}\left(\sum_{j=1}^{N_i}Y_{ij} \mid A_i=a, N_i  \right) \right] \\
     &+  \mathbb{E}(N_i^2) \sum_{k\neq i} \mathbb{E} \left[ \operatorname{Var}\left( \sum_{l=1}^{N_k}Y_{kl} \mid A_k, N_k \right)\mid A_i=a \right]  \\
     =&\overline{N}^2 \left\{ (M-1) \, CV^2 + (M-1)^2 \right\} \theta_a + \overline{N}^2 (1+ CV^2)(M-1)\left\{q\theta_1 + (1-q)\theta_0 \right\}  \\
     =& (M-1)^2\overline{N}^2 \theta_a  + \underbrace{ (M-1)\overline{N}^2\left[ CV^2\, \theta_a^2 + (1+\, CV^2) \left\{q\theta_1 + (1-q)\theta_0\right\}\right]}_{\mathcal{C}_1},
\end{align*}
where \(\theta_a =\operatorname{Var}\left(\sum_{j=1}^{N_i}Y_{ij} \mid A_i=a, N_i\right) =p_a(1-p_a)\overline{N}\left[ 1 + \left\{ (1+CV^2) \overline{N}  - 1 \right\}\rho_a \right] \). Because the term \(\mathcal{C}_1 = O_p(M)\), and assume \(\rho_1 = \rho_0 = \rho\), and due to the arm-invariant cluster size, the asymptotic variance in Theorem 1 for \(W_D\) reduces to 
\begin{align*}
\sigma_D^2  & = \left\{\frac{p_1(1-p_1)}{qM\overline{N}}  + \frac{p_0(1-p_0)}{(1-q)M\overline{N}} \right\} \left[ 1+ \left\{ (1+CV^2) \right\}\overline{N} -1 \right]\rho + o_p(M^{-1}) \\
&= \left\{\frac{p_1(1-p_1)}{qM\overline{N}}  + \frac{p_0(1-p_0)}{(1-q)M\overline{N}} \right\}  \,\text{VIF}   + o_p(M^{-1}).   
\end{align*}

For CRT, it is asymptotically equivalent to the variance expression of risk difference estimator in CRTs discussed in \citet{cornfield1978symposium} and \citet{donner1981randomization} with an independent correlation structure. 
 Similarly for \(\sigma_R^2\), we have 
\begin{align*}
    \sigma_R^2 &= \left\{\frac{2\,\{1/(1-\pi_{\text{tie}})\}}{\,1-\left(\{1/(1-\pi_{\text{tie}})\}W_D\right)^2}\right\}^{\!2} \times \left\{\frac{p_1(1-p_1)}{qM\overline{N}}  + \frac{p_0(1-p_0)}{(1-q)M\overline{N}} \right\}  \,\text{VIF}   + o_p(M^{-1})  \nonumber  \\
    & = \frac{(p_1+p_0 - 2p_1p_0)^2}{4p_1^2p_0^2(1-p_1)^2(1-p_0)^2} \times \left\{\frac{p_1(1-p_1)}{qM\overline{N}}  + \frac{p_0(1-p_0)}{(1-q)M\overline{N}} \right\}  \,\text{VIF}   + o_p(M^{-1}) \nonumber \\
    & = \frac{1}{M \overline{N}} \frac{(p_1+ p_0 - 2p_1p_0)^2}{4p_1p_0(1-p_1)(1-p_0)} \left\{ \frac{1}{qp_0(1-p_0)} + \frac{1}{(1-q) p_1(1-p_1)} \right\} \, \text{VIF} + o_p(M^{-1}).
\end{align*}

Under the null hypothesis $W_D=0$, i.e., $p_1=p_0$, we have
\begin{align*}
    \sigma_R^2 =\frac{1}{M \overline{N}} \left\{ \frac{1}{qp_0(1-p_0)} + \frac{1}{(1-q) p_1(1-p_1)} \right\} \, \text{VIF}  + o_p(M^{-1}),
\end{align*}
which is the variance expressions for studying log odds ratio in CRTs \citep{shih1997sample,rutterford2015methods}. 

\section{Connection between Theorem 1 and the asymptotic variance derived from clustered U-statistics} \label{supp:sec:u-stat}

In this section, we will show that the asymptotic variance in Theorem \ref{theorem:u-var} is asympotically equivalent to the variance proposed in \citet{zhang2021inference}.

Following the method in \citet{zhang2021inference}, the two kernels are defined as \(\phi_{\text{win}}(\bm{Y}_{ij}, \bm{Y}_{kl}) = \mathbb{I} (\bm{Y}_{ij} \succ \bm{Y}_{kl} \mid A_i=1, A_k=0 ) \) and \(\phi_{\text{loss}}(\bm{Y}_{ij},\bm{Y}_{kl}) = \mathbb{I}(\bm{Y}_{ij} \prec \bm{Y}_{kl} \mid A_i=1, A_k=0)\), with two U-statistics defined as 
\begin{align*}
U_{\text{win}} &= \frac{1}{q(1-q)M^2} \sum_{i=1}^{M}\sum_{k=1}^{M} \sum_{j=1}^{N_i} \sum_{k=1}^{N_k} \phi_{\text{win}}(\bm{Y}_{ij},\bm{Y}_{kl}),\\
U_{\text{loss}} &= \frac{1}{q(1-q)M^2} \sum_{i=1}^{M}\sum_{k=1}^{M} \sum_{j=1}^{N_i} \sum_{k=1}^{N_k} \phi_{\text{loss}}(\bm{Y}_{ij},\bm{Y}_{kl}),
\end{align*}
and \(\widehat{W}_R = U_{\text{win}}/U_{\text{loss}}\). By \citet{zhang2021inference} and  \citet{lee2005generalized}, as \(M\rightarrow\infty\), \(\log(U_{\text{win}}/U_{\text{loss}})\) converges to the normal distribution with mean \(\log(W_R)\) and variance 
\begin{align*}
    \zeta_R^2 = \frac{\operatorname{Var}(U_{\text{win}})}{\overline{N}^4\pi_{\text{win}}^2} - \frac{2 \ \operatorname{Cov} (U_{\text{win}},U_{\text{loss}}) }{\overline{N}^4\pi_{\text{win}}\pi_{\text{loss}}} + \frac{\operatorname{Var}(U_{\text{loss}})}{\overline{N}^4\pi_{\text{loss}}^2}.
\end{align*}
From \citet{zhang2021inference}, we know that 
\begin{eqnarray*}
&&\operatorname{Var}(U_{\text{win}})\\
&=& \frac{\overline{N}^2}{qM}\operatorname{Var} \left[\sum_{j=1}^{N_i}  \left\{\mathbb{P}_{\bm{Y}_{kl}}\left(\bm{Y}_{ij} \succ \bm{Y}_{kl}\mid A_i=1,A_k=0 \right)  - \mathbb{P}_{\bm{Y}_{ij},\bm{Y}_{kl}} \left(\bm{Y}_{ij} \succ \bm{Y}_{kl} \mid A_i=1,A_k=0 \right) \right\} \right]  \\
&+& \frac{\overline{N}^2}{(1-q)M}\operatorname{Var}\left[ \sum_{j=1}^{N_i}  \left\{\mathbb{P}_{\bm{Y}_{ij}}\left(\bm{Y}_{ij} \succ \bm{Y}_{kl} \mid A_i=1,A_k=0\right)  - \mathbb{P}_{\bm{Y}_{ij},\bm{Y}_{kl}} \left(\bm{Y}_{ij} \succ \bm{Y}_{kl} \mid A_i=1,A_k=0 \right) \right\} \right], \\
&&\operatorname{Var}(U_{\text{loss}}) \\
&=& \frac{\overline{N}^2}{qM}\operatorname{Var} \left[\sum_{j=1}^{N_i}  \left\{\mathbb{P}_{\bm{Y}_{kl}}\left(\bm{Y}_{ij} \prec \bm{Y}_{kl} \mid A_i=1,A_k=0\right)  - \mathbb{P}_{\bm{Y}_{ij},\bm{Y}_{kl}} \left(\bm{Y}_{ij} \prec \bm{Y}_{kl} \mid A_i=1,A_k=0 \right) \right\} \right]  \\
& + &\frac{\overline{N}^2}{(1-q)M}\operatorname{Var}\left[ \sum_{j=1}^{N_i}  \left\{\mathbb{P}_{\bm{Y}_{ij}}\left(\bm{Y}_{ij} \prec \bm{Y}_{kl} \mid A_i=1,A_k=0 \right)  - \mathbb{P}_{\bm{Y}_{ij},\bm{Y}_{kl}} \left(\bm{Y}_{ij} \prec \bm{Y}_{kl} \mid A_i=1,A_k=0 \right) \right\} \right], \\
&&\operatorname{Cov}(U_{\text{win}},U_{\text{loss}}) \\
&=& \frac{\overline{N}^2}{qM}\operatorname{Cov} \Bigg[ \sum_{j=1}^{N_i}  \left\{\mathbb{P}_{\bm{Y}_{kl}}\left(\bm{Y}_{ij} \succ \bm{Y}_{kl} \mid A_i=1,A_k=0 \right)  - \mathbb{P}_{\bm{Y}_{ij},\bm{Y}_{kl}} \left(\bm{Y}_{ij} \succ \bm{Y}_{kl} \mid A_i=1,A_k=0 \right) \right\}, \\
&& \sum_{j=1}^{N_i}  \left\{\mathbb{P}_{\bm{Y}_{kl}}\left(\bm{Y}_{ij} \prec \bm{Y}_{kl} \mid A_i=1,A_k=0\right)  - \mathbb{P}_{\bm{Y}_{ij},\bm{Y}_{kl}} \left(\bm{Y}_{ij} \prec \bm{Y}_{kl} \mid A_i=1,A_k=0 \right) \right\}
    \Bigg]\\
& +& \frac{\overline{N}^2}{(1-q)M} \operatorname{Cov}\Bigg[ \sum_{j=1}^{N_i}  \left\{\mathbb{P}_{\bm{Y}_{ij}} \left(\bm{Y}_{ij} \succ \bm{Y}_{kl} \mid A_i=1,A_k=0 \right)  - \mathbb{P}_{\bm{Y}_{ij},\bm{Y}_{kl}} \left(\bm{Y}_{ij} \prec \bm{Y}_{kl} \mid A_i=1,A_k=0 \right) \right\},  \\
&& \sum_{j=1}^{N_i}  \left\{\mathbb{P}_{\bm{Y}_{ij}}\left(\bm{Y}_{ij} \prec \bm{Y}_{kl} \mid A_i=1,A_k=0\right)  - \mathbb{P}_{\bm{Y}_{ij},\bm{Y}_{kl}} \left(\bm{Y}_{ij} \prec \bm{Y}_{kl} \mid A_i=1,A_k=0 \right) \right\} 
    \Bigg].   
\end{eqnarray*}

It is noted that  \(\overline{N}^2\pi_{\text{win}} = \mathbb{E}(U_{\text{win}})\), and \(\overline{N}^2\pi_{\text{loss}} = \mathbb{E}(U_{\text{loss}})\). To simplify the above equations, for $a\in \{0,1\}$, we denote
\begin{align*}
    &\mathcal{W}_i^{(a)} = \sum_{j=1}^{N_i} \left\{m_a^{\text{win}}(\bm{Y}_{ij}) - \pi_{\text{win}}\right\},\qquad \mathcal{L}_i^{(a)} =  \sum_{j=1}^{N_i} \left\{m_a^{\text{loss}}(\bm{Y}_{ij}) - \pi_{\text{loss}}\right\},\\
       &  m_1^{\text{win}}(\bm y) = \mathbb{P}_{\bm{Y}_{kl}} (\bm y \succ \bm{Y}_{kl} \mid A_i=1, A_k=0),\\
    & m_0^{\text{win}}(\bm y) = \mathbb{P}_{\bm{Y}_{kl}} ( \bm{Y}_{kl} \succ \bm y \mid A_i=1, A_k=0),\\
    &  m_1^{\text{loss}}(\bm y) = \mathbb{P}_{\bm{Y}_{kl}} (\bm y \prec \bm{Y}_{kl} \mid A_i=1, A_k=0),\\
    & m_0^{\text{loss}}(\bm y) = \mathbb{P}_{\bm{Y}_{kl}} ( \bm{Y}_{kl} \prec \bm y \mid A_i=1, A_k=0).
\end{align*}
Then
\begin{align*}
    & \zeta_R^2 = \frac{1}{qM\overline{N}^2}  \mathsf{\Xi}_1 +\frac{1}{(1-q)M\overline{N}^2} \mathsf{\Xi}_0,\\
    & \mathsf{\Xi}_a  = \operatorname{Var} \left\{\left( \frac{\mathcal{W}_i^{(a)}}{\pi_{\text{win}}} - \frac{\mathcal{L}_i^{(a)}}{\pi_{\text{loss}}}   \right)    \right\},\\
    & \frac{\mathcal{W}_i^{(a)}}{\pi_{\text{win}}} - \frac{\mathcal{L}_i^{(a)}}{\pi_{\text{loss}}} = \sum_{j=1}^{N_i} \left\{ \frac{m_a^{\text{win}}(\bm{Y}_{ij}) - \pi_{\text{win}} }{\pi_{\text{win}}} - \frac{m_a^{\text{loss}}(\bm{Y}_{ij}) - \pi_{\text{loss}} }{\pi_{\text{loss}}}  \right\}
\end{align*}

If we denote \(m_a^s = m_a^{\text{win}} - m_a^{\text{loss}}\), and \(m_a^t = m_a^{\text{win}} + m_{a}^{\text{loss}}\), we can show that \(\mathbb{E}(m_a^s) = W_D\) and \(\mathbb{E}(m_a^{t}) = 1-\pi_{\text{tie}}\). Note that $\pi_{\text{win}}=\frac{1-\pi_{\text{tie}}+W_D}{2}$ and $\pi_{\text{loss}}=\frac{1-\pi_{\text{tie}}-W_D}{2}$, so we have
\[
\frac{m_a^{\text{win}}}{\pi_{\text{win}}} - \frac{m_a^{\text{loss}}}{\pi_{\text{loss}}} = \frac{2}{(1-\pi_{\text{tie}})^2 - W_D^2} \left\{(1-\pi_{\text{tie}})(m_a^{\text{win}} - m_a^{\text{loss}}) - W_D (m_a^{\text{win}} + m_a^{\text{loss}}) \right\}.
\]
Therefore, 
\[
\mathsf{\Xi}_a  = \left\{\frac{2}{(1-\pi_{\text{tie}})^2 - W_D^2} \right\}^2 \operatorname{Var} \left[\sum_{j=1}^{N_i}  \left\{ (1-\pi_{\text{tie}}) \widetilde{m}_a^{s} (\bm{Y}_{ij}) - W_D \widetilde{m}_a^{t} (\bm{Y}_{ij}) \right\}   \right],
\]
where \(\widetilde{m}_a^{s} = m_a^s - W_D\), and \(\widetilde{m}_a^t = m_a^t - (1-\pi_{\text{tie}})\).

Note that the order of $\operatorname{Var}(W_D)$ is $M^{-3}$ by Theorem 1, and we only keep the leading term of order \(M^{-1}\) in \(\mathsf{\Xi}_a\), then we can simplify the above as:
\[
\operatorname{Var} \left[  \sum_{j=1}^{N_i} \left\{ (1-\pi_{\text{tie}}) \widetilde{m}_a^{s} (\bm{Y}_{ij}) - W_D \widetilde{m}_a^{t} (\bm{Y}_{ij}) \right\}   \right] = (1-\pi_{\text{tie}})^2 \operatorname{Var}\left\{ \sum_{j=1}^{N_i} m_a^s(\bm{Y}_{ij})  \right\} + o_p(1) .
\]
Since the empirical distribution of the treatment arm is 
\[\sum_{j=1}^{N_i} m_1^{s}(\bm{Y}_{ij}) = \frac{1}{(1-q)M\overline{N}}\sum_{j=1}^{N_i} \sum_{k=1}^{M} \sum_{l=1}^{N_k} \mathbb{I}(A_k=0) s_{ij,kl} + o_p(M^{-1}) = M^{-1}\overline{N}^{-1} (S_i\mid A_i=1) +  o_p(M^{-1}). \]
Then, we obtain \(\operatorname{Var}\left\{ \sum_{j=1}^{N_i} m_1^s(\bm{Y}_{ij}) \right\}  = M^{-2}\overline{N}^{-2}\sigma_1^2 +o_p(M^{-2}) \). Similarly, we can obtain  \(\operatorname{Var}\left\{ \sum_{j=1}^{N_i}m_0^{s}(\bm{Y}_{ij})\right\}\) and show that
\begin{align*}
    & \frac{1}{qM\overline{N}^2} \operatorname{Var}\left( \sum_{j=1}^{N_i}\widetilde{m}_1^s (\bm{Y}_{ij}) \right) + \frac{1 }{(1-q)M\overline{N}^2} \operatorname{Var}\left(  \sum_{j=1}^{N_i}\widetilde{m}_0^s (\bm{Y}_{ij}) \right) \\
    & = \frac{\sigma_1^2}{qM^3\overline{N}^4} + \frac{\sigma_0^2}{(1-q)M^3\overline{N}^4} + o_p(M^{-2})
\end{align*}
Combining all the above, we have
\begin{align*}
    \zeta_R^2 = \left\{ \frac{2(1-\pi_{\text{tie}}) }{(1-\pi_{\text{tie}})^2 - W_D^2} \right\}^2 \frac{1}{M^3\overline{N}^4}\left( \frac{\sigma_1^2}{q} + \frac{\sigma_0^2}{1-q} \right) + o_p(M^{-3}) = \sigma_R^2 + o_p(M^{-3}).
\end{align*}

\section{Proof of Theorem 2} \label{supp:sec:theorem2}

First, we introduce the following lemma, which establishes the asymptotic equivalence between the probability of observing a tied pair across treatment arms and the overall tie probability under the transitive property. 

\begin{lemma} \label{lem:tie_equ}
Let $n = \sum_{i=1}^{M}N_i$ individuals be partitioned into tie blocks $g=1,\dots,G$ of sizes $t_g\ge 1$ induced solely by the outcome-based tie rule of the win statistic, with $\sum_{g=1}^G t_g = n$. Consider a 2-arm design with $n_1$ treated and $n_0=n-n_1$ controlled individuals, assigned either (i) by complete randomization with fixed $(n_1,n_0)$ or (ii) by independent Bernoulli with probability $q = E(A_i)$ assignment with $q\in(0,1)$. Define
\[
\pi_{\text{tie}}
= \mathbb{E}\left(\frac{\sum_{g=1}^G t_g^{1} t_g^{0}}{n_1 n_0}\right),
\qquad
p_T
= \mathbb{E} \left\{\frac{\sum_{g=1}^G \binom{t_g}{2}}{\binom{n}{2}}
= \frac{\sum_{g=1}^G t_g(t_g-1)}{n(n-1)} \right\},
\]
where $t_g^1$ and $t_g^0$ denote the treated and controlled counts in block $g$. Then:
\begin{enumerate}
    \item Under complete randomization, $\pi_{\text{tie}}=p_T$ exactly for any finite $n$.
    \item Under Bernoulli assignment, $\pi_{\text{tie}}=p_T+o_p(1)$ and hence $\pi_{\text{tie}}\to p_T$ in probability as $n\to\infty$.
\end{enumerate}
\end{lemma}

\begin{proof}
We condition throughout on the block structure $\{t_g\}$, which is independent of treatment allocation by design.

We first consider the complete randomization case. Given $(n_1,n_0)$, the treated count $t_g^1$ in block $g$ follows a hypergeometric distribution with population size $n$. Then, its expectation and variance are
\[
\mathbb{E}(t_g^1)=t_g\frac{n_1}{n},\qquad
\mathrm{Var}(t_g^1)=t_g\frac{n_1}{n}\Big(1-\frac{n_1}{n}\Big)\frac{n-t_g}{n-1}.
\]
Since $t_g^0=t_g-t_g^1$, we have
\[
\mathbb{E}(t_g^1 t_g^0)
=\mathbb{E}\big\{t_g^1(t_g-t_g^1)\big\}
=\mathbb{E} \left\{ \frac{n_1 n_0}{n(n-1)}\,t_g(t_g-1) \right\}.
\]
Summing over all blocks and dividing by $n_1n_0$ gives exact equality, that is, 
\[
\pi_{\text{tie}}
= \mathbb{E} \left\{\frac{\sum_{g=1}^G t_g(t_g-1)}{n(n-1)}\right\}
=\mathbb{E} \left\{\frac{\sum_{g=1}^G \binom{t_g}{2}}{\binom{n}{2}}  \right\}
= p_T.
\]

Next, we consider the Bernoulli assignment. Let $A_i\stackrel{\text{i.i.d.}}{\sim}\mathrm{Bernoulli}(q)$ denote treatment indicators. For block $g$, we have
\[
\mathbb{E}(t_g^1 t_g^0)
= t_g\,\mathbb{E}(t_g^1)-\mathbb{E}\{(t_g^1)^2\}
= q(1-q)(t_g^2-t_g).
\]
Therefore,
\[
\mathbb{E}\left(\sum_{g=1}^G t_g^1 t_g^0 \right)
= q(1-q) \mathbb{E} \left\{\sum_{g=1}^G t_g(t_g-1) \right\}.
\]
Since $n_1n_0/n^2 \to q(1-q)$ and $n(n-1)/n^2 \to 1$ as $M\to\infty$, and cluster sizes \(N_i\) are finite, through Taylor expansion and  Slutsky's theorem, we can show that
\[
\pi_{\text{tie}}
= \mathbb{E}\left\{\frac{\sum_{g=1}^G t_g(t_g-1)}{n(n-1)}\cdot \frac{q(1-q)}{n_1n_0/n^2} \right\}
= p_T + o_p(1).
\]
So, $\pi_{\text{tie}}\stackrel{\cal P}{\longrightarrow} p_T$.

This completes the proof.
\end{proof}

The correct definition of $p_T$ uses $\binom{t_g}{2}$ and $\binom{n}{2}$, representing the probability that a uniformly sampled unordered pair from the full cohort lies in the same tie block. The key difference between \(\pi_{\text{tie}}\) and \(p_T\) arises from the estimation of the treatment assignment probability \(n_1/n\) and its variance, which is \(O_p(M^{-1})\), and thus \(\pi_{\text{tie}} = p_T + o_p(1)\) as \(M\rightarrow \infty\).

Next, we prove Theorem \ref{theorem:var_single}.
\begin{proof}
For each individual $j$ in cluster $i$, consider the pairwise-count decomposition
\[
W_{ij} = \sum_{k=1}^M \sum_{l=1}^{N_k} \mathbb{I}\{\mathbf{Y}_{ij} \succ \mathbf{Y}_{kl}\}, \quad
L_{ij} = \sum_{k=1}^M \sum_{l=1}^{N_k} \mathbb{I}\{\mathbf{Y}_{ij} \prec \mathbf{Y}_{kl}\}, \quad
T_{ij} = \sum_{k=1}^M \sum_{l=1}^{N_k} \mathbb{I}\{\mathbf{Y}_{ij} = \mathbf{Y}_{kl}\} - 1,
\]
which satisfies $W_{ij} + L_{ij} + T_{ij} = n - 1$.  
Let $R_{ij}$ denote the mid-rank of $\mathbf{Y}_{ij}$ among all $n$ observations. Then
\[
R_{ij} = W_{ij} + 1 + \frac{T_{ij}}{2}, \qquad
\phi_{ij} = W_{ij} - L_{ij} = 2R_{ij} - (n+1).
\]
We have already known that \citep{kruskal1952use}:
\[
\mathbb{E}(R_{ij}^2) = \frac{(n+1)(2n+1)}{6} - \frac{1}{12n} \sum_{g=1}^G (t_g^3 - t_g).
\]
Since $\mathbb{E}[R_{ij}] = (n+1)/2$, it follows that
\begin{equation}
\mathrm{Var}(R_{ij})
= \frac{N^2 - 1}{12} - \frac{1}{12n} \sum_{g=1}^G (t_g^3 - t_g).
\label{eq:sigmaR}
\end{equation}

It is noted that $\sum_{g=1}^G t_g=n$, when $M\rightarrow \infty$, 
\begin{align*}
p_T = \mathbb{E} \left\{ \frac{\sum_{g=1}^{G} t_g(t_g-1) }{n(n-1)}   \right\} = \mathbb{E} \left\{ \frac{\sum_{g=1}^{G} t_g^2 }{n^2} \right\} + o_p(1) .
\end{align*}
Therefore \( \{\sum_{g=1}^{G} (t_g/n)^2\}^2 \stackrel{\mathcal{P}}{\rightarrow} p_T^2 \) in probability. 

Note that by Cauchy-Schwarz inequality,
\[
0 \leq \frac{\sum_{g=1}^{G}t_g^3 }{n^3} - \left( \frac{\sum_{g=1}^{G} t_g^2}{n^2} \right)^2 \leq \left\{ (\max_g t_g )/n \right\}^2.
\]
Since \(\max_g t_g = O_p(1)\), a uniformly bounded block sizes, then we have \(\sum_{g=1}^{G} (t_g/n)^3 - [\sum_{g=1}^{G} (t_g/n)^2]^2 = O_p(n^{-2}) \). Therefore
\begin{align*}
    \frac{1}{12n} \sum_{g=1}^G (t_g^3 - t_g) & = \frac{1}{12} \left[ \frac{\sum_{g=1}^{G} t_g^3}{n} -1\right] \\
    & = \frac{1}{12}\left(n^2 \sum_{g=1}^{G} \left(\frac{t_g}{n}\right)^{3} -1\right)\\
    &= \frac{1}{12}\Bigg(n^2\left[\sum_{g=1}^{G}\left(\frac{t_g}{n}\right)^{2}\right]^{2} - 1\Bigg)
   + \frac{n^2}{12}\left\{\sum_{g=1}^{G}\left(\frac{t_g}{n}\right)^{3}
   - \left[\sum_{g=1}^{G}\left(\frac{t_g}{n}\right)^{2}\right]^{2}\right\}\\
&= \frac{n^2 p_T^{2} - 1}{12}
   + \frac{n^2}{12}\left\{\sum_{g=1}^{G}\left(\frac{t_g}{n}\right)^{3}
   - \left[\sum_{g=1}^{G}\left(\frac{t_g}{n}\right)^{2}\right]^{2}\right\} + O_p(1) \\
   & = \frac{n^2 p_T^{2} - 1}{12} + O_p(1).
\end{align*}

Then, $\mathrm{Var}(R_{ij})$ in (\ref{eq:sigmaR}) can be simplified as
\begin{equation}
\mathrm{Var}(R_{ij})  = \frac{n^2 - 1}{12} - \frac{n^2 p_T^{2} - 1}{12} + O_p(1) = \frac{n^2(1-p_T^2)}{12} + O_p(1),
\label{eq:sigmaR1}
\end{equation}
which implies $12 \mathrm{Var}(R_{ij}) / n^2 = 1 - p_T^2 + o_p(1)$. In the special case of equal tie-block size $t$ (so $G = n/t$), we have $\mathrm{Var}(R_{ij}) = (n^2 - t^2)/12$ and
\[
p_T = \frac{\sum_{g=1}^G t_g(t_g - 1)}{n(n-1)} = \frac{t - 1}{n - 1},
\] we can obtain the same result in (\ref{eq:sigmaR1}).
Therefore, by Lemma \ref{lem:tie_equ}, we have 
\begin{eqnarray}
\mathrm{Var}(R_{ij})=\frac{n^2}{12}(1-p_T^2)+O_p(1)=\frac{n^2}{12}(1-\pi_{\text{tie}}^2)+O_p(1). \label{rank_V}
\end{eqnarray}

The cluster-level rank contrast is
\[
S_i = \sum_{j=1}^{N_i} \phi_{ij} = 2 \sum_{j=1}^{N_i} R_{ij} - (n+1) N_i.
\]
Because $\mathbb{E}(R_{ij}) = (n+1)/2$, we have $\mathbb{E}(S_i) = 0$ and
\[
\mathrm{Var}(S_i) = 4\mathrm{Var}\left( \sum_{j=1}^{N_i} R_{ij} \right)
= 4 \left\{ \sum_{j=1}^{N_i} \mathrm{Var}(R_{ij}) + \sum_{j \neq j'} \mathrm{Cov}(R_{ij}, R_{ij'}) \right\}.
\]
The rank ICC can be written as 
\[
\rho = \frac{\mathrm{Cov}(R_{ij}, R_{ij'})}{\mathrm{Var}(R_{ij})}, \quad j \neq j',
\]
and using $\mathrm{Var}(R_{ij})$ from \eqref{eq:sigmaR}, we obtain
\begin{align*}
    \mathbb{E}(S_i^2) &= 4\,\mathrm{Var}(R_{ij}) \{ N_i + \rho N_i (N_i - 1) \} \\ \label{rank_second}
    & = 4 \, \mathrm{Var}(R_{ij})\overline{N} \left\{1+ \rho \overline{N}^2(CV^2 +1) - \rho \overline{N} \right\} \\
    & = 4\overline{N} \, \mathrm{Var}(R_{ij}) \text{VIF},
\end{align*}
where \(\text{VIF} = 1+ \rho \left\{\overline{N}(CV^2 +1)-1 \right\} \). We've already known that \(\sigma_a^2=s_a-\mu_a^2\), where \(s_a=\mathbb E(S_i^2\mid A_i=a)\). Using the fact that \(\sum_i^M S_i=0\Rightarrow q\mu_1+(1-q)\mu_0=0\) and the identity
\[
\mathbb E(\overline S_1-\overline S_0)=\mu_1-\mu_0
=\frac{M\,n_1n_0}{M_1M_0}\,W_D,
\]
we get \(\mu_1= (1-q)\,M\overline N^{\,2}\,W_D\), and \(\mu_0= -q\,M\overline N^{\,2}\,W_D\)
(as \(n_1 \rightarrow_p qM\overline N\), \(n_0 \rightarrow_p (1-q)M\overline N\)).
Hence
\[
\frac{\sigma_1^2}{q}+\frac{\sigma_0^2}{1-q}
=\frac{s_1}{q}+\frac{s_0}{1-q}
-\Big(\frac{\mu_1^2}{q}+\frac{\mu_0^2}{1-q}\Big)
=\frac{s_1}{q}+\frac{s_0}{1-q}
- M^{2}\overline N^{\,4}\!\left\{\frac{(1-q)^2}{q}+\frac{q^2}{1-q}\right\}W_D^2.
\]
Following the asymptotic variance of \(\widehat{W}_D\) in  \eqref{wd_asymp}, we can show that
\[
\sigma_D^2
= \frac{1}{M^{3}\overline N^{\,4}}
\left(\frac{s_1}{q}+\frac{s_0}{1-q}\right)
-\left\{\frac{(1-q)^2}{q}+\frac{q^2}{1-q}\right\}\frac{W_D^2}{M}.
\]
Assume the following conditions hold:
\begin{itemize}
\item[(H1)] Common ICC \(\rho \in [0,1]\) across arms. \label{H1}
\item[(H2)] 
The first two moments of $N_i$ are the same across arms, namely,
\[
\mathbb{E}(N_i\mid A_i=a)=\overline N,\qquad
\operatorname{Var}(N_i\mid A_i=a)=\overline N^{\,2}\,CV^2,
\qquad a\in\{0,1\}.
\] \label{H2}
\item[(H3)]
There exists $\delta>0$ such that
\[
\sup_{M}\frac{1}{M}\sum_{i=1}^M \mathbb{E}\big(\,|S_i|^{2+\delta}\,\big)<\infty,
\qquad
\frac{1}{M}\sum_{i=1}^M \mathbb{E}\!\left(S_i^2\,
\mathbb{I}\{|S_i|>\varepsilon\sqrt{M}\}\right)\to 0
\ \ \text{for all }\varepsilon>0.
\]   (Moment/Lindeberg condition for cluster scores.)\label{H3}
\end{itemize}

Under (H1) - (H3), 
\(s = \mathbb{E}(S_i^2) = qs_1 + (1-q)s_0\), and \(\sigma_1^2 - \sigma_0^2 = O_p(M^{-1})\), we can show that 
\begin{align*}
s_1 - s_0 &= (\mu_1^2 - \mu_0^2) + O_p(M^{-1}) =  (1-2q) (M\overline{N}^2 W_D)^2 + O_p(M^{-1}),\\
s &= \mathbb{E}(S_i^2) = \frac{M^2 \overline{N}^3}{3} (1 - \pi_{\text{tie}}^2) \, \text{VIF} + O_p(M^2).
\end{align*}
Then
\[
\frac{s_1}{q}+\frac{s_0}{1-q}
=s\!\left(\frac{1}{q}+\frac{1}{1-q}\right) + \frac{(1-2q)^2}{q(1-q)} (M\overline{N}^2W_D)^2 + o_p(M^2).
\]
Combining the above, we can show that as \(\sigma_D^2\) is asymptotically equivalent to
\begin{align*}
\sigma_D^2
& = \frac{1-\pi_{\text{tie}}^2}{3\,M\,\overline N}\,
\left(\frac{1}{q}+\frac{1}{1-q}\right)\, \text{VIF}
-\left(\frac{(1-q)^2}{q}+\frac{q^2}{1-q}\right)\frac{W_D^2}{M} + \frac{(1-2q)^2}{Mq(1-q)}W_D^2
+ o_p(M^{-1}) \\
& = \nu_D^2 + o_p(M^{-1})
\end{align*}
Similarly, we can show that \(\nu_R^2 = \left[ \frac{2\{1/(1 - \pi_{\text{tie}})\}}{1-(\{1/(1 - \pi_{\text{tie}})\} W_D)^2}  \right]^2 \nu_D^2 \), and \(\nu_O^2 = \left[ \frac{2}{1-W_D^2}\right]^2 \nu_D^2  \).
\end{proof}

For an individually randomized trial, we have the following remark.
\begin{remark} \label{supp:rmk:irt}
    For IRT with \(N\) individuals  \(\text{VIF} = 1\), the variances in Theorem 2 for \(W_D\),  \(\log (W_R)\), and \(\log(W_O)\) become
\begin{align*}
    \nu_{I,D}^2 & = \frac{1-\pi_{\text{tie}}^2}{3\,N}\,
\left(\frac{1}{q}+\frac{1}{1-q}\right)
-\frac{W_D^2}{M}\\
 \nu_{I,R}^2 & = \left[ \frac{2\{1/(1 - \pi_{\text{tie}})\}}{1-(\{1/(1 - \pi_{\text{tie}})\} W_D)^2}  \right]^2 \nu_{I,D}^2\\
\nu_{I,O}^2 & = \left[ \frac{2}{1-W_D^2}\right]^2 \nu_{I,D}^2.
\end{align*}
\end{remark}
Under the null with \(\Delta = 0\) (\(W_D = 0, \, \log (W_R) = 0, \, \log (W_O) = 0 \)), or the contiguous alternative with \(W_D = O_p(M^{-1/2})\), the asymptotic variance for \(\log(\widehat{W}_R)\) becomes
\begin{align*}
    \nu_{I,R}^2 &= \frac{1-\pi_{\text{tie}}^2}{3\,N}\,
\left(\frac{1}{q}+\frac{1}{1-q}\right) 4(1-\pi_{\text{tie}})^{-2} \\
& = \frac{4(1+\pi_{\text{tie}})}{3Nq(1-q)(1-\pi_{\text{tie}})},
\end{align*}
which is exactly the variance for \(\log(\widehat{W}_R)\) proposed in \citet{yu2022sample}. The test statistic \(\log (\widehat{W}_D)/\widehat{\nu}_{I,R}^2 \) where  \(\widehat{\nu}_{I,R}^2\) is the finite-sample estimator of \(\nu_{I,R}^2\) and is also equivalent to that in \citet{yu2022sample}.

\subsection{Optimal efficiency under randomization} \label{supp:sec:optimalq}
We take the derivative of \(\sigma_D^2\) over \(q\) and then set it equal to zero, that is,
\[ \frac{d \, \nu_D^2}{d\,q} = \frac{1-\pi_{\text{tie}}^2}{3M\overline{N}} \text{VIF} \left(-q^{-2} + (1-q)^{-2} \right) = 0. \]
This gives us \(q = \frac{1}{2}\). By taking second derivative, we can also show that 
\[ \frac{d^2\, \nu_D^2}{d \, q^2} = \frac{1-\pi_{\text{tie}}^2}{3M\overline{N}} \text{VIF} \left( \frac{2}{q^3} + \frac{2}{(1-q)^3} \right) > 0\]
since \(q \in (0, 1)\) and $\pi_{\text{tie}} \in (0,1)$. Thus, \(\nu_D^2\) is strictly convex and \(q=\frac{1}{2}\) is global minimum. Therefore, the optimal efficiency for the win-statistics is reached with balanced randomization.

\section{Connection between Theorem 2 and the Mann-Whitney U-statistics for clustered data} \label{supp:sec:mann_withney}

In this section, we will prove that the asymptotic variance provided in Theorem \ref{theorem:var_single} is asymptotically equivalent to Mann-Whitney U-statistics for clustered data for continuous outcome proposed by \citet{rosner1999use} under the null hypothesis.

\begin{figure}
    \centering
    \includegraphics[width=\linewidth]{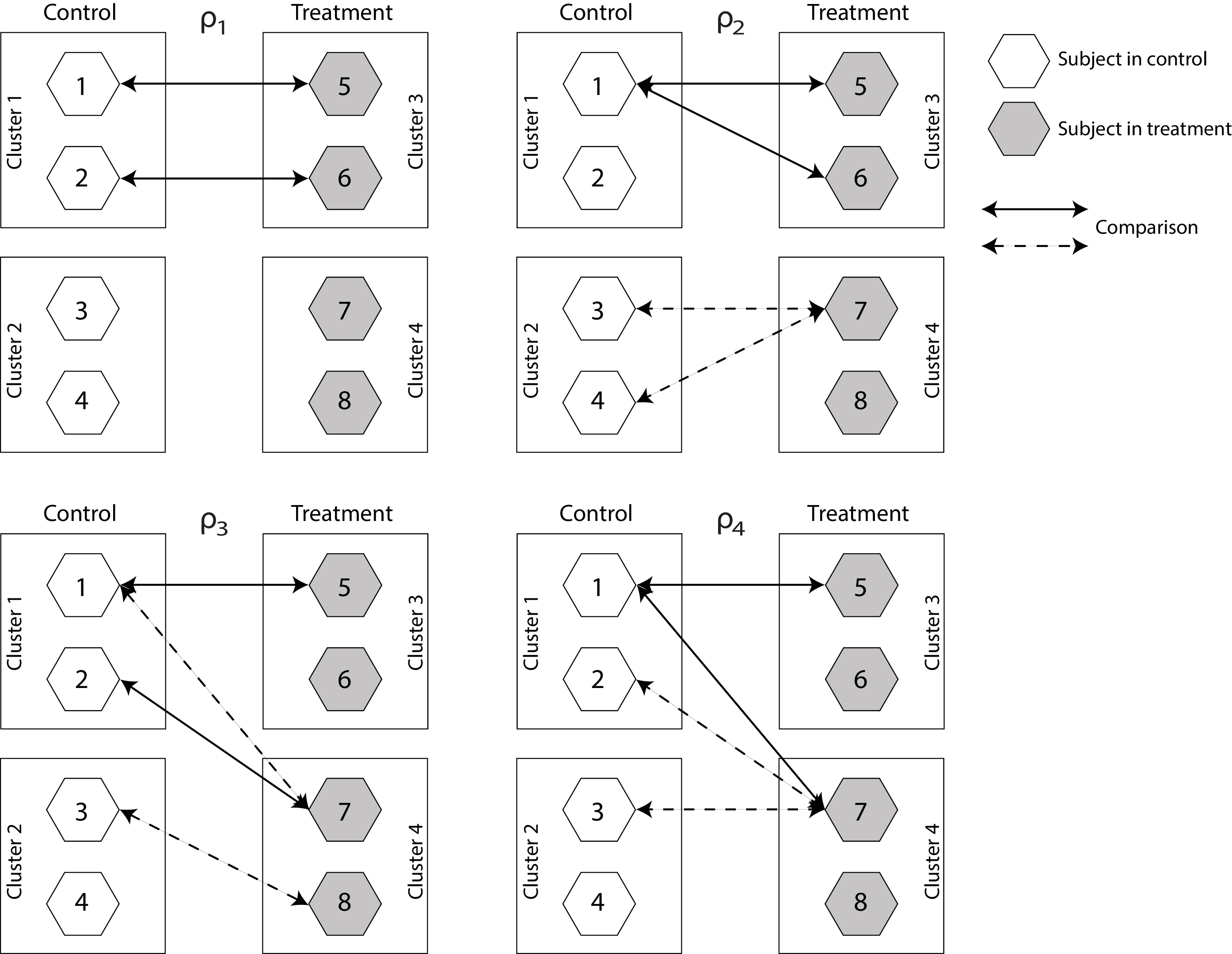}
    \caption{Pairwise correlations across arms and individuals in the CRT. Different line styles represent distinct pairwise comparisons corresponding to $\rho_1$--$\rho_4$, as defined by the correlation structures between units $U_{ij,k\ell}$ under various within- and between-arm configurations.}
    \label{fig:comp_rho}
\end{figure}

Following the \citet{rosner1999use}, where the \(W_r = \sum_{i=1}^{M}\sum_{k=1}^{M} \sum_{j=1}^{N_i} \sum_{l=1}^{N_k} U(Y_{ij},Y_{kl}) \), and \(U(Y_{ij},Y_{kl}) = U_{ij,kl}= \mathbb{I} (Y_{ij} \succ Y_{kl} \mid A_i=1, A_k=0 ) + \frac{1}{2} \mathbb{I}(Y_{ij}=Y_{kl} \mid A_i=1, A_k=0 )\). In this ways, we can find that \(W-L = 2 W_r  - n_1n_0\), thus \(\operatorname{Var}(W_D) = 4 \operatorname{Var}(W_r/n_1 n_0) \). Following \citet{rosner1999use}, define the correlations across-arms, for \(A_i=1, A_k=0\) in the way of 
\(\rho_1=\mathrm{Corr}\!\big(U_{ij_1,k\ell_1},U_{ij_2,k\ell_2}\big)\),
\(\rho_2=\mathrm{Corr}\!\big(U_{ij,k\ell_1},U_{ij,k\ell_2}\big)=\mathrm{Corr}\!\big(U_{ij_1,kl},U_{ij_2,kl}\big)\), and 
\(\rho_3=\mathrm{Corr}\!\big(U_{ij_1,k_1\ell_1},U_{ij_2,k_2\ell_2}\big) = \mathrm{Corr} \!\big(U_{i_1j_1,k\ell_1},U_{i_2j_2,k\ell_2}\big)\), 
\(\rho_4=\mathrm{Corr}\!\big(U_{ij,k_1\ell_1},U_{ij,k_2\ell_2}\big)=\mathrm{Corr}\!\big(U_{i_1j_1,kl},U_{i_2j_2,kl}\big)\) as illustrated in Figure \ref{fig:comp_rho}, 
where \(\rho_1,\rho_2,\rho_3\) correspond to cluster sharing with no shared individual, and \(\rho_4\) corresponds to sharing exactly one individual. For continuous outcome, with mid-rank kernel, the within cluster rank ICC defined in the main paper can be written as
\[\rho = \frac{\operatorname{Cov \{\mathcal{F}^{*}(Y_{ij}) , \mathcal{F}^{*}(Y_{ij'}) \} }}{\operatorname{Var} \{ \mathcal{F}^{*}(Y_{ij}) \}  } = \frac{\operatorname{Cov} (R_{ij}, R_{ij'} ) }{\operatorname{Var}(R_{ij})}.\]

The variance of \(W_r\) following equation (6) in \citet{rosner1999use} is 
\begin{align*}
 \operatorname{Var}(W_r)  =& \Big[n_1n_0+(T_2-n_1)(C_2-n_0)\rho_1\\
 &+\left\{n_1(C_2-n_0)+(T_2-n_1)n_0\right\}\rho_2 \\
     & + \left\{(T_2-n_1)(n_0^2-C_2)+(C_2-n_0)(n_1^2-T_2)\right\}\rho_3\\
     & +\left\{n_1(n_0^2-C_2)+(n_1^2-T_2)n_0\right\}\rho_4\Big]\, \frac{1-\pi_{\text{tie}}}{4},
\end{align*}
where \(T_2=\sum_{i=1}^M A_iN_i^2,\quad C_2=\sum_{i=1}^M(1-A_i)N_i^2\). Considering varying cluster size, we have 
\begin{align*}
T_2 - n_1 &= q M \,\overline N\,\{\overline N\,(1+\mathrm{CV}^2) - 1\} \;+\; o_p(M),\\
C_2 - n_0 &= (1-q) M \,\overline N\,\{\overline N\,(1+\mathrm{CV}^2) - 1\} \;+\; o_p(M), \\
n_0^2 - C_2 &= (1-q)M\,\overline N^{\,2}\bigl\{(1-q)M - (1+\mathrm{CV}^2)\bigr\} \;+\; o_p(M^2), \\
n_1^2 - T_2 &= qM\,\overline N^{\,2}\bigl\{qM - (1+\mathrm{CV}^2)\bigr\} \;+\; o_p(M^2). 
\end{align*}
Due to \((T_2-n_1)(C_2-n_0) = O_p(M^{2}) \), and \(n_1(C_2-n_0)+(T_2-n_1)n_0 = O_p(M^{2})\), the contribution of \((T_2-n_1)(C_2-n_0) /(n_1n_0)^2\) and \(\left\{n_1(C_2-n_0)+(T_2-n_1)n_0 \right\}/(n_1n_0)^2 \) are all \(o_p(M^{-1})\). Through keeping the leading term we have
\begin{align*}
    \frac{\operatorname{Var}(W_r)}{(n_1n_0)^2} = \frac{1-\pi_{\text{tie}}}{4} \frac{1}{q(1-q)M} \left\{ \frac{\overline{N} (1+CV^2) -1 }{\overline{N}} \rho_3 +\frac{\rho_4}{\overline{N}} \right\} + o_p(M^{-1}).
\end{align*}

Under the null, \(Y_{ij} \perp Y_{kl}\) and the treated and control distributions coincide. 
Since \(\mathbb{E}(U_{ij,kl}\mid Y_{ij},A_i=1,A_k=0) = \mathbb{P}(Y_{kl} < Y_{ij} \mid  Y_{ij} ) + \frac{1}{2}\mathbb{P}(Y_{kl} = Y_{ij} \mid Y_{ij} ) = R_{ij}/n  \). Thus, we can show that 
\[\mathbb{E}(U_{ij,kl} ) = \frac{1}{2}, \quad \mathbb{E}(U_{ij,kl}^2 ) =  \frac{1}{2} - \frac{\pi_{\text{tie}}}{4}. \]
The variance can be shown as 
\[\operatorname{Var}(U_{ij,kl}) = \frac{1-\pi_{\text{tie}}}{4}.\]
By the law of total covariance \(\operatorname{Cov}(U_{ij_1,k_1\ell_1},U_{ij_2,k_2\ell_2}) = n^{-2} \operatorname{Cov}(R_{ij_1},R_{ij_2})\), and from \eqref{rank_V}, we can show that
\begin{align*}
\rho_3 &= \frac{\operatorname{Cov}(U_{ij_1,k_1\ell_1} , U_{ij_2,k_2\ell_2})  }{\sqrt{\operatorname{Var}(U_{ij_1,k_1\ell_1})\operatorname{Var}(U_{ij_2,k_2\ell_2})}} = \frac{n^{-2}\operatorname{Cov} (R_{ij_1},R_{ij_2}) }{\sqrt{\operatorname{Var}(U_{ij_1,k_1\ell_1})\operatorname{Var}(U_{ij_2,k_2\ell_2})}} \\
& = \frac{n^{-2}\rho \operatorname{Var}(R_{ij_1})}{\sqrt{\operatorname{Var}(U_{ij_1,k_1\ell_1})\operatorname{Var}(U_{ij_2,k_2\ell_2})}} = \frac{ \frac{1-\pi_{\text{tie}}^2}{12} }{ \frac{1-\pi_{\text{tie}}}{4} }\rho = \frac{1+\pi_{\text{tie}}}{3} \rho.    
\end{align*}
Similarly, by the law of total covariance: \(\operatorname{Cov}(U_{ij,k_1\ell_1},U_{ij,k_2\ell_2}) = n^{-2} \operatorname{Var}(R_{ij})\), therefore,
\[\rho_4 = \frac{\operatorname{Cov}(U_{ij,k_1\ell_1} U_{ij,k_2\ell_2}) }{\sqrt{\operatorname{Var}(U_{ij,k_1\ell_1}) \operatorname{Var}(U_{ij,k_2\ell_2})} } = \frac{n^{-2}\operatorname{Var}(R_{ij})}{ \operatorname{Var}(U_{ij,k_1\ell_1}) } = \frac{ \frac{1-\pi_{\text{tie}}^2}{12} }{ \frac{1-\pi_{\text{tie}}}{4} } = \frac{1+\pi_{\text{tie}}}{3}. \]
With \(\rho_3 = \frac{1+\pi_{\text{tie}}}{3} \rho \), and \(\rho_4 = \frac{1+\pi_{\text{tie}}}{3}\) we can get
\[ \frac{\operatorname{Var}(W_r)}{(n_1n_0)^2} = \frac{1-\pi_{\text{tie}}^2}{12} \frac{1}{q(1-q)M} \left\{ \frac{\overline{N}(1+CV^2)-1}{\overline{N}}\rho + \frac{1}{\overline{N}} \right\} + o_p(M^{-1}), \]
and as a result,
\begin{align*}
    \operatorname{Var}\!\left(\frac{W-L}{n_1 n_0}\right)
&= 4\,\operatorname{Var}\!\left(\frac{W_r}{n_1 n_0}\right) \nonumber\\
& = \frac{1-\pi_{\text{tie}}^2}{3} \frac{1}{q(1-q)M} \frac{1+ \left\{(1+CV^2)\overline{N} -1 \right\}\rho }{\overline{N}}  + o_p(M^{-1}) \\
& = \nu_D^2 + o_p(M^{-1}).
\end{align*}
Therefore, when the outcome is continuous, the asymptotic variance of the clustered Mann--Whitney \(U\)-test statistic \citep{rosner1999use} coincides with the variance expression derived in Theorem \ref{theorem:var_single}.

\section{Proof of Theorem 3} \label{supp:sec:theorem3}
In this section, we will prove the asymptotic variance provided in Theorem \ref{theorem:var_comp}. With a composite endpoint, if we relax the transitive ties as in a single non-survival endpoint, we have
\begin{align*}
\text{Var}(R_{ij}) &= \mathbb{E}(R_{ij}^2) - \left\{\mathbb{E}(R_{ij})\right\}^2 \\
    & = \mathbb{E}(W_{ij}^2) + \mathbb{E}(W_{ij}T_{ij}) + \frac{1}{4}\mathbb{E}(T_{ij}^2) + 2\mathbb{E}(W_{ij})  + \mathbb{E}(T_{ij})+ 1 - \mathbb{E}(R_{ij}). 
\end{align*}
Next, we will focus on the \(\mathbb{E}(R_{ij}^2)\) and beside the assumptions H1 - H3, we assume 
\begin{itemize}
    \item[(H4)] The one- and two-way probabilities, $p_W, p_T, p_{WW}, p_{WT}, p_{TT}$, converge to finite limits and the tie rate satisfies $1-\pi_{\text{tie}}$ bounded away from $0$, such that $\operatorname{Var}(R_{ij})$ and $\rho^{*}$ are bounded and nondegenerate. \label{H4}
\end{itemize}
We can show that the expectation of each term is as follows:
\begin{align*}
    \mathbb{E}\left(W_{ij}^2\right) &= \mathbb{E} \left\{\sum_{(k,l)\neq (i,j)} \mathbb{I}(Y_{ij} \succ Y_{kl})^2\right\} + \mathbb{E}\left\{\sum_{(k,l)\neq (i,j), (m,n)\neq (i,j)} \mathbb{I}(Y_{ij} \succ Y_{kl}) \mathbb{I}(Y_{ij} \succ Y_{mn})\right\} \\
    & = (M\overline{N}- 1)  p_W + (M\overline{N} - 1)(M\overline{N}-2)p_{WW}.
\end{align*}
Similarly, we have
\begin{align*}
\mathbb{E}(T_{ij}^2) = (M\overline{N} -1) p_T + (M\overline{N}-1)(M\overline{N}-2)p_{TT},
\end{align*}
and
\begin{align*}
\mathbb{E}(W_{ij}T_{ij}) &= E\left\{\sum_{(k,l)\neq (i,j), (m,n)\neq (i,j)} \mathbb{I}(Y_{ij}\succ Y_{kl}) \mathbb{I}(Y_{ij}=Y_{mn})   \right\} \\
& = (M\overline{N} - 1) (M\overline{N}-2)p_{WT}.
\end{align*}
Combine all, we have
\begin{align} \label{exp_sqr}
&\mathbb{E}(R_{ij}^2)\nonumber \\
=& 1 + (M\overline{N}-1)\left(3p_W + \frac{5}{4}p_T \right) + (M\overline{N} - 1)(M\overline{N}-2)\left(p_{WW} + p_{WT} + \frac{1}{4} p_{TT}\right).
\end{align}
It is noted that \(\mathbb{E}(R_{ij}) = (M\overline{N}+1)/2\), and for ICC \(\rho^{*}\), we have \(\mathbb{E}(R_{ij}R_{ik})= \rho^{*} \text{Var}(R_{ij}) + \{\mathbb{E}(R_{ij})\}^2\). 
Then, we obtain that
\begin{align*}
    s  = &4\left\{ \mathbb{E}(R_{ij}^2) - \left( \frac{M\overline{N}+1}{2} \right)^2 \right\} \mathbb{E}\left\{ N_i  + \rho^{*}N_i(N_i-1) \right\} \\
     =& 4\left\{ \mathbb{E}(R_{ij}^2) - \left( \frac{M\overline{N}+1}{2} \right)^2 \right\} \left[ \overline{N} +\rho^{*} \left\{ \overline{N}^2 \left(1+CV^2 \right) - \overline{N}  \right\}  \right] \\
     =& 4 \left\{  1 + (M\overline{N}-1)\left(3p_W + \frac{5}{4}p_T \right) + (M\overline{N} - 1)(M\overline{N}-2)\left(p_{WW} + p_{WT} + \frac{1}{4} p_{TT}\right) - \frac{(M\overline{N}+1)^2}{4} \right\} \\
    & \times \left[ \overline{N} +\rho^{*} \left\{ \overline{N}^2 \left(1+CV^2 \right) - \overline{N}  \right\}  \right]\\
     = &4\overline{N} \, \text{VIF}^{*}\left\{ 1+(M\overline{N}-1)P + (M\overline{N}-1)(M\overline{N}-2) Q - \frac{(M\overline{N}+1)^2}{4}  \right\},
\end{align*}
where \(P = 3p_W + \frac{5}{4}p_T\), \(Q = p_{WW}+p_{WT}+ \frac{1}{4}p_{TT}\), \(G = \rho^{*}\overline{N}^2\{M+(M-1)\text{CV}^2\} + (1-\rho^{*})M\overline{N}  \), and \(\text{VIF}^{*} = 1 + \left\{ (1+CV^2) \overline{N} -1  \right\}\rho^{*}  \). Plug into \(\sigma_D^2\), we have
\begin{align*}
    \sigma_D^2  =& \frac{1}{M^3\overline{N}^4}\left(\frac{1}{q} + \frac{1}{1-q}\right)s - \frac{W_D^2}{M} + o_p(M^{-1}) \\
    =& \frac{4}{M^3\overline{N}^3}\left(\frac{1}{q} + \frac{1}{1-q}\right)  \text{VIF}^{*}\left\{ 1+(M\overline{N}-1)P + (M\overline{N}-1)(M\overline{N}-2) Q - \frac{(M\overline{N}+1)^2}{4}  \right\} \\
    & - \frac{W_D^2}{M} + o_p(M^{-1}) \\
    =& \widetilde{\nu}_D^2 + o_p(M^{-1}).
\end{align*}
The variances \(\widetilde{\nu}_R^2\) and \(\widetilde{\nu}_O^2\) can be derived similarly via the Delta method as 
\[\widetilde{\nu}_R^2 = \left[ \frac{2\{1/(1 - \pi_{\text{tie}})\}}{1-(\{1/(1 - \pi_{\text{tie}})\} W_D)^2}  \right] \widetilde{\nu_D}^2 \quad {\rm and} \quad \widetilde{\nu}_O^2 = \left[ \frac{2}{1-W_D^2}\right]^2 \widetilde{\nu_D}^2.  \quad\]

\section{Proof of Remark \ref{remk:equiv}: Equivalence of variances in Theorem \ref{theorem:var_single} and Theorem \ref{theorem:var_comp}} \label{supp:sec:theorem2_equiv_3}
The key difference between the two variance forms lies in the evaluation of the second-order rank moment $\mathbb{E}(R_{ij}^2)$. In Theorem 2, this moment is derived under the assumption of scalar outcomes and standard midrank assignment, yielding a simplified expression based on the tie group structure and overall sample size. In contrast, Theorem \ref{theorem:var_comp} generalizes to prioritized composite outcomes, expressing $\mathbb{E}(R_{ij}^2)$ in terms of pairwise and triplet comparison probabilities. To demonstrate that the two expressions coincide under appropriate structural conditions in Remark \ref{remk:equiv}, we derive an identity linking squared midranks to tie group sizes. Suppose the set of $n = M\overline{N}$ individuals is partitioned into $G$ tie groups indexed by $g = 1, \dots, G$, ordered from least to most favorable. Each group $g$ consists of $t_g$ mutually tied individuals who dominate all members of groups $h < g$ and are strictly dominated by those in groups $h > g$. Let $L_g = \sum_{h < g} t_h$ denote the number of individuals strictly less favorable than group $g$, so that the integer ranks assigned to group $g$ are $\{L_g + 1, \dots, L_g + t_g\}$. Under the midrank convention, each member of group $g$ is assigned the same midrank, that is,
\[
r_g = 1 + L_g + \frac{t_g - 1}{2}.
\]
Based on this structure, the pairwise and higher-order comparison probabilities from equation \eqref{exp_sqr} can be expressed as functions of the tie groups:
\begin{align*}
    p_T &= \frac{1}{n(n-1)} \sum_{g=1}^{G} t_g (t_g - 1), \qquad p_W = \frac{1}{n(n-1)} \sum_{g=1}^{G} t_g L_g,\\
    p_{TT} &= \frac{1}{n(n-1)(n-2)} \sum_{g=1}^{G} t_g (t_g - 1)(t_g - 2),\\
    p_{WT} &= \frac{1}{n(n-1)(n-2)} \sum_{g=1}^{G} t_g L_g (t_g - 1),\\
    p_{WW} &= \frac{1}{n(n-1)(n-2)} \sum_{g=1}^{G} t_g L_g (L_g - 1).
\end{align*}
Substituting these into equation \eqref{exp_sqr}, we obtain
\begin{align*}
    \mathbb{E}(R_{ij}^2)
    &= 1 + \frac{1}{n} \sum_{g=1}^{G} \left\{ 3t_g L_g + \frac{5}{4} t_g(t_g - 1) \right\} \\
    &\quad + \frac{1}{n} \sum_{g=1}^{G} \left\{ t_g L_g (L_g - 1) + t_g L_g (t_g - 1) + \frac{1}{4} t_g (t_g - 1)(t_g - 2) \right\} \\
    &= 1 + \frac{1}{n} \sum_{g=1}^{G} t_g \left\{ L_g^2 + L_g(t_g - 1) + 2L_g \right\}
     + \frac{1}{n} \sum_{g=1}^{G} t_g \left\{ (t_g - 1) + \frac{(t_g - 1)^2}{4} \right\} \\
    &= \frac{1}{n} \sum_{g=1}^{G} t_g r_g^2.
\end{align*}
We now show that the average squared midrank admits the following expression:
\[
\frac{1}{n} \sum_{g=1}^G t_g r_g^2 
= \frac{(n+1)(2n+1)}{6} - \frac{1}{12n} \sum_{g=1}^G t_g (t_g^2 - 1).
\]
Within tie group $g$, let $m = 1, \dots, t_g$ index the relative position of an individual in its group. Then, the integer rank assigned to individual $m$ in group $g$ is $L_g + m$ and the sum of squared integer ranks within group $g$ is
\[
\sum_{m=1}^{t_g} (L_g + m)^2 
= t_g r_g^2 + \sum_{m=1}^{t_g} \left( (L_g + m) - r_g \right)^2,
\]
where the cross-term vanishes because $\sum_{m=1}^{t_g} \left((L_g + m) - r_g\right) = 0$ by construction of the midrank as the group average. The second term is the sum of squared deviations from the mean of $t_g$ consecutive integers, a standard result of which is equal to $t_g(t_g^2 - 1)/12$. Thus,
\[
t_g r_g^2 = \sum_{m=1}^{t_g} (L_g + m)^2 - \frac{t_g(t_g^2 - 1)}{12}.
\]

Summing over all groups $g$ and noting that the union of all rank intervals $\{L_g + 1, \dots, L_g + t_g\}$ covers $\{1, \dots, N\}$, we obtain that
\[
\sum_{g=1}^G t_g r_g^2 = \sum_{r=1}^n r^2 - \frac{1}{12} \sum_{g=1}^G t_g (t_g^2 - 1).
\]
Using the identity $\sum_{r=1}^N r^2 = n(n+1)(2n+1)/6$ and dividing both sides by $n$, we have
\[
\frac{1}{n} \sum_{g=1}^G t_g r_g^2 
= \frac{(n+1)(2n+1)}{6} - \frac{1}{12n} \sum_{g=1}^G t_g (t_g^2 - 1).
\]
The proof is completed.

\subsection{Variance of win-statistics under IRT} \label{supp:sec:theorem3_irt}
For IRT with \(N\) individuals, following Section \ref{supp:sec:theorem3}, we have the following lemma:
\begin{lemma} \label{lmm:comp}
The asymptotic variance for  \(W_D\), \(\log W_R\) and \(\log W_O\)  for i.i.d data with \(N\) individuals are
\begin{align*}
    \widetilde{\nu}_{I,D}^2 & = \frac{4}{N^3}\left(\frac{1}{q} + \frac{1}{1-q}\right)  \text{VIF}^{*}\left\{ 1+(N-1)P + (N-1)(N-2) Q - \frac{(N+1)^2}{4}  \right\} \frac{W_D^2}{M}\\
   \widetilde{\nu}_{I,R}^2 & = \left[ \frac{2\{1/(1 - \pi_{\text{tie}})\}}{1-(\{1/(1 - \pi_{\text{tie}})\} W_D)^2}  \right] \widetilde{\nu}_{I,D}^2 \\
   \widetilde{\nu}_{I,O}^2 &= \left[ \frac{2}{1-W_D^2}\right]^2 \widetilde{\nu}_{I,D}^2.
\end{align*}
\end{lemma}
Lemma \ref{lmm:comp} provides an extension of \citet{yu2022sample} to especially consider the composite endpoint, where the non-transitive properties may break the assumption for Theorem \ref{theorem:var_single}.

Following the above equivalence in Remark \ref{remk:equiv}, which is also held for IRT following Gehan's rule \citep{gehan1965generalized}, we can show that \(\nu^2_{I,\star} + o_p(M^{-1}) = \widetilde{\nu}^2_{I,\star}\) for a single non-survival endpoint. 

\section{Implementation considerations for composite survival endpoints under a generative perspective} \label{supp:sec:generating_composite}

In this section, we derive relative quantities in Theorem \ref{theorem:var_comp} under a specified data-generating model in detail. For illustration, we consider semi-competing risk outcomes, where hospitalization and death are generated from a Gumbel--Hougaard copula model. Each cluster \(i \in \{1,\dots,M\}\) is independently randomized to treatment \(A_i \sim \mathrm{Bernoulli}(q)\) with \(q=E(A_i)\), and each individual \((i,j)\) within cluster \(i\) is associated with a multiplicative frailty \(\gamma_{i}\sim\mathrm{Gamma}\!\left(\text{shape}=\nu,\text{rate}=\nu\right)\), independently across clusters and treatment \(A_i\). Let \(f_\gamma(\gamma)=\nu^{\nu}\gamma^{\nu-1}e^{-\nu\gamma}/\Gamma(\nu)\) denote the corresponding \(\mathrm{Gamma}(\nu,\nu)\) density. Conditional on \((A_i,\gamma_{i})\), the individual-specific hazards are modeled as
\[
\lambda_{ij}^D(A_i)=\gamma_{i}\lambda_0^D e^{-\eta_D A_i},\qquad
\lambda_{ij}^H(A_i)=\gamma_{i}\lambda_0^H e^{-\eta_H A_i},\qquad
\lambda_{ij}^C(A_i)=\lambda_0^C e^{-\eta_C A_i},
\]
and analogously for individual \((k,l)\) with parameters \((A_k,\gamma_{k})\). Consider two distinct individuals \((i,j)\) and \((k,l)\) with cluster assignments \((A_i=a_i,A_k=a_k)\). Let \(T_{ij}^D,T_{ij}^H,C_{ij}\) and \(T_{kl}^D,T_{kl}^H,C_{kl}\) denote their death, hospitalization, and censoring times for \( (i,j)\) and \((k,l)\), respectively. The probability that none of the four events \((D_{ij},D_{kl},C_{ij},C_{kl})\) has occurred by time \(t\) is
\[
S_{\mathrm{all}}\!\left(t\mid a_i,a_k\right)
=\exp\!\left[-\left\{\lambda_{ij}^D(a_i)+\lambda_{kl}^D(a_k)+\lambda_{ij}^C(a_i)+\lambda_{kl}^C(a_k)\right\}t\right].
\]
We adopt a prioritized comparison rule in which death is compared first, and hospitalization is compared only when both deaths are undetermined (i.e., \(\min\!\left\{T_{ij}^D,T_{kl}^D\right\}>\min\!\left\{C_{ij},C_{kl}\right\}\)). Accordingly, the mortality-tier win probability for individual \((i,j)\) over \((k,l)\), conditional on treatment \((a_i,a_k)\), is
\begin{align*}
O_{ij,kl}\!\left(a_i,a_k\right)
&=\mathbb{P}\!\left\{T_{kl}^D<\min\!\left(T_{ij}^D,C_{ij},C_{kl}\right)\,\middle|\,A_i=a_i,A_k=a_k\right\}\\
&=\int_{0}^{\infty}
\underbrace{\lambda_{kl}^D\!\left(a_k\right)}_{\substack{\text{death hazard}}}
\;\underbrace{S_{\mathrm{all}}\!\left(t\mid a_i,a_k\right)}_{\substack{\text{no event}}}\,dt\\
&=\frac{\lambda_{kl}^D\!\left(a_k\right)}
{\lambda_{ij}^D\!\left(a_i\right)+\lambda_{kl}^D\!\left(a_k\right)+\lambda_{ij}^C\!\left(a_i\right)+\lambda_{kl}^C\!\left(a_k\right)}.
\end{align*}

When death is undetermined for both individuals, hospitalization determines the comparison. Under a Gumbel--Hougaard copula with parameter \(\phi\ge1 \) for each individual's joint \((T_{ij}^H,T_{ij}^D)\) times, under treatment \(a_i\), we denote
\begin{align*}
S_{ij}^{HD}\!\left(t_1,t_2\mid a_i\right)
&=\mathbb{P}\!\left(T_{ij}^H>t_1,\,T_{ij}^D>t_2 \mid A_i=a_i\right)
=\exp\!\left(-\left[\left\{\lambda_{ij}^H(a_i)t_1\right\}^{\phi}+\left\{\lambda_{ij}^D(a_i)t_2\right\}^{\phi}\right]^{1/\phi}\right),\\
h_{ij}^H\!\left(t_1,t_2\mid a_i\right)
&=\mathbb{P}\!\left(T_{ij}^H = t_1\mid T_{ij}^H>t_1,\,T_{ij}^D>t_2,\,A_i=a_i\right)\\
&=\lambda_{ij}^H(a_i)\!\left\{\lambda_{ij}^H(a_i)t_1\right\}^{\phi-1}
\!\left[\left\{\lambda_{ij}^H(a_i)t_1\right\}^{\phi}+\left\{\lambda_{ij}^D(a_i)t_2\right\}^{\phi}\right]^{1/\phi-1},\\
S_{ij}^C\!\left(t\mid a_i\right)&=\exp\!\left\{-\lambda_{ij}^C(a_i)t\right\},
\end{align*}
and similarly for \((k,l)\) under \(a_k\). To simplify, writing \(S_{ij}^{HD}\!\left(t\mid a_i\right)=S_{ij}^{HD}\!\left(t,t\mid a_i\right)\) and \(h_{ij}^H\!\left(t\mid a_i\right)=h_{ij}^H\!\left(t,t\mid a_i\right)\). Then, the hospitalization-tier win probability for \((i,j)\) over \((k,l)\) is
\begin{align*}
H_{ij,kl}\!\left(a_i,a_k\right)
&=\mathbb{P}\!\left\{\min\!\left(T_{ij}^D,T_{kl}^D\right)>\min\!\left(C_{ij},C_{kl}\right),\ 
T_{kl}^H<\min\!\left(T_{ij}^H,C_{ij},C_{kl}\right)
\,\middle|\,A_i=a_i,A_k=a_k\right\}\\
&=\int_{0}^{\infty}\!\!\int_{0}^{t_2}
\underbrace{h_{kl}^H\!\left(t_1,t_2\mid a_k\right)}_{\substack{\text{hospitalization hazard }}}
\underbrace{S_{ij}^{HD}\!\left(t_1,t_2\mid a_i\right)S_{kl}^{HD}\!\left(t_1,t_2\mid a_k\right)}_{\substack{\text{both alive and not hospitalized}}}
\underbrace{S_{ij}^C\!\left(t_2\mid a_i\right)S_{kl}^C\!\left(t_2\mid a_k\right)}_{\substack{\text{both uncensored}}}\,dt_1\,dt_2.
\end{align*}
Therefore, the population win ratio comparing intervention versus control is given as:
\[
W_R
=\frac{\displaystyle \int_{0}^{\infty}\!\!\int_{0}^{\infty}\!\!\left\{O_{ij,kl}\!\left(1,0\right)+H_{ij,kl}\!\left(1,0\right)\right\}
f_\gamma\!\left(\gamma_{i}\right)f_\gamma\!\left(\gamma_{k}\right)\,d\gamma_{i}\,d\gamma_{k}}
{\displaystyle \int_{0}^{\infty}\!\!\int_{0}^{\infty}\!\!\left\{O_{ij,kl}\!\left(0,1\right)+H_{ij,kl}\!\left(0,1\right)\right\}
f_\gamma\!\left(\gamma_{i}\right)f_\gamma\!\left(\gamma_{k}\right)\,d\gamma_{i}\,d\gamma_{k}}.
\]
For a fixed pair of individuals \((i,j)\) and \((k,l)\), the pairwise win, loss and tie probabilities are given as:
\begin{align*}
p_W = \mathbb{P}\big(\bm{Y}_{ij}\succ \bm{Y}_{kl}\big) 
&= \sum_{a_i=0}^1\sum_{a_k=0}^1 \Bigg[
   q^{a_i}(1-q)^{1-a_i}\,q^{a_k}(1-q)^{1-a_k}  \\
&\quad\times \iint\! \Big\{ A_{ij,kl}(a_i,a_k)+H_{ij,kl}(a_i,a_k) \Big\}
f_\gamma(\gamma_{i})\,f_\gamma(\gamma_{k})\,d\gamma_{i}\,d\gamma_{k} \Bigg], \\[0.8em]
p_L = \mathbb{P}\big(\bm{Y}_{ij}\prec \bm{Y}_{kl}\big) 
&= \sum_{a_i=0}^1\sum_{a_k=0}^1 \Bigg[
   q^{a_i}(1-q)^{1-a_i}\,q^{a_k}(1-q)^{1-a_k} \\
&\quad\times \iint\! \Big\{ A_{kl,ij}(a_k,a_i)+H_{kl,ij}(a_k,a_i) \Big\} f_\gamma(\gamma_{i})\,f_\gamma(\gamma_{k})\,d\gamma_{i}\,d\gamma_{k} \Bigg],
\end{align*}
and $p_T = 1 - p_W - p_L$.

Similarly, for three distinct individuals \((i,j)\), \((k,l)\), and \((m,r)\), the triplet probabilities are
\allowdisplaybreaks
\begin{align*}
p_{WW}
=& \mathbb{P}\!\left(\bm{Y}_{ij}\succ\bm{Y}_{kl},\ \bm{Y}_{ij}\succ\bm{Y}_{mr}\right)\\
=& \sum_{a_i=0}^{1}\sum_{a_k=0}^{1}\sum_{a_m=0}^{1}
   q^{a_i}\!\left(1-q\right)^{1-a_i}\,
   q^{a_k}\!\left(1-q\right)^{1-a_k}\,
   q^{a_m}\!\left(1-q\right)^{1-a_m} \\
& \times \int_{0}^{\infty}\!\!\int_{0}^{\infty}\!\!\int_{0}^{\infty}
   \Bigl[ O_{ij,kl}\!\left(a_i,a_k;\gamma_i,\gamma_k\right)
        + H_{ij,kl}\!\left(a_i,a_k;\gamma_i,\gamma_k\right) \Bigr]\\
& \times \Bigl[ O_{ij,mr}\!\left(a_i,a_m;\gamma_i,\gamma_m\right)
               + H_{ij,mr}\!\left(a_i,a_m;\gamma_i,\gamma_m\right) \Bigr] \times f_\gamma\!\left(\gamma_i\right)\,f_\gamma\!\left(\gamma_k\right)\,f_\gamma\!\left(\gamma_m\right)\,
   d\gamma_i\,d\gamma_k\,d\gamma_m,\\[6pt]
p_{WT}
=& \mathbb{P}\!\left(\bm{Y}_{ij}\succ\bm{Y}_{kl},\ \bm{Y}_{ij}=\bm{Y}_{mr}\right)\\
=& \sum_{a_i=0}^{1}\sum_{a_k=0}^{1}\sum_{a_m=0}^{1}
   q^{a_i}\!\left(1-q\right)^{1-a_i}\,
   q^{a_k}\!\left(1-q\right)^{1-a_k}\,
   q^{a_m}\!\left(1-q\right)^{1-a_m} &\\
& \times \int_{0}^{\infty}\!\!\int_{0}^{\infty}\!\!\int_{0}^{\infty}
   \Bigl[ O_{ij,kl}\!\left(a_i,a_k;\gamma_i,\gamma_k\right)
        + H_{ij,kl}\!\left(a_i,a_k;\gamma_i,\gamma_k\right) \Bigr]\\
& \times \Bigl\{ 1
   - \Bigl[ O_{ij,mr}\!\left(a_i,a_m;\gamma_i,\gamma_m\right)
           + H_{ij,mr}\!\left(a_i,a_m;\gamma_i,\gamma_m\right) \Bigr]\\
   &- \Bigl[ O_{mr,ij}\!\left(a_m,a_i;\gamma_m,\gamma_i\right)
           + H_{mr,ij}\!\left(a_m,a_i;\gamma_m,\gamma_i\right) \Bigr]
   \Bigr\} \times f_\gamma\!\left(\gamma_i\right)\,f_\gamma\!\left(\gamma_k\right)\,f_\gamma\!\left(\gamma_m\right)\,
   d\gamma_i\,d\gamma_k\,d\gamma_m,\\[6pt]
p_{TT}
=& \mathbb{P}\!\left(\bm{Y}_{ij}=\bm{Y}_{kl},\ \bm{Y}_{ij}=\bm{Y}_{mr}\right)\\
=& \sum_{a_i=0}^{1}\sum_{a_k=0}^{1}\sum_{a_m=0}^{1}
   q^{a_i}\!\left(1-q\right)^{1-a_i}\,
   q^{a_k}\!\left(1-q\right)^{1-a_k}\,
   q^{a_m}\!\left(1-q\right)^{1-a_m} &\\
& \times \int_{0}^{\infty}\!\!\int_{0}^{\infty}\!\!\int_{0}^{\infty}
   \Bigl\{ 1
   - \Bigl[ O_{ij,kl}\!\left(a_i,a_k;\gamma_i,\gamma_k\right)
           + H_{ij,kl}\!\left(a_i,a_k;\gamma_i,\gamma_k\right) \Bigr]\\
   &- \Bigl[ O_{kl,ij}\!\left(a_k,a_i;\gamma_k,\gamma_i\right)
           + H_{kl,ij}\!\left(a_k,a_i;\gamma_k,\gamma_i\right) \Bigr]
   \Bigr\}\\
& \times \Bigl\{ 1
   - \Bigl[ O_{ij,mr}\!\left(a_i,a_m;\gamma_i,\gamma_m\right)
           + H_{ij,mr}\!\left(a_i,a_m;\gamma_i,\gamma_m\right) \Bigr]\\
   & - \Bigl[ O_{mr,ij}\!\left(a_m,a_i;\gamma_m,\gamma_i\right)
           + H_{mr,ij}\!\left(a_m,a_i;\gamma_m,\gamma_i\right) \Bigr]
   \Bigr\}\times f_\gamma\!\left(\gamma_i\right)\,f_\gamma\!\left(\gamma_k\right)\,f_\gamma\!\left(\gamma_m\right)\,
   d\gamma_i\,d\gamma_k\,d\gamma_m.
\end{align*}

Closed-form evaluation of the above probabilities is generally infeasible because \(O_{ij,kl}\!\left(a_i,a_k;\gamma_i,\gamma_k\right)\) and \\\(H_{ij,kl}\!\left(a_i,a_k;\gamma_i,\gamma_k\right)\) are nonlinear in the underlying hazards and must be averaged over the joint distribution of \(\left(A_i,A_k,\gamma_i,\gamma_k\right)\) \\(or \(\left(A_i,A_k,A_m,\gamma_i,\gamma_k,\gamma_m\right)\) for triplets). In practice, these quantities are approximated via {Monte Carlo integration. Specifically, we draw \(\left(A_i,A_k\right)\) (or \(\left(A_i,A_k,A_m\right)\)) independently from \(\mathrm{Bernoulli}\!\left(q\right)\), and draw \(\left(\gamma_i,\gamma_k\right)\) (or \(\left(\gamma_i,\gamma_k\, \gamma_m\right)\)) independently from \(\mathrm{Gamma}\!\left(\text{shape}=\nu,\text{rate}=\nu\right)\). For each simulated replication, we compute \(O_{ij,kl}\) and \(H_{ij,kl}\) at the sampled values and then average the resulting probabilities across a large number of Monte Carlo draws. This yields consistent numerical approximations with standard errors that decrease at the usual \(O\!\left(B^{-1/2}\right)\) rate as the number of replicates \(B\) increases.

\subsection{Monte Carlo approximation of the generalized rank ICC}
\label{supp:sec:rho_star_computation}

The generalized rank ICC \(\rho^{*}\) can be approximated under the generative model by first evaluating the conditional probability of superiority score \(\mathcal{F}^{**}\) for individuals generated in correlated clusters and then calculating the empirical within-cluster correlation of these scores. To preserve the definition in Section \ref{sec:win_stat}, both the focal outcomes and their independent comparators are generated from the pooled trial population, with treatment assignment drawn according to the planned allocation probability \(q\). Specifically, generate a focal sample consisting of \(M_{\rho}\) independent clusters. For cluster \(i\), draw \(A_i\sim\mathrm{Bernoulli}(q)\), draw the shared frailty \(\gamma_i\), and then generate \(m_{\rho}\geq2\) individual composite outcomes
\(\bm{Y}_{i1},\ldots,\bm{Y}_{im_{\rho}}\) under the event-time and censoring model described above. Individuals in the same focal cluster share \(A_i\) and \(\gamma_i\), so their simulated outcomes retain the within-cluster dependence induced by the generative model. Independently of the focal sample, generate a reference sample \(\bm{Y}'_1,\ldots,\bm{Y}'_{B_{\mathrm{ref}}}\)
from the pooled participant-level marginal distribution. For each reference individual \(b\), treatment \(A'_b\sim\mathrm{Bernoulli}(q)\), frailty, event times, and censoring times are generated independently of the focal clusters and of the other reference individuals. The reference outcomes should therefore be viewed as individuals drawn from new independent clusters. For focal individual \((i,j)\), approximate the conditional probability of superiority score by
\[
\widehat{\mathcal{F}}^{**}_{ij}=\frac{1}{B_{\mathrm{ref}}}\sum_{b=1}^{B_{\mathrm{ref}}}\left[\mathbb{I}\!\left\{\bm{Y}_{ij}\succ\bm{Y}'_b\right\}+\frac{1}{2}\mathbb{I}\!\left\{\bm{Y}_{ij}=\bm{Y}'_b\right\}\right],
\]
where every comparison uses the same prespecified prioritized comparison rule as the planned analysis, and an unresolved comparison due to censoring is treated as a tie. Equivalently, one may calculate the signed comparison score
\[
\widehat{G}_{ij}=\frac{1}{B_{\mathrm{ref}}}\sum_{b=1}^{B_{\mathrm{ref}}}s\!\left(\bm{Y}_{ij},\bm{Y}'_b\right)=2\widehat{\mathcal{F}}^{**}_{ij}-1.
\]
Because correlation is invariant to a common positive affine transformation, the empirical intracluster correlation of \(\widehat{G}_{ij}\) is identical to that of \(\widehat{\mathcal{F}}^{**}_{ij}\). Thus, either score can be used computationally. For completeness, let \(U_{ij}\) denote either
\(\widehat{\mathcal{F}}^{**}_{ij}\) or \(\widehat{G}_{ij}\), and define the cluster-weighted empirical mean
\[
\overline{U}_{\rho}=\frac{1}{M_{\rho}}\sum_{i=1}^{M_{\rho}}\frac{1}{m_{\rho}}\sum_{j=1}^{m_{\rho}}U_{ij}.
\]
The marginal score variance and within-cluster score covariance are estimated by
\begin{align*}
\widehat{V}_{\rho}
&=
\frac{1}{M_{\rho}}
\sum_{i=1}^{M_{\rho}}
\frac{1}{m_{\rho}}
\sum_{j=1}^{m_{\rho}}
\left(U_{ij}-\overline{U}_{\rho}\right)^2,\\
\widehat{C}_{\rho}
&=
\frac{1}{M_{\rho}}
\sum_{i=1}^{M_{\rho}}
\frac{1}{m_{\rho}(m_{\rho}-1)}
\sum_{\substack{j,j'=1\\j\neq j'}}^{m_{\rho}}
\left(U_{ij}-\overline{U}_{\rho}\right)
\left(U_{ij'}-\overline{U}_{\rho}\right).
\end{align*}
The generalized rank ICC is then approximated by \(\widehat{\rho}^{*}=\frac{\widehat{C}_{\rho}}{\widehat{V}_{\rho}}\), provided that \(\widehat{V}_{\rho}>0\). When \(m_{\rho}=2\), which is sufficient under the assumed exchangeable within-cluster dependence, the covariance estimator simplifies to
\[
\widehat{C}_{\rho}=\frac{1}{M_{\rho}}\sum_{i=1}^{M_{\rho}}\left(U_{i1}-\overline{U}_{\rho}\right)\left(U_{i2}-\overline{U}_{\rho}\right).
\]
The focal sample preserves the shared treatment and frailty within each cluster, whereas the independent reference sample approximates the pooled comparator distribution appearing in the definition of \(\mathcal{F}^{**}\). Consequently, \(\widehat{\rho}^{*}\) estimates the trial-wide pooled generalized rank ICC used in \(\mathrm{VIF}^{*}\), rather than an arm-specific outcome-scale ICC. The values of \(M_{\rho}\) and \(B_{\mathrm{ref}}\) should be sufficiently large that both the empirical correlation and the individual superiority scores are numerically stable.

\section{Additional details for simulations with ordinal outcomes} \label{supp:sec:ordinal_est}

In this section, we provide a derivation of pairwise win, loss, and tie probabilities for an ordinal endpoint under a proportional odds mixed-effect model with cluster-level random intercepts. Let \(K \ge 2\) denote the number of ordered categories, and define the logistic cumulative distribution function \(\Lambda(x) = \left(1 + e^{-x}\right)^{-1}\). For each cluster \(i\), the random intercept follows \(b_i \sim \mathcal{N}\!\left(0, \sigma_b^2\right)\). Conditional on treatment assignment \(A_i\) and random effect \(b_i\), the proportional-odds model is specified as
\[
\mathbb{P}\!\left(Y_{ij} \le k \mid A_i, b_i\right)
= \Lambda\!\left(\theta_k - \beta A_i - b_i\right), \qquad k = 1, \dots, K - 1,
\]
where \(\theta_1 < \cdots < \theta_{K-1}\). Thus, the category probability for \(a \in \{0,1\}\), is
\[
p_a\!\left(k \mid b\right)
= \mathbb{P}\!\left(Y_{ij} = k \mid A_i = a, b_i\right)
= \Lambda\!\left(\theta_k - \beta a - b_i\right)
- \Lambda\!\left(\theta_{k-1} - \beta a - b_i\right), \quad k = 1, \dots, K.
\]
For two distinct individuals \((Y_{ij},Y_{kl})\), given \(\left(b_i, b_k\right)\) and \(\left(A_i, A_k\right) = (1, 0)\), the corresponding conditional win, loss, and ties probabilities are:
\begin{align*}
\mathbb{P}\!\left(Y_{ij} > Y_{kl} \mid b_i, b_k, A_i = 1, A_k = 0\right)
&= \sum_{k=1}^{K} p_1\!\left(k \mid b_i\right)
   \sum_{s=1}^{k-1} p_0\!\left(s \mid b_k\right),\\
\mathbb{P}\!\left(Y_{ij} < Y_{kl} \mid b_i, b_k, A_i = 1, A_k = 0\right)
&= \sum_{k=1}^{K} p_1\!\left(k \mid b_i\right)
   \sum_{s=k+1}^{K} p_0\!\left(s \mid b_k\right),\\
\mathbb{P}\!\left(Y_{ij} = Y_{kl} \mid b_i, b_k, A_i = 1, A_k = 0\right)
&= \sum_{k=1}^{K} p_1\!\left(k \mid b_i\right)\,p_0\!\left(k \mid b_k\right).
\end{align*}

Let \(\varphi_{b}\!\left(b\right)\) denote the probability density of distribution \(\mathcal{N}\!\left(0, \sigma_b^2\right)\). Then, the marginal category probabilities under treatment \(a\) are:
\[
\pi_a\!\left(k\right)
= \int_{-\infty}^{\infty}
\!\left\{
\Lambda\!\left(\theta_k - \beta a - b\right)
- \Lambda\!\left(\theta_{k-1} - \beta a - b\right)
\right\}
\varphi_{b}\!\left(b\right)\,db,
\quad k = 1, \dots, K.
\]
Define the control cumulative masses \(F_0^{-}\!\left(k\right) = \sum_{s=1}^{k-1} \pi_0\!\left(s\right)\) and \(F_0\!\left(k\right) = \sum_{s=1}^{k} \pi_0\!\left(s\right)\). Then, the marginal pairwise probabilities win, loss and ties are 
\begin{align*}
    \mathbb{P}\!\left(Y_{ij} > Y_{kl} \mid A_i = 1, A_k = 0\right)
    &= \sum_{k=1}^{K} \pi_1\!\left(k\right)\,F_0^{-}\!\left(k\right),\\
    \mathbb{P}\!\left(Y_{ij} < Y_{kl} \mid A_i = 1, A_k = 0\right) &= \sum_{k=1}^{K} \pi_1\!\left(k\right)\left\{1 - F_0\!\left(k\right)\right\},\\
     \mathbb{P}\!\left(Y_{ij} = Y_{kl} \mid A_i = 1, A_k = 0\right) &= \sum_{k=1}^{K} \pi_1\!\left(k\right)\,\pi_0\!\left(k\right) \\
    &= 1 - \sum_{k=1}^{K} \pi_1\!\left(k\right)\,F_0^{-}\!\left(k\right) - \sum_{k=1}^{K} \pi_1\!\left(k\right)\left\{1 - F_0\!\left(k\right)\right\}.
\end{align*}

To calibrate \(\theta_k\), \(k=1,\dots, K-1\) to a prescribed marginal probabilities for control \(\pi_0(k)\) with cumulative sums \(S_k = \sum_{s=1}^{k} \pi_0(s)\), we solve
\[
\int_{-\infty}^{\infty}
\!\Lambda\!\left(\theta_k - b\right)\varphi_{b}\!\left(b\right)\,db = S_k, \qquad k = 1, \dots, K - 1,
\]
which admits a unique solution because
\[
\frac{d}{d\theta}
\int_{-\infty}^{\infty}
\!\Lambda\!\left(\theta - b\right)\varphi_{b}\!\left(b\right)\,db
= \int_{-\infty}^{\infty}
\!\Lambda'\!\left(\theta - b\right)\varphi_{b}\!\left(b\right)\,db > 0,
\]
where \(\Lambda'\!\left(x\right) = \frac{e^{-x}}{\left(1 + e^{-x}\right)^{2}}\). For numerical evaluation of \(\int_{-\infty}^{\infty} \Lambda\!\left(\eta -b\right)\varphi_{b}\!\left(b\right)\,db\), we use Gauss-Hermite quadrature for \(b = \sigma_b \sqrt{2}\,z\), giving
\[
\int_{-\infty}^{\infty}\!
\Lambda\!\left(\eta - b\right)\varphi_{b}\!\left(b\right)\,db
= \frac{1}{\sqrt{\pi}}
\int_{-\infty}^{\infty}\!
\Lambda\!\left(\eta - \sigma_b\sqrt{2}\,z\right)e^{-z^{2}}\,dz
\approx \frac{1}{\sqrt{\pi}}
\sum_{r=1}^{R} w_r\,\Lambda\!\left(\eta - \sigma_b\sqrt{2}\,z_r\right),
\]
where \(\{z_r, w_r\}_{r=1}^{R}\) are the \(R\)-point Gauss-Hermite nodes and weights. As a results, for \(k = 1, \dots, K\),
\[ \pi_1\!\left(k\right) \approx \frac{1}{\sqrt{\pi}} \sum_{r=1}^{R} w_r \left\{\Lambda\!\left(\theta_k - \beta - \sigma_b\sqrt{2}\,z_r\right) - \Lambda\!\left(\theta_{k-1} - \beta - \sigma_b\sqrt{2}\,z_r\right) \right\}.
\]
Finally, since clusters are independently randomized with \(\mathbb{P}\!\left(A = 1\right) = q\),
then
\begin{align*}
&\mathbb{P}\!\left(Y_{ij} \gtrless Y_{kl}, A_i = 1, A_k = 0\right)
= q(1 - q)\,
\mathbb{P}\!\left(Y_{ij} \gtrless Y_{kl} \mid A_i = 1, A_k = 0\right),\\
& \mathbb{P}\!\left(Y_{ij} = Y_{kl}, A_i = 1, A_k = 0\right)
= q(1 - q)\sum_{k=1}^{K}\pi_1\!\left(k\right)\pi_0\!\left(k\right).    
\end{align*}

\section{Additional simulation results}
\label{supp:sec:additional_simulation}

This section reports additional simulation results that complement Section \ref{sec:simu} of the main manuscript. Whereas the main manuscript focuses on \(\Delta=\log(W_R)\) and reports empirical type I error and power calibration for the win ratio, here we present the corresponding results for the alternative win measures \(\Delta=W_D\) and \(\Delta=\log(W_O)\), together with further diagnostics for relative bias, Monte Carlo standard deviation (MCSD), average estimated standard error (ASE), design-based standard error (DSE), power prediction error, and the required number of clusters.

All simulations in this section follow the same design grid as the main manuscript. Clusters were randomized to treatment or control in a \(1:1\) ratio, and the nominal two-sided significance level was fixed at \(\alpha=0.05\). The number of clusters \(M\) varied over the grid used in the main text for the power simulations; for the type~I error simulations, this grid was augmented with very small values of \(M\) to assess small-sample calibration. The cluster-size designs comprised both fixed cluster sizes and discrete uniform cluster-size distributions, allowing the effect of cluster-size heterogeneity to be evaluated. For each simulated dataset, we computed \(\widehat W_D\), \(\log(\widehat W_R)\), and \(\log(\widehat W_O)\), together with the replication-specific standard errors from Theorem \ref{theorem:u-var}.

For a design configuration with \(B\) Monte Carlo replications, let \(\widehat{\Delta}_{\star b}\) and \(\widehat{\sigma}^{(1)}_{\star b}\) denote the estimate and the standard error estimate from Theorem \ref{theorem:u-var} in replication \(b\), respectively, for \(\star\in\{D,R,O\}\). We define
\begin{align*}
\overline{\widehat{\Delta}}_{\star}&=
\frac{1}{B}\sum_{b=1}^{B}\widehat{\Delta}_{\star b},\\
\mathrm{MCSD}_{\star}&=
\left\{\frac{1}{B-1}\sum_{b=1}^{B}\left(\widehat{\Delta}_{\star b}-\overline{\widehat{\Delta}}_{\star}\right)^2
\right\}^{1/2},\\
\mathrm{ASE}_{\star}&=\frac{1}{B}\sum_{b=1}^{B}\widehat{\sigma}^{(1)}_{\star b}.
\end{align*}
The design-based standard error, namely, the standard error implied by the closed-form design-stage variance formula, is computed once for each design configuration as
\[
\mathrm{DSE}_{\star}=
\begin{cases}
\sqrt{\nu_{\star}^{2}},
&
\text{for a scalar ordinal endpoint, using Theorem \ref{theorem:var_single}},\\[3pt]
\sqrt{\widetilde{\nu}_{\star}^{2}},
&
\text{for a composite survival endpoint, using Theorem \ref{theorem:var_comp}},
\end{cases}
\]
after substituting the true design inputs from the data-generating mechanism. Thus, the ASE is the Monte Carlo average of the estimated standard errors used for inference, whereas the DSE is the design-stage standard error used in the analytic power calculation. Empirical rejection probabilities were obtained from the Wald \(z\)-test and the Wald \(t\)-test with \(M-2\) degrees of freedom. Predicted power was obtained by substituting the design input into the corresponding power expression. We define the power prediction error as
\[
\mathrm{PPE}_{\star}=100\left\{\widehat{\mathrm{Power}}_{\mathrm{emp},\star}-\mathrm{Power}_{\mathrm{pred},\star}\right\},
\]
reported in percentage points. A positive value indicates conservative formula-predicted power.

\subsection{Additional type I error results}
\label{supp:sec:typeI_additional}

Figure~\ref{supp:fig:typeI_WD} displays the empirical type~I error for \(\Delta=W_D\), which is consistent with the \(\log(W_R)\) results reported in the main manuscript. For ordinal outcomes, the empirical type~I error of the Wald \(z\)-test ranges from approximately \(4.96\%\) to \(8.09\%\), with the largest values arising when \(M\) is very small, whereas the Wald \(t\)-test ranges from approximately \(4.57\%\) to \(5.33\%\) and thus exhibits substantially improved small-\(M\) calibration. For composite survival endpoints, the Wald \(z\)-test ranges from approximately \(5.10\%\) to \(8.24\%\) and the Wald \(t\)-test from approximately \(4.65\%\) to \(5.42\%\). For \(W_D\), therefore, the principal finite-sample concern is the use of the normal reference distribution when the number of independent clusters is small; the \(t\)-reference distribution largely eliminates this liberal behavior.

\begin{figure}[!htbp]
\centering
\includegraphics[width=\textwidth]{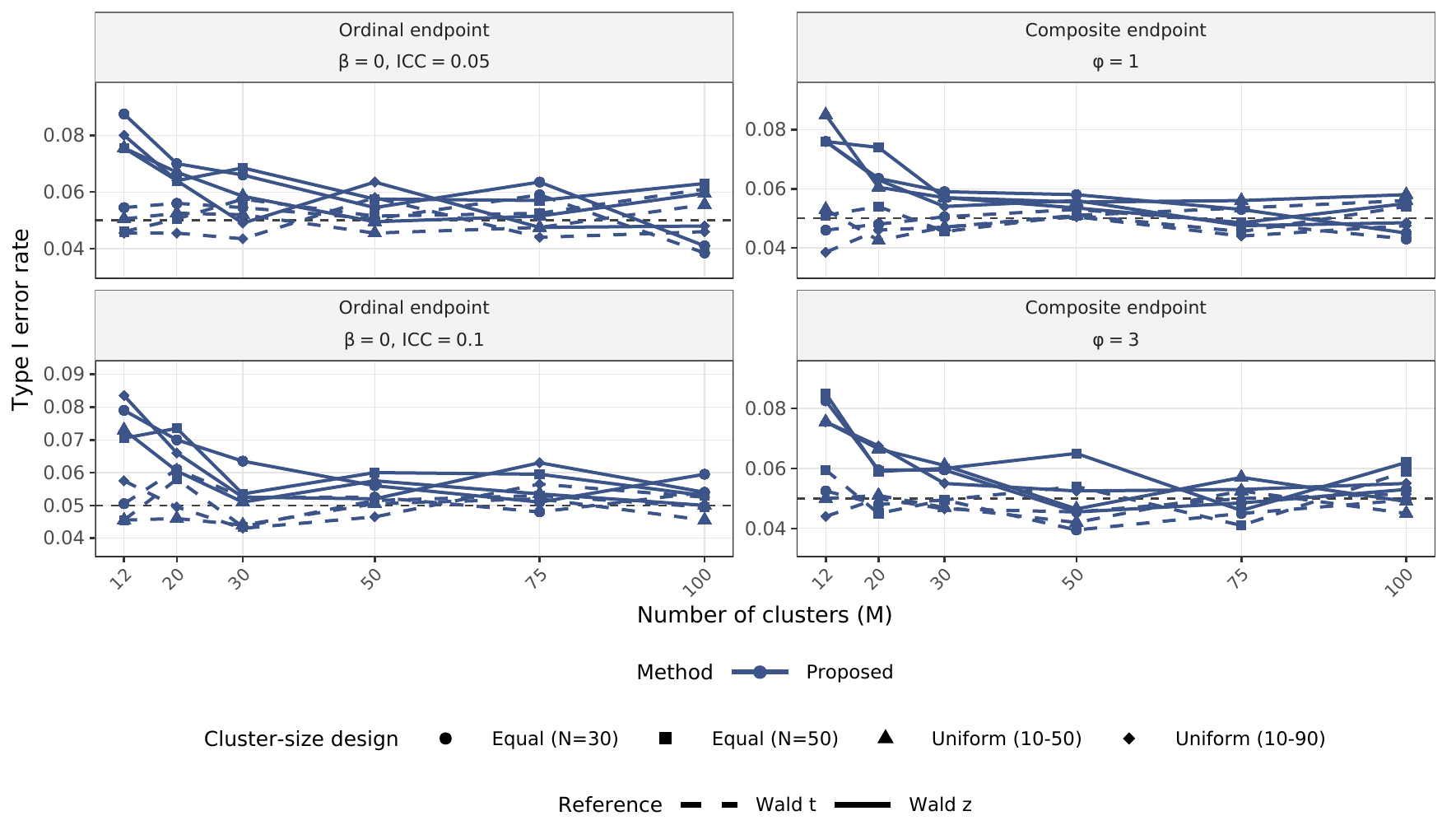}
\caption{Empirical type~I error for \(\Delta=W_D\) under the proposed Wald \(z\)- and Wald \(t\)-tests for scalar ordinal and composite survival endpoints. The horizontal dashed line marks the nominal \(5\%\) level.}
\label{supp:fig:typeI_WD}
\end{figure}

Figure~\ref{supp:fig:typeI_logWO} presents the analogous type~I error results for \(\Delta=\log(W_O)\), which follow the same pattern. For ordinal outcomes, the Wald \(z\)-test ranges from approximately \(4.87\%\) to \(7.63\%\) and the Wald \(t\)-test from approximately \(4.18\%\) to \(5.13\%\); for composite survival endpoints, the corresponding ranges are approximately \(4.98\%\) to \(7.63\%\) and \(4.29\%\) to \(5.23\%\). These results again indicate that the proposed cluster-score variance estimator is well calibrated once the reference distribution is adjusted for the number of clusters: the \(z\)-test approaches the nominal level as \(M\) increases, whereas the \(t\)-test is preferable when \(M\) is small.

\begin{figure}[!htbp]
\centering
\includegraphics[width=\textwidth]{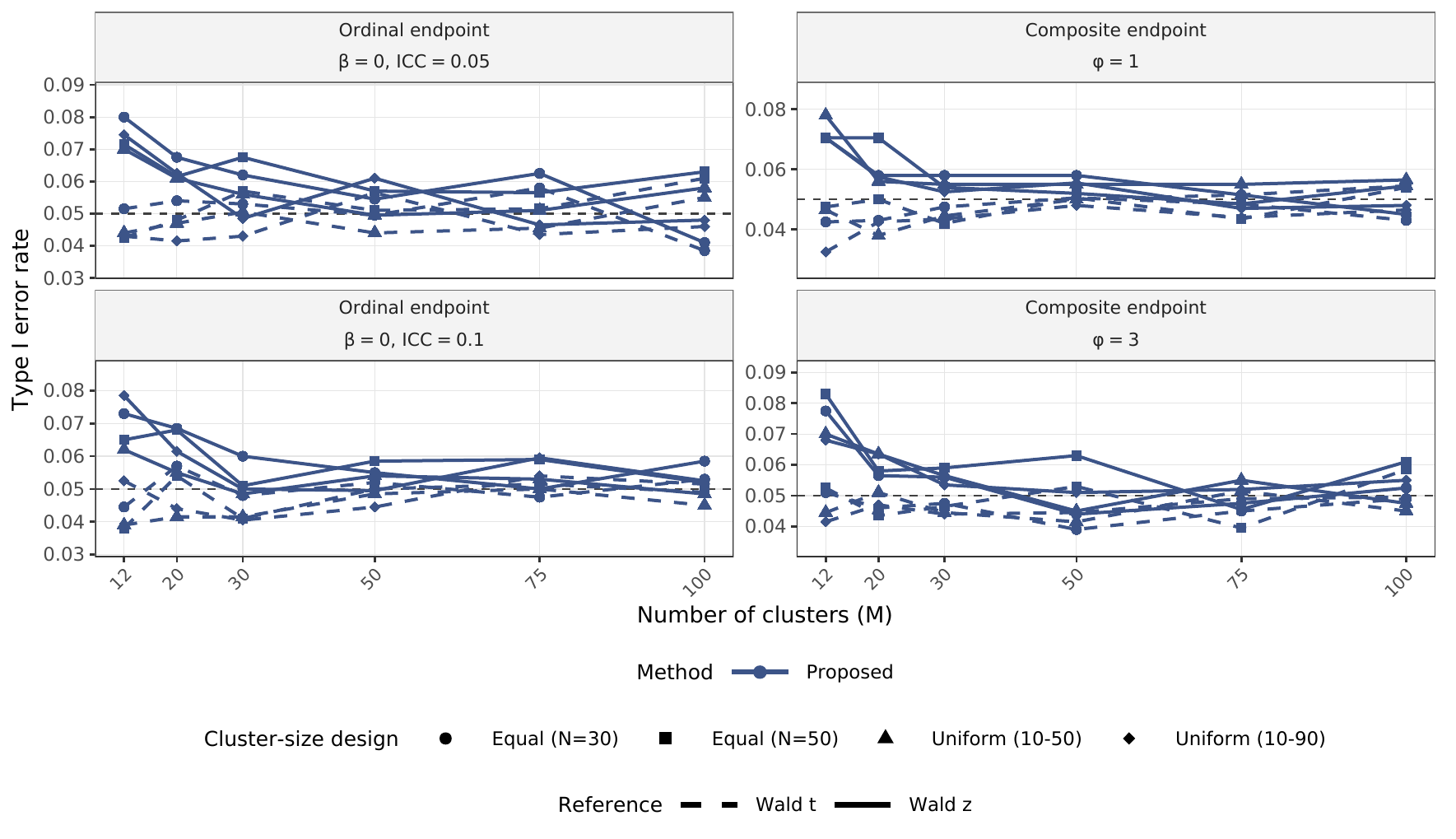}
\caption{Empirical type~I error for \(\Delta=\log(W_O)\) under the proposed Wald \(z\)- and Wald \(t\)-tests for scalar ordinal and composite survival endpoints. The horizontal dashed line marks the nominal \(5\%\) level.}
\label{supp:fig:typeI_logWO}
\end{figure}

Taken together, Figures~\ref{supp:fig:typeI_WD} and \ref{supp:fig:typeI_logWO} indicate that the type~I error pattern is governed primarily by the number of clusters rather than by the choice of win measure. The normal approximation is mildly liberal for very small \(M\), whereas the \(t\)-reference distribution yields type~I error close to the nominal level. This reinforces the recommendation in the main text to use the Wald \(t\)-test for CRTs with a small number of clusters.

\subsection{Additional power calibration results}
\label{supp:sec:power_additional}

Figure~\ref{supp:fig:power_WD} presents representative power curves for \(\Delta=W_D\), for which the empirical and predicted curves follow the same monotone trend in \(M\). For ordinal outcomes, empirical power ranges from approximately \(8.7\%\) in the weakest and smallest designs to above \(99\%\) in the strongest and largest designs, and the predicted power preserves this ordering, with a median absolute power prediction error of approximately \(2.4\) percentage points under the Wald \(z\)-test. The absolute power prediction error is largest in the smaller designs and decreases as the number of clusters increases. The same trend holds for composite survival endpoints, although the approximation is somewhat more conservative, with a median absolute power prediction error of approximately \(3.0\%\) and the largest discrepancies occurring under the weakest composite alternatives. This additional conservativeness is anticipated, because the composite comparison rule involves censoring, partial rankings, and triplet probabilities, and hence induces a more complex finite-sample structure than the scalar ordinal endpoint.

\begin{figure}[!htbp]
\centering
\includegraphics[width=\textwidth]{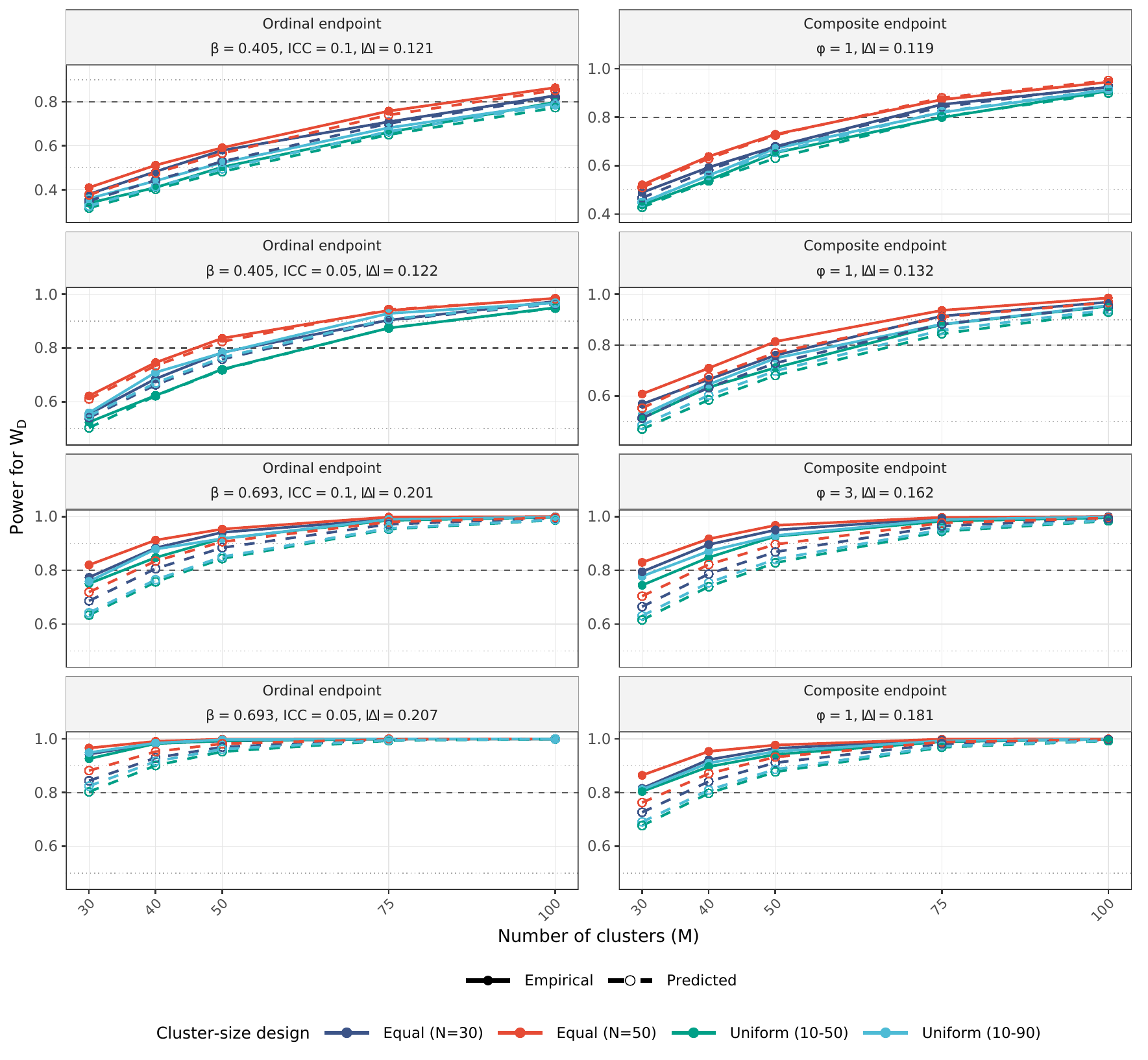}
\caption{Representative power curves for \(\Delta=W_D\). Solid lines denote empirical power, and dashed lines denote predicted power based on Theorem \ref{theorem:var_single} for ordinal endpoints or Theorem \ref{theorem:var_comp} for composite endpoints. The horizontal dashed line marks \(80\%\) power. Rows correspond to selected alternatives, and columns distinguish the ordinal and composite endpoint settings.}
\label{supp:fig:power_WD}
\end{figure}

Figure~\ref{supp:fig:power_logWO} gives the corresponding power curves for \(\Delta=\log(W_O)\), whose behavior closely resembles that of \(W_D\). For ordinal outcomes, the median absolute discrepancy between empirical and predicted Wald \(z\)-test power is approximately \(2.0\) percentage points, with most of the difference concentrated in the small-to-moderate \(M\) range; for composite endpoints, the median discrepancy is approximately \(2.7\) percentage points. As in Figure~\ref{supp:fig:power_WD}, empirical power increases with \(M\), with the treatment effect size, and with the average cluster size, while cluster-size heterogeneity reduces efficiency. These patterns confirm that the proposed power formula captures the leading design behavior for the win odds as well as for the win difference and win ratio.

\begin{figure}[!htbp]
\centering
\includegraphics[width=\textwidth]{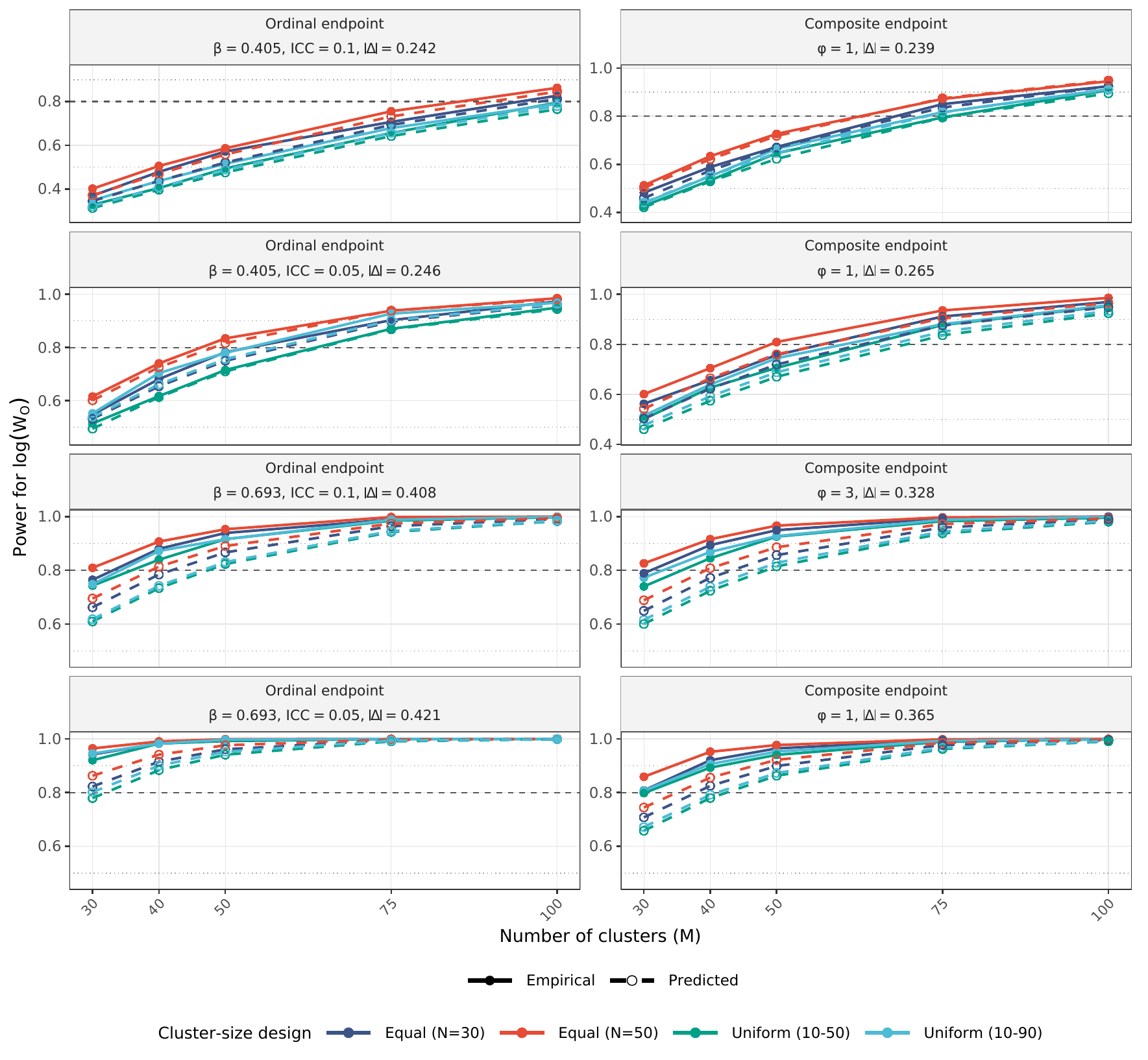}
\caption{Representative power curves for \(\Delta=\log(W_O)\). Solid lines denote empirical power, and dashed lines denote predicted power based on Theorem \ref{theorem:var_single} for ordinal endpoints or Theorem \ref{theorem:var_comp} for composite endpoints. The horizontal dashed line marks \(80\%\) power. Rows correspond to selected alternatives, and columns distinguish the ordinal and composite endpoint settings.}
\label{supp:fig:power_logWO}
\end{figure}

Figures \ref{supp:fig:power_WD} and \ref{supp:fig:power_logWO} demonstrate that the conclusions obtained for \(\log(W_R)\) in the main text extend to the remaining two win summaries. The formula is most accurate for ordinal endpoints with moderately large \(M\) and is slightly more conservative for composite endpoints, particularly under weaker alternatives and unequal cluster sizes. Such conservativeness is desirable at the planning stage, as it guards against overstating power in small-to-moderate CRTs.

\subsection{Relative bias diagnostics}
\label{supp:sec:bias_diagnostics}

Figure \ref{supp:fig:rel_bias_all} reports the relative bias of the three win-statistic estimators across the non-null scenarios. These diagnostics help distinguish two potential sources of power prediction error: bias in the point estimator and conservativeness of the design-stage variance approximation. For ordinal outcomes, the relative bias is small for all three estimands, ranging approximately from \(-3.6\%\) to \(4.9\%\) across the displayed settings, depending on the estimand, and is centered near zero.

For composite survival endpoints, the relative bias can appear larger in percentage terms, particularly under very weak effects for which the denominator \(|\Delta|\) is small. The smallest composite effects, for instance, are close to \(0.02\) on the \(W_D\) scale and close to \(0.04\)--\(0.05\) on the log-ratio scales, so that a small absolute difference translates into a comparatively large percentage. The maximum absolute bias nonetheless remains modest, approximately \(0.014\) for \(W_D\), \(0.029\) for \(\log(W_O)\), and \(0.038\) for \(\log(W_R)\). The large relative values in weak composite settings should therefore be interpreted as a scale artifact rather than as evidence of substantial estimator bias; once the small denominator is taken into account, the estimators remain centered on their intended targets.

\begin{figure}[!htbp]
\centering
\includegraphics[width=0.8\textwidth]{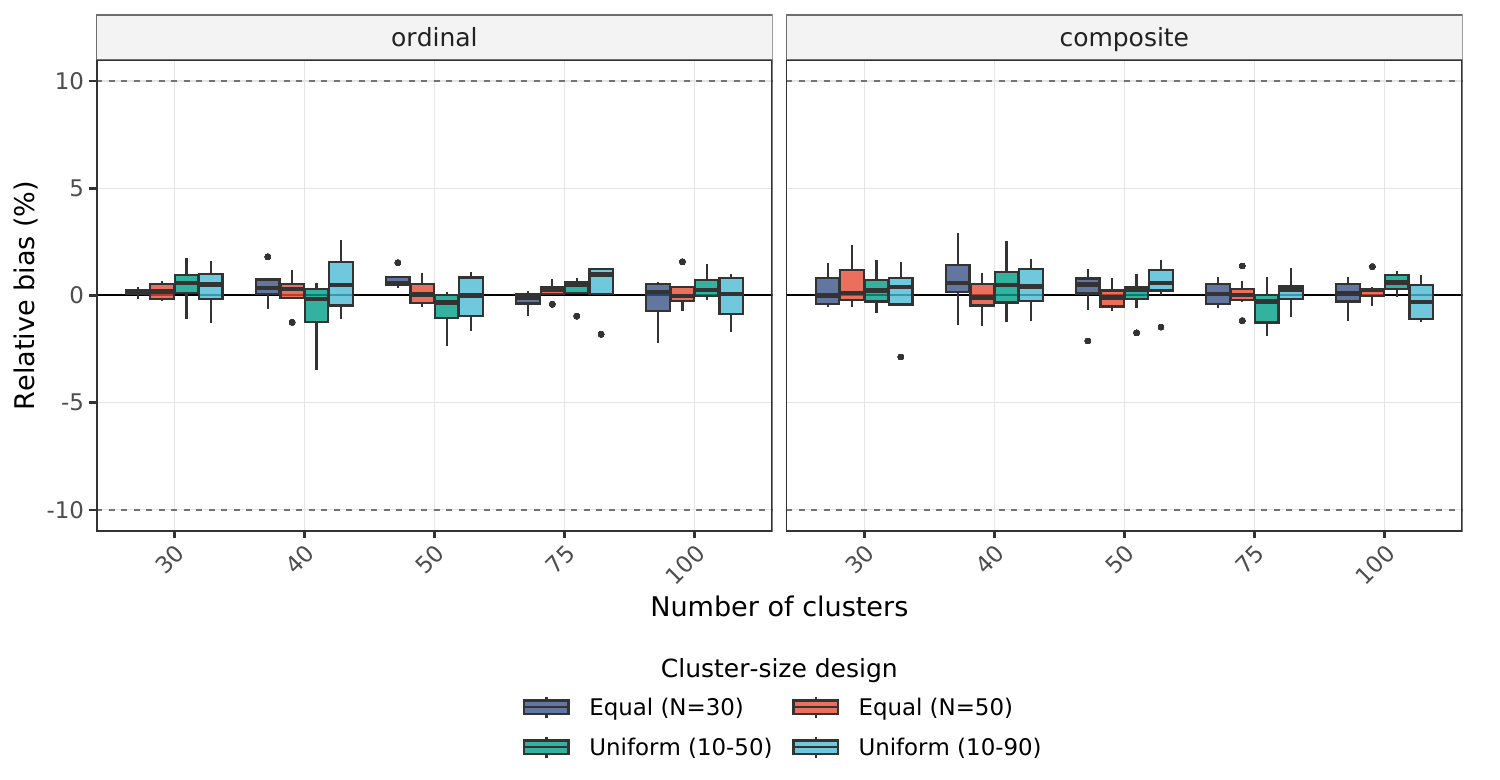}\\[1ex]
\includegraphics[width=0.8\textwidth]{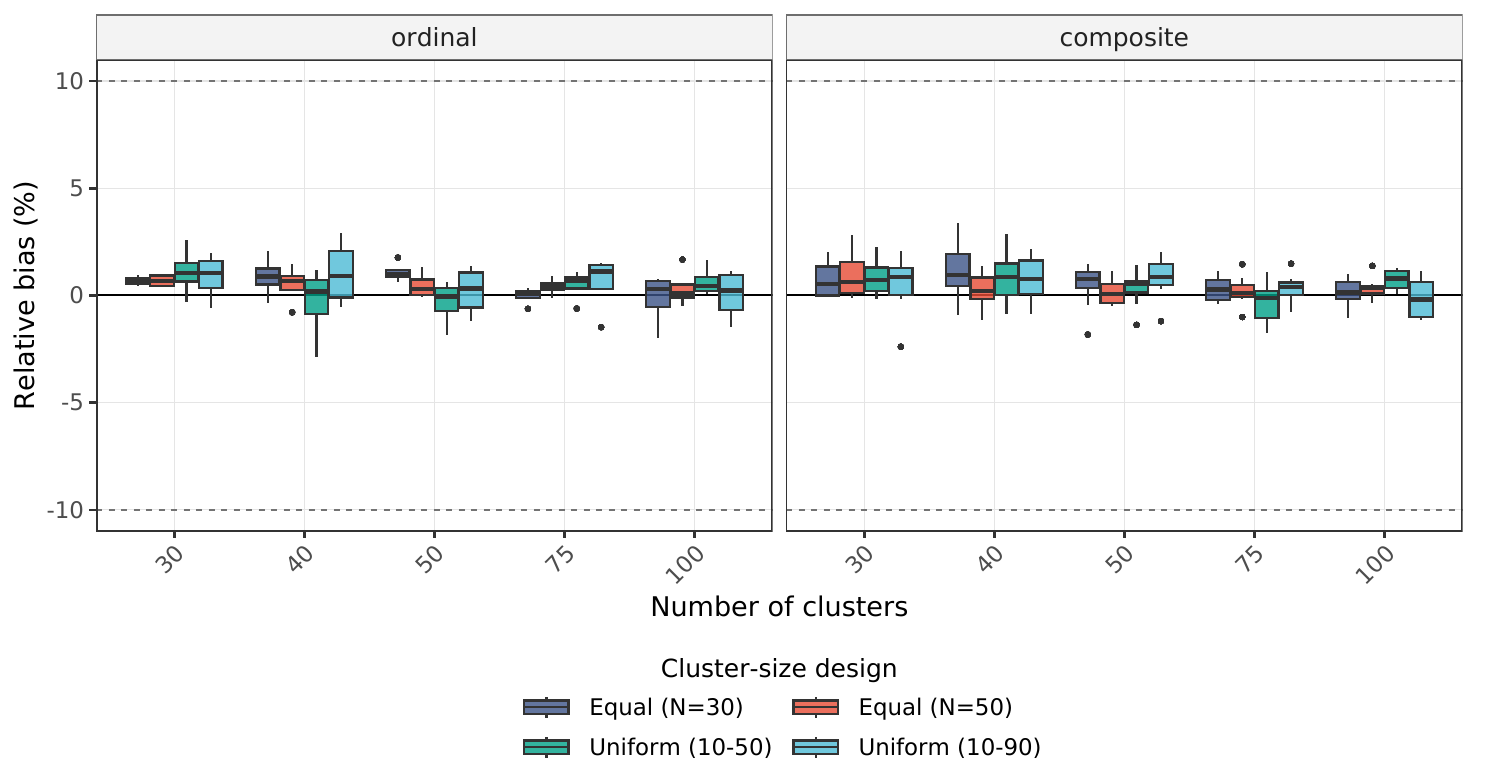}\\[1ex]
\includegraphics[width=0.8\textwidth]{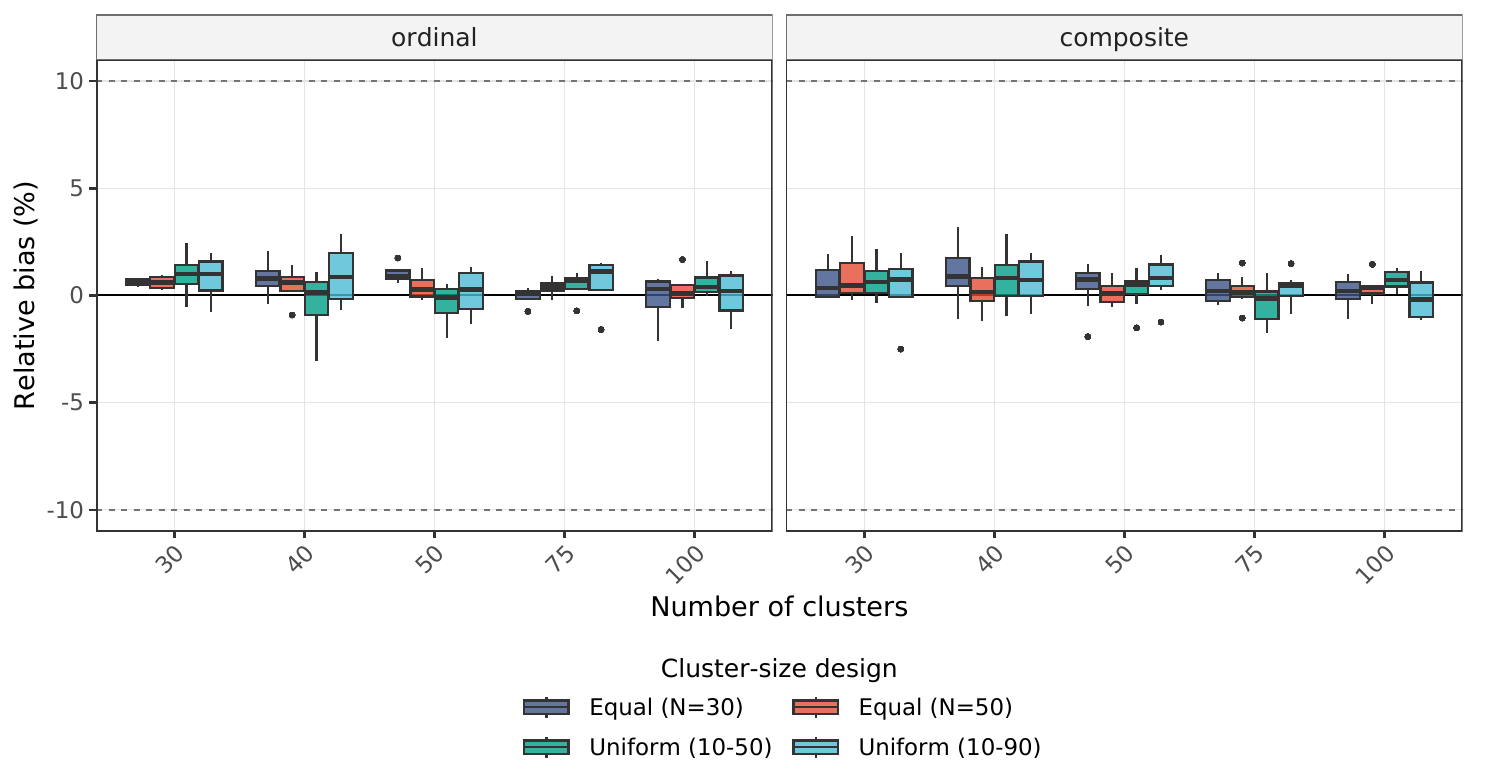}
\caption{Relative bias diagnostics for the three win estimands across the non-null simulation settings, defined as the Monte Carlo mean minus the truth divided by the absolute truth. From top to bottom, the panels correspond to \(\Delta=W_D\), \(\Delta=\log(W_R)\), and \(\Delta=\log(W_O)\).}
\label{supp:fig:rel_bias_all}
\end{figure}

\subsection{Standard error diagnostics}
\label{supp:sec:se_diagnostics}

Figure \ref{supp:fig:ase_ratio_all} compares the two standard-error summaries with the MCSD. The ASE/MCSD ratio evaluates the finite-sample calibration of the replication-specific standard-error estimator from Theorem~\ref{theorem:u-var}. The DSE/MCSD ratio evaluates the closed-form design-based standard error from Theorem \ref{theorem:var_single} for ordinal endpoints and Theorem \ref{theorem:var_comp} for composite endpoints.

\begin{figure}[!htbp]
\centering
\includegraphics[width=0.8\textwidth]{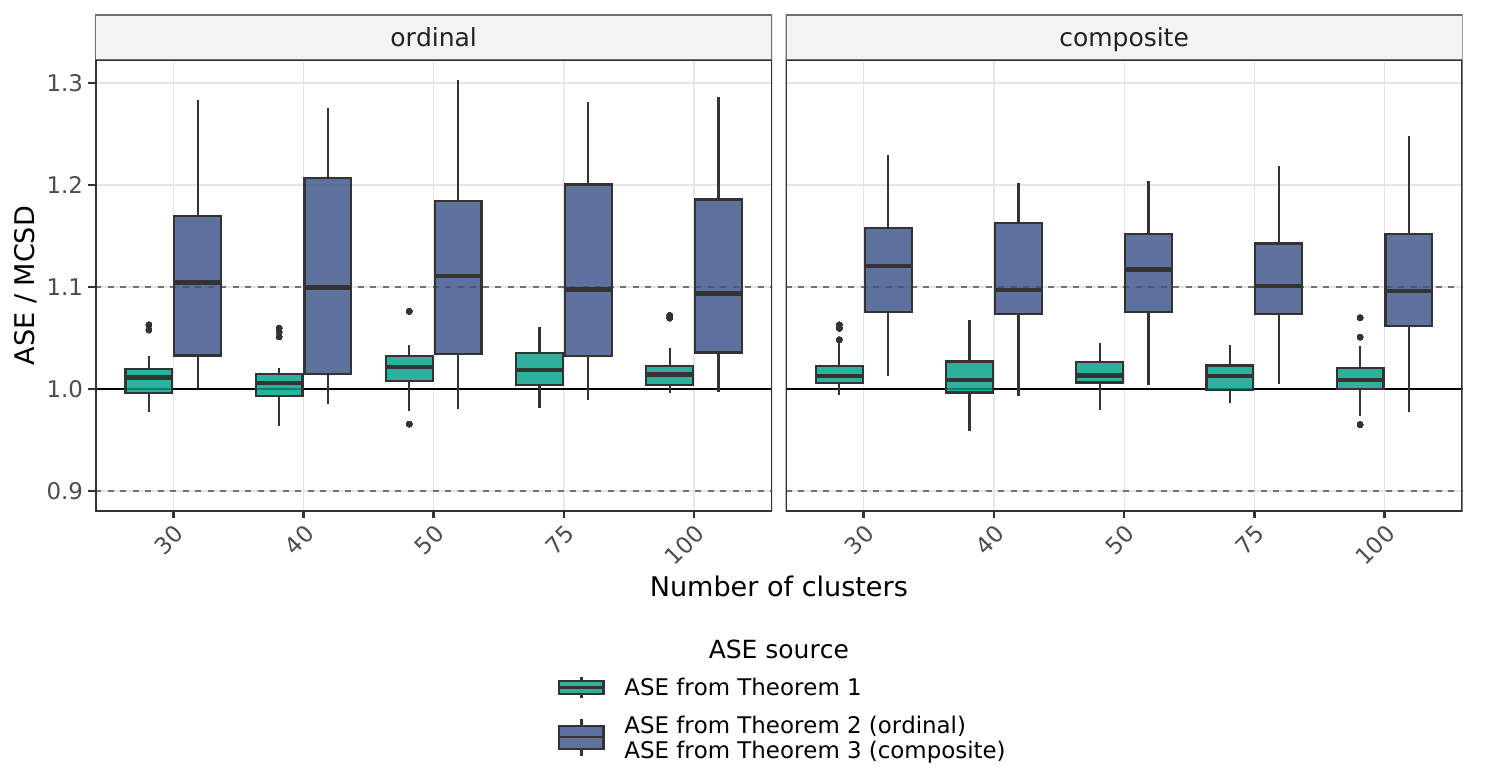}\\[1ex]
\includegraphics[width=0.8\textwidth]{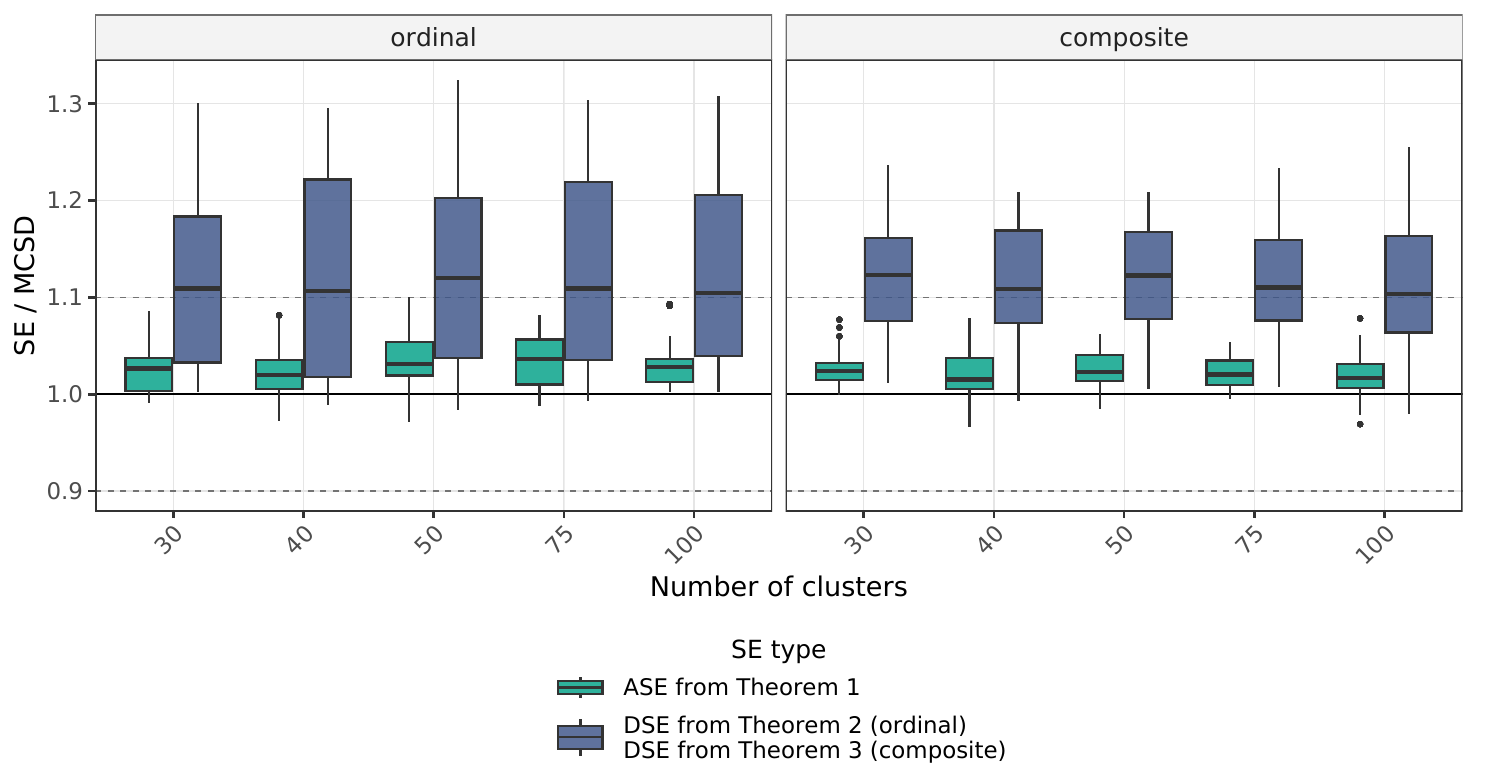}\\[1ex]
\includegraphics[width=0.8\textwidth]{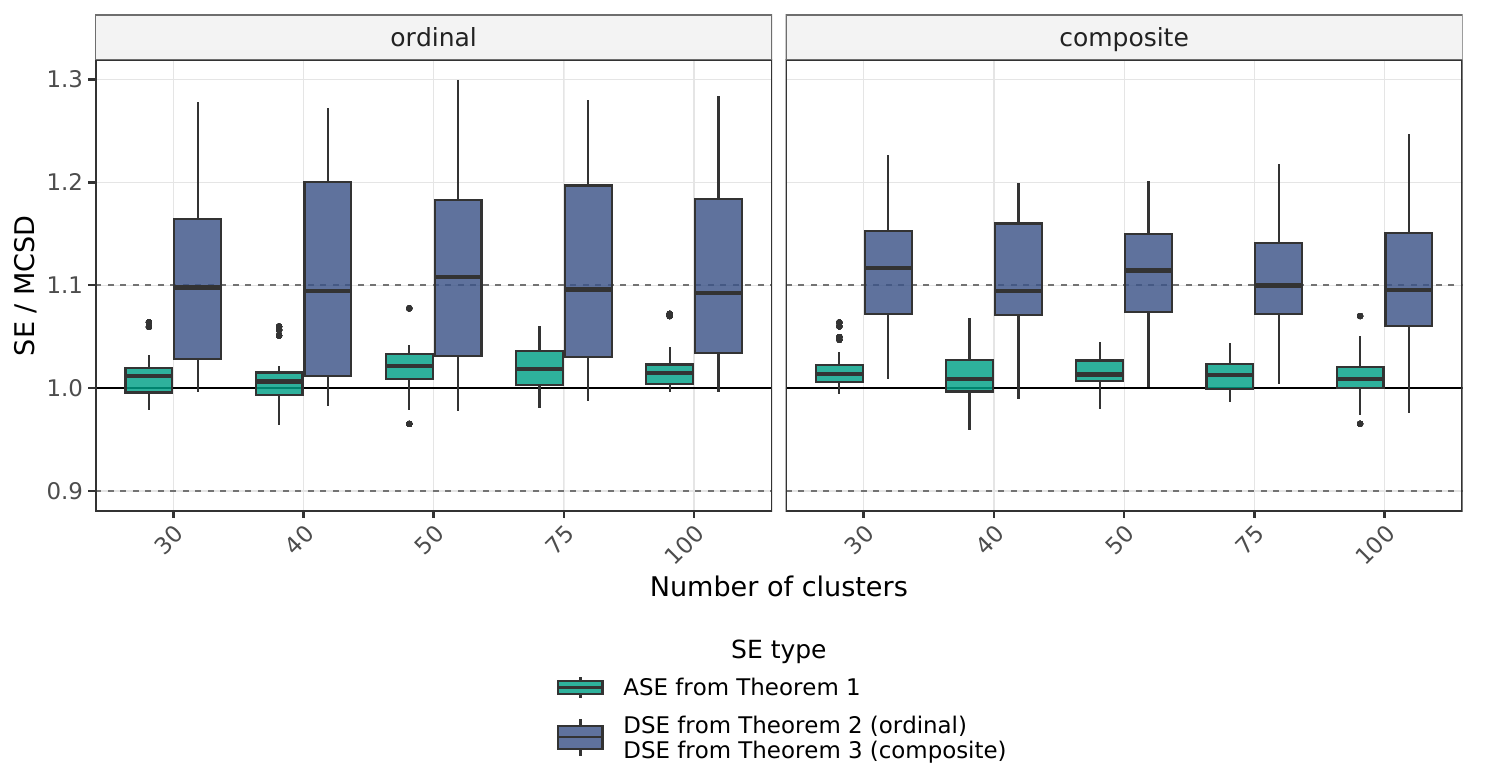}
\caption{ASE/MCSD diagnostics for the three win estimands. From top to bottom, the panels correspond to \(\Delta=W_D\), \(\Delta=\log(W_R)\), and \(\Delta=\log(W_O)\). The direct cluster-score ASE refers to the variance estimator of Theorem~\ref{theorem:u-var}, whereas the formula ASE refers to the design-stage variance approximation of Theorem~\ref{theorem:var_single}.}
\label{supp:fig:ase_ratio_all}
\end{figure}

Figure \ref{supp:fig:ase_ratio_all} clarifies why formula-predicted power is slightly conservative in some designs, most notably for composite endpoints. The ASE from Theorem \ref{theorem:u-var} closely tracks the MCSD, indicating that the standard error estimator used for inference is well calibrated. The DSE is generally larger than the MCSD because it is a smoother design stage approximation based on low-dimensional planning inputs. Because formula-predicted power is calculated using the DSE, a DSE/MCSD ratio above one produces a positive power prediction error, especially where the power curve is steep.

\subsection{Required number of clusters}
\label{supp:sec:requiredM}

Figures~\ref{supp:fig:reqM_ordinal} and \ref{supp:fig:reqM_composite} summarize the required number of clusters for the ordinal and composite endpoint settings, respectively. In each heatmap, the rows are ordered by effect size, so that stronger effects correspond to smaller required \(M\).

For ordinal outcomes, Figure~\ref{supp:fig:reqM_ordinal} shows that the required \(M\) decreases with increasing effect size and increases with the rank ICC and with cluster-size heterogeneity. Across the displayed ordinal designs, the required number of clusters ranges approximately from \(50\) in the most favorable settings to more than \(600\) under weak effects and stronger dependence. This behavior is precisely that implied by Theorem~\ref{theorem:var_single}: a larger \(|\Delta|\) increases the noncentrality parameter, whereas a larger rank ICC and unequal cluster sizes inflate the variance.

\begin{figure}[!htbp]
\centering
\includegraphics[width=0.8\textwidth]{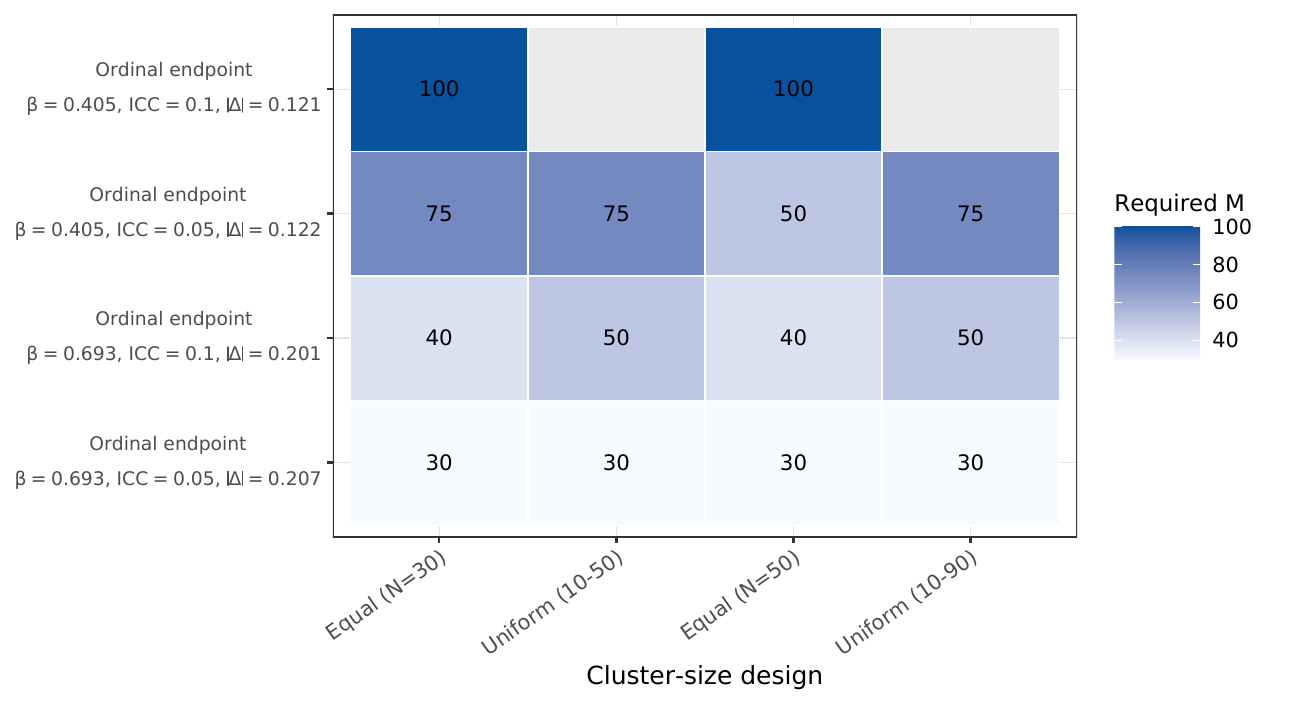}\\[1ex]
\includegraphics[width=0.8\textwidth]{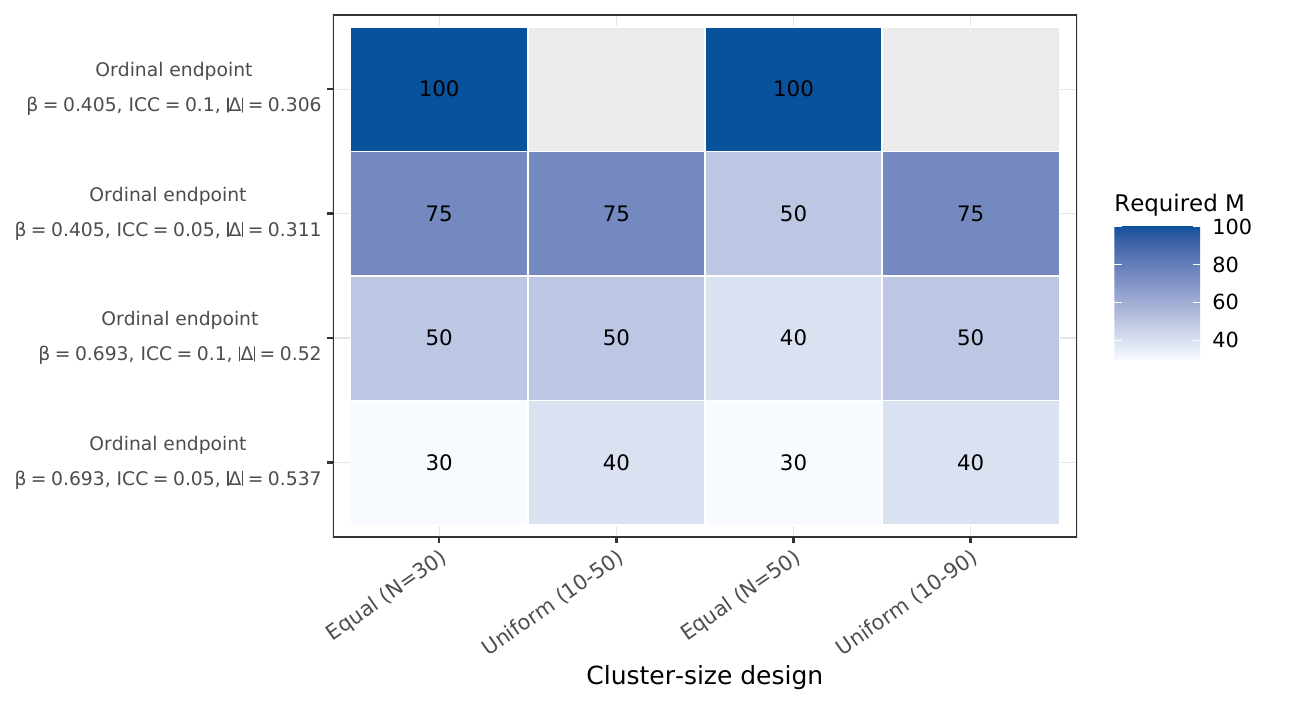}\\[1ex]
\includegraphics[width=0.8\textwidth]{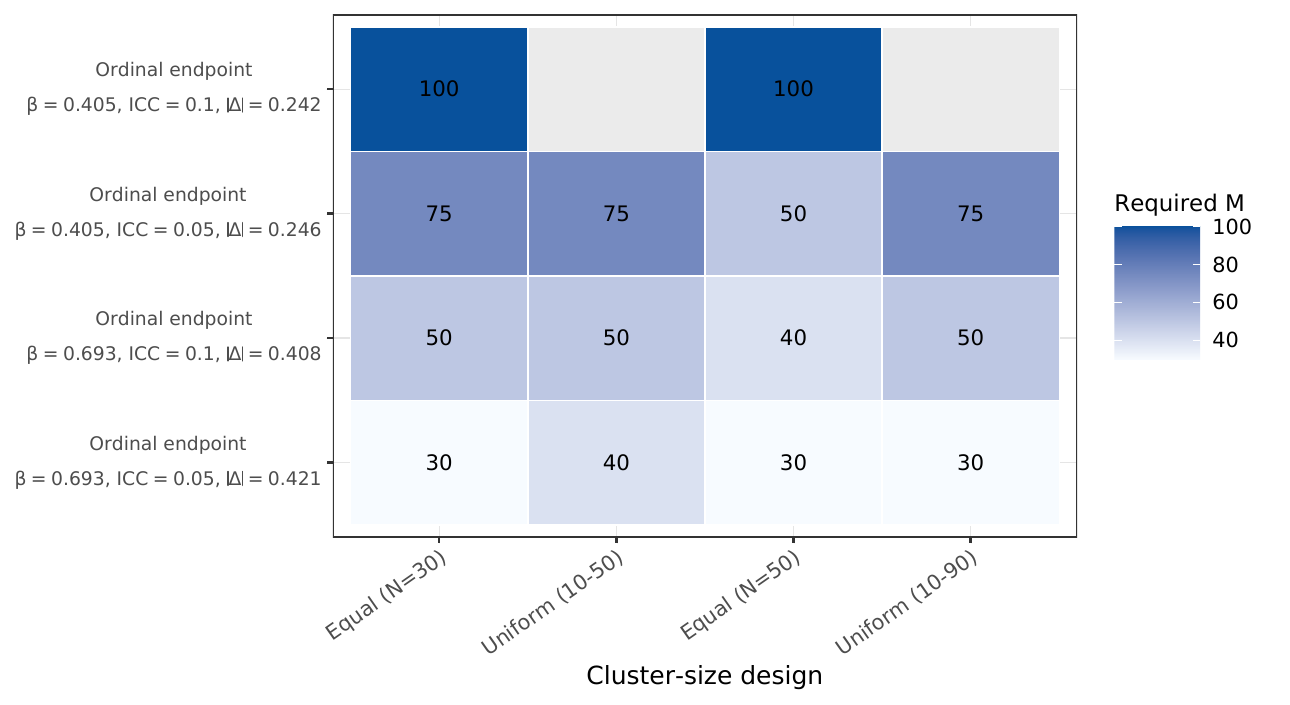}
\caption{Required number of clusters for ordinal outcomes. From top to bottom, the panels correspond to \(\Delta=W_D\), \(\Delta=\log(W_R)\), and \(\Delta=\log(W_O)\). Rows are ordered by effect size, and each heatmap entry gives the required \(M\) for the corresponding cluster-size design and estimand.}
\label{supp:fig:reqM_ordinal}
\end{figure}

For composite survival endpoints, Figure~\ref{supp:fig:reqM_composite} exhibits the same qualitative pattern, but the required number of clusters is generally larger than in the ordinal setting, ranging approximately from \(158\) to more than \(700\) across the displayed composite designs. This larger requirement reflects the additional uncertainty introduced by censoring, partial rankings, and the triplet probabilities \((p_{WW},p_{WT},p_{TT})\) that enter the composite variance. In particular, two settings with similar marginal win or tie probabilities may still require different numbers of clusters, because they differ in their anchor-wise dependence through the term \(Q=p_{WW}+p_{WT}+\frac{1}{4}p_{TT}\).

\begin{figure}[!htbp]
\centering
\includegraphics[width=0.6\textwidth]{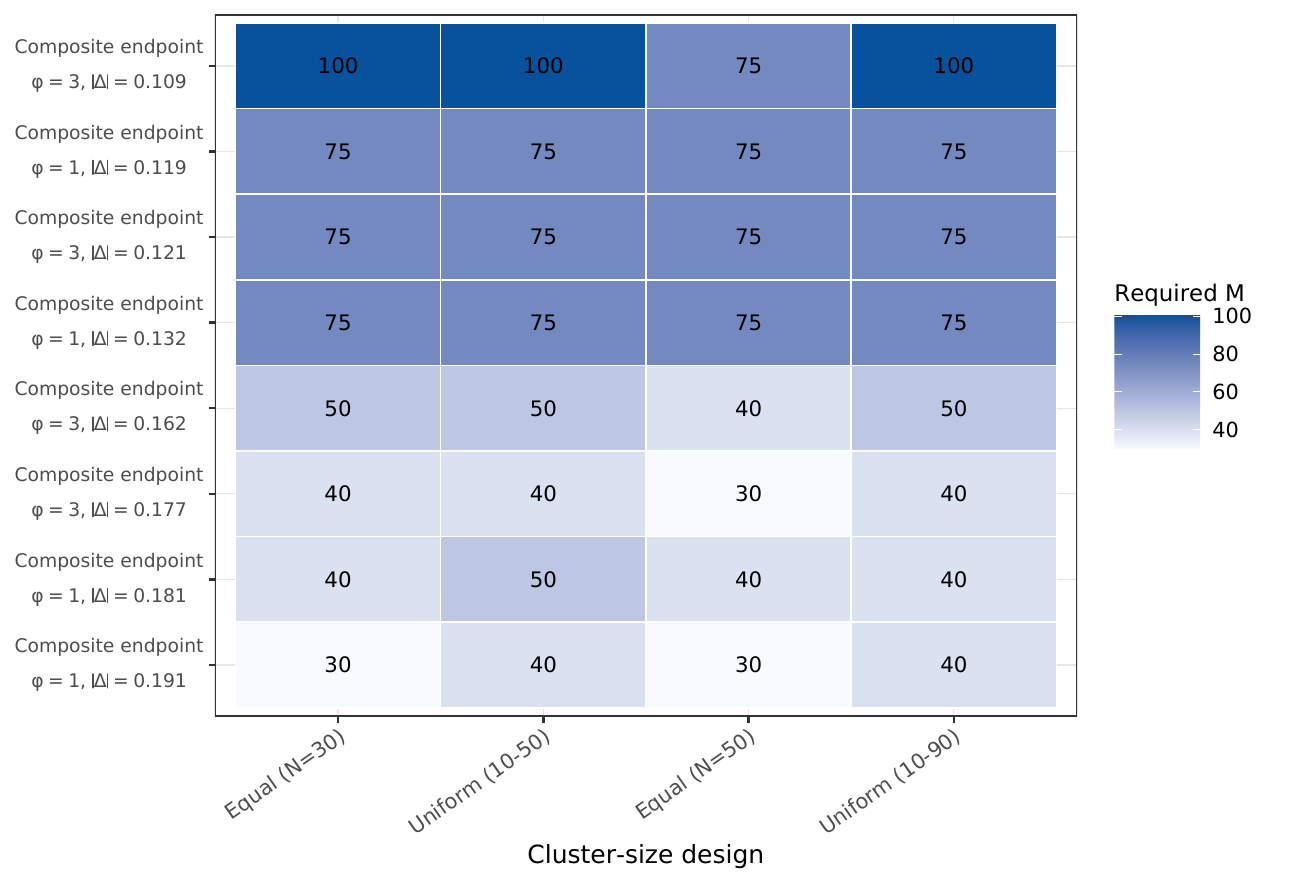}\\[1ex]
\includegraphics[width=0.6\textwidth]{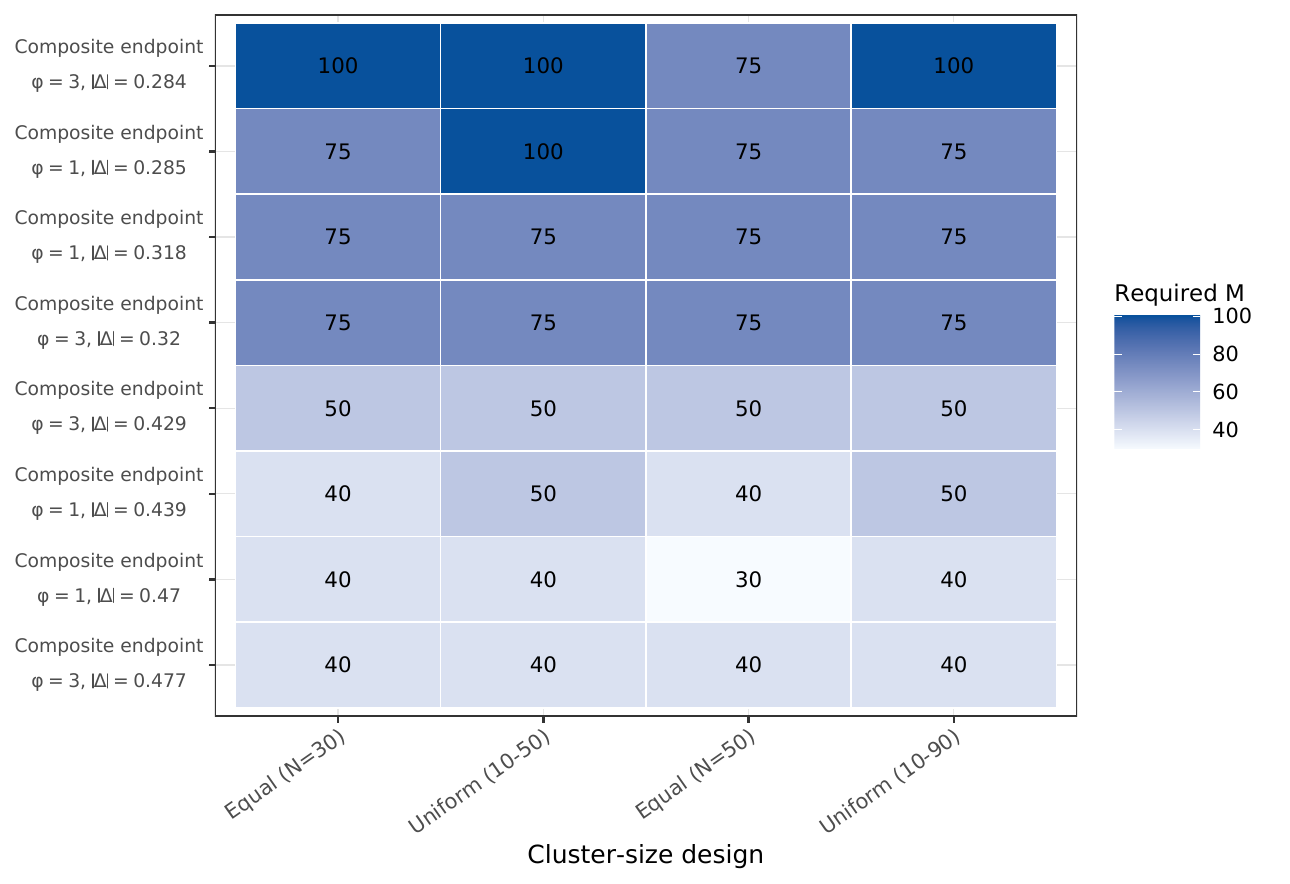}\\[1ex]
\includegraphics[width=0.6\textwidth]{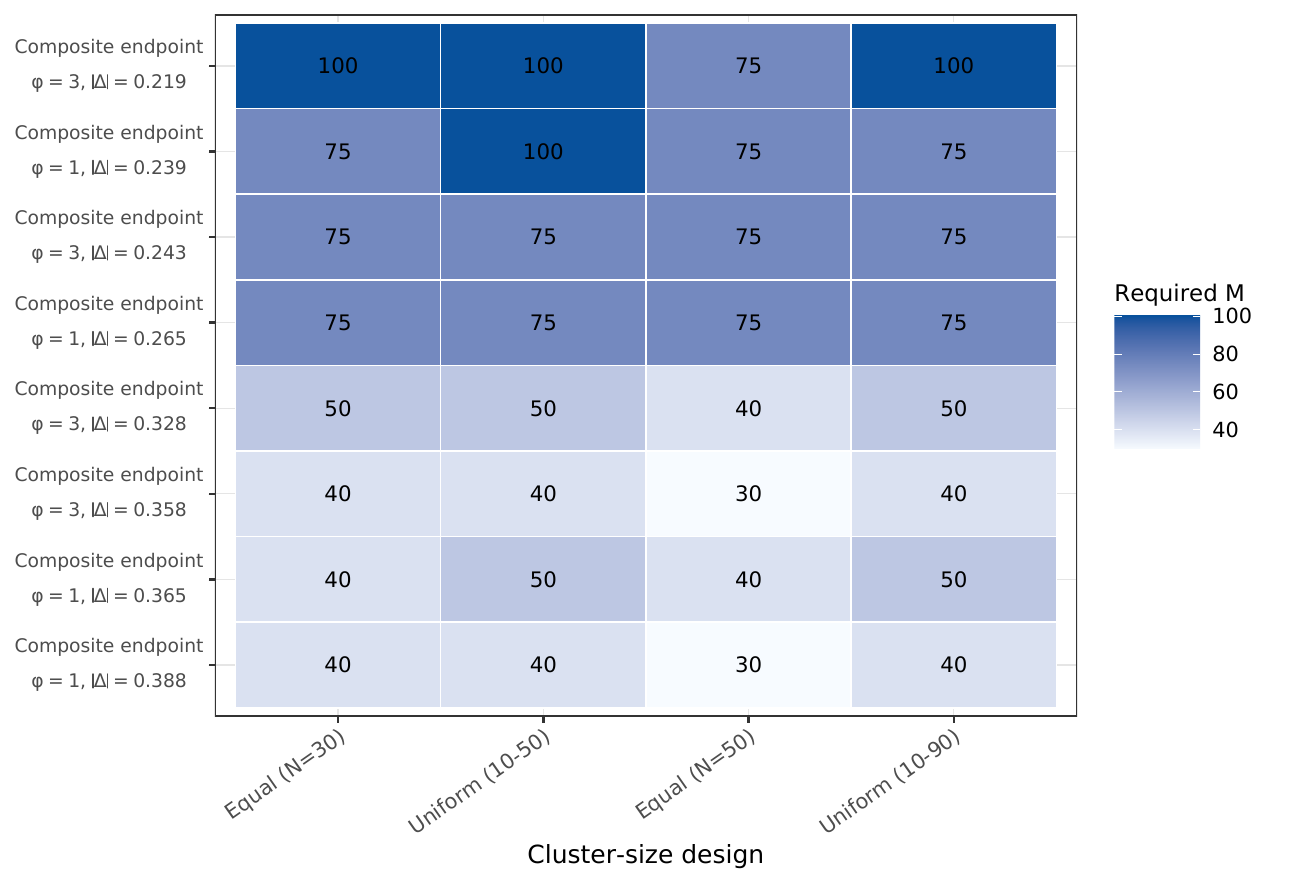}
\caption{Required number of clusters for composite survival endpoints. From top to bottom, the panels correspond to \(\Delta=W_D\), \(\Delta=\log(W_R)\), and \(\Delta=\log(W_O)\). Rows are ordered by effect size, and each heatmap entry gives the required \(M\) for the corresponding cluster-size design and estimand.}
\label{supp:fig:reqM_composite}
\end{figure}

The required-\(M\) heatmaps reinforce the design interpretation of the proposed formulas. Increasing the target effect size reduces the required number of clusters, whereas increasing the rank dependence, the tie probability, the censoring-induced fraction of unresolved comparisons, or the cluster-size heterogeneity increases it. The same monotone behavior holds for \(W_D\), \(\log(W_R)\), and \(\log(W_O)\), confirming that the framework is not specific to any single win summary.

\subsection{Sensitivity of the required number of clusters to the tie probability and rank ICC}
\label{supp:sec:tie_sensitivity}

Because the tie probability \(\pi_{\mathrm{tie}}\) and the generalized rank ICC \(\rho^{*}\) may not be known precisely before trial initiation, we conducted an additional design sensitivity analysis under the prioritized composite survival setting used in the main simulation. Rather than assigning arbitrary values to \(\pi_{\mathrm{tie}}\), we varied the censoring distribution while holding the event time model fixed. This provides a coherent assessment because censoring jointly determines the frequency of unresolved comparisons and the comparison moments entering the proposed variance formula. Clusters were randomized according to \(A_i\sim\operatorname{Bernoulli}(0.5)\). A shared cluster frailty \(\gamma_i\sim\operatorname{Gamma}(7.5,7.5)\) was used to induce dependence among participants in the same cluster, where the Gamma distribution was parameterized by its shape and rate. Conditional on \((A_i,\gamma_i)\), the hospitalization, death, and censoring hazards were
\[
\lambda_{ij}^{H}(A_i)=\gamma_i\lambda_0^{H}\exp(-\eta_H A_i),
\qquad
\lambda_{ij}^{D}(A_i)=\gamma_i\lambda_0^{D}\exp(-\eta_D A_i),
\qquad
\lambda_{ij}^{C}(A_i)=\lambda_0^{C}\exp(-\eta_C A_i).
\]
The hospitalization and death times for the same participant were generated from a Gumbel--Hougaard copula with parameter \(\phi\in\{1,3\}\). Thus, \(\phi\) controls dependence between the hospitalization and death components within a participant, whereas the shared frailty controls dependence among participants in the same cluster. We fixed \(\lambda_0^{H}=0.10\),  \(\lambda_0^{D}=0.08\), \(\eta_H=\eta_D=0.30\), and  \(\eta_C=0.15\), and varied the baseline censoring hazard over \(\lambda_0^{C} \in \{0.005,0.015,0.030,0.060,0.100,0.150\}\). The value \(\lambda_0^{C}=0.030\) corresponds to the censoring setting used in the main simulation. Death remained the first priority component, and hospitalization was compared only when the death comparison was unresolved. Because all event times were continuous, exact equality of the event times had probability zero. Consequently, ties in this setting arose almost entirely when censoring prevented either component from producing a decisive comparison. For each combination of \(\phi\) and \(\lambda_0^{C}\), \(2{,}000{,}000\) independently generated treatment and control pairs were used to estimate \(\pi_{\mathrm{win}}\), \(\pi_{\mathrm{loss}}\), \(\pi_{\mathrm{tie}}=1-\pi_{\mathrm{win}}-\pi_{\mathrm{loss}}\), and \(\log(W_R)=\log\left(\frac{\pi_{\mathrm{win}}}{\pi_{\mathrm{loss}}}\right)\).  Separate Monte Carlo samples of \(2{,}000{,}000\) pooled triplets were used to estimate the pair and triplet comparison moments entering \(P\) and \(Q\). An additional \(2{,}000{,}000\) draws were used to calculate the generalized rank ICC induced by each data generating mechanism. The induced values of \(\rho^{*}\) ranged from 0.035 to 0.109, supporting the planning grid \(\rho^{*}\in\{0.03,0.06,0.09,0.12\}\).  To separate uncertainty about this global dependence parameter from uncertainty due to censoring, every censoring scenario was evaluated at each value of the planning grid for \(\rho^{*}\), while retaining the corresponding values of \(\log(W_R)\), \(P\), and \(Q\) generated under that censoring scenario. The primary display used \(N_i\sim\operatorname{Uniform}\{10,\ldots,90\}\),  for which \(\overline N=50\) and \(\mathrm{CV}=0.468\). We calculated the required even number of clusters \(M\) for 80\% power at a two sided type I error of $0.05$ using the proposed composite endpoint sample size formula and the Wald \(t\) reference distribution with \(M-2\) degrees of freedom.

As the baseline censoring hazard increased, \(\pi_{\mathrm{tie}}\) increased monotonically. When \(\phi=1\), \(\pi_{\mathrm{tie}}\) increased from 0.031 under \(\lambda_0^{C}=0.005\) to 0.480 under \(\lambda_0^{C}=0.150\). When \(\phi=3\), it increased from 0.047 to 0.589. At the censoring setting used in the main simulation, \(\lambda_0^{C}=0.030\), the tie probabilities were 0.159 for \(\phi=1\) and 0.228 for \(\phi=3\). In contrast, the treatment effect was nearly unchanged across the censoring scenarios, with \(\log(W_R)\) ranging only from $0.282$ to $0.295$. Figure \ref{supp:fig:tie_sensitivity} shows the resulting relationship between \(\pi_{\mathrm{tie}}\) and the required number of clusters for \(\phi=3\), which produced the wider range of tie probabilities. At the planning value \(\rho^{*}=0.09\), the required number of clusters increased from 76 to 156 as \(\pi_{\mathrm{tie}}\) increased from 0.047 to 0.589. The corresponding increase for \(\phi=1\) was from 74 to 120 as \(\pi_{\mathrm{tie}}\) increased from 0.031 to 0.480. At the baseline censoring hazard, the required number of clusters increased from 36 to 102 for \(\phi=1\), and from 42 to 116 for \(\phi=3\), when the planning value of \(\rho^{*}\) increased from 0.03 to 0.12. These results demonstrate that uncertainty in both \(\pi_{\mathrm{tie}}\) and \(\rho^{*}\) can materially affect the required number of clusters. Because \(\log(W_R)\) remained nearly constant, the increase in \(M\) was driven primarily by the loss of decisive comparisons and the associated changes in the second moment quantities entering the planning variance. 

\begin{figure}[tbp]
\centering
\includegraphics[width=0.95\textwidth]{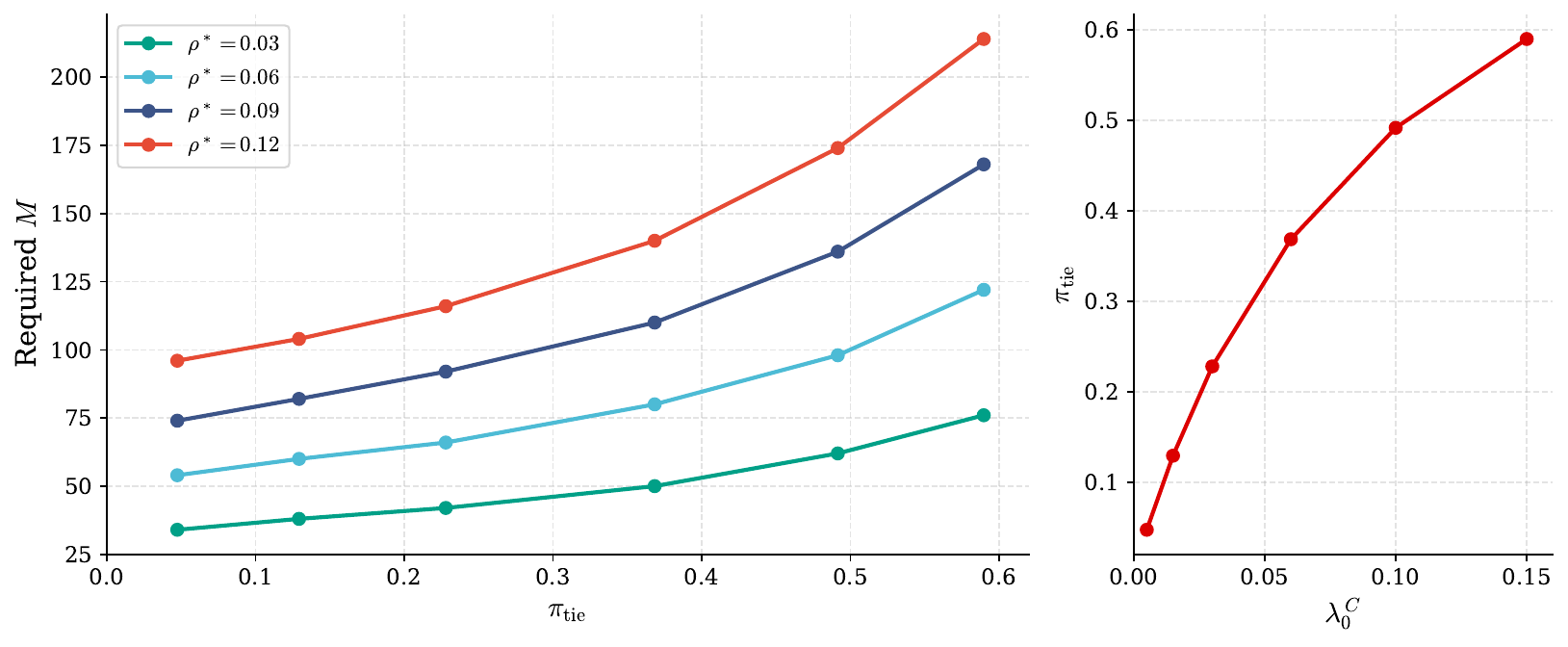}
\caption{Sensitivity of the required number of clusters \(M\) to the censoring induced tie probability \(\pi_{\mathrm{tie}}\) for \(\phi=3\). Colored curves correspond to planning values \(\rho^{*}\in\{0.03,0.06,0.09,0.12\}\). The event time parameters were held fixed while the baseline censoring hazard was varied over \(\lambda_0^{C}\in\{0.005,0.015,0.030,0.060,0.100,0.150\}\). At each point, the win ratio and the pair and triplet comparison moments were recalculated under the corresponding censoring scenario. Required \(M\) was calculated for \(N_i\sim\operatorname{Uniform}\{10,\ldots,90\}\), 80\% power, a two sided type I error of 0.05, and the Wald \(t\) reference distribution with \(M-2\) degrees of freedom.}
\label{supp:fig:tie_sensitivity}
\end{figure}

\section{Example Code for the \texttt{WinCRT} Package}

We introduce the \texttt{WinCRT} package (\url{https://github.com/fancy575/WinCRT}), which implements win-statistics inference for CRT with either hierarchical (composite) endpoints or a single endpoint. Inference is provided via Wald tests based on either a $t$-test with degrees of freedom $M-2$ or a \(z\)-test reference.

The main function has the following declaration:
\begin{verbatim}
wincrt <- function(data, id, trt, cluster, outcome, tier)
\end{verbatim}
with arguments:
\begin{itemize}
  \item \texttt{data}: A \texttt{data.frame} (in long format for composite outcomes) .
  \item \texttt{id}: Name of the subject identifier column.
  \item \texttt{trt}: Name of the treatment/arm indicator \(1=\) treatment, \(0=\) control.
  \item \texttt{cluster}: Name of the cluster identifier column.
  \item \texttt{outcome}: Name of the numeric outcome column. For survival-style endpoints, this is a time; for single non-survival endpoints, it is the scalar outcome.
  \item \texttt{tier}: Name of the tier code column. Use \texttt{0} for censoring rows; use \texttt{\(1,\dots, K\)} for event tiers, where larger values indicate higher priority. For a single endpoint, set \texttt{tier = 1} for all rows.
\end{itemize}

We show \texttt{wincrt()} using the included dataset \texttt{scData}, which is a survival data in long format with tier $=0$ for censoring and tier $=1,2$ for increasing clinical priority (e.g., hospitalization, then death).
\begin{spacing}{0.9}
\begin{verbatim}
library(WinCRT)
data(scData)

fit <- wincrt(
  data    = df, id      = "id", trt     = "z", cluster = "cluster", 
  outcome = "outcome", tier    = "tier"
)

summary(fit, test = "z", estimand = "all", alpha = 0.05, 
alternative = "two.sided")
\end{verbatim}

Example output:
\begin{verbatim}
Win Ratio Summary (Z-test, alpha=0.05, alternative=two.sided)
Clusters: M1=15, M0=15, M=30
Subjects: n1=418, n0=502
Totals (between arms): Wins=92499, Losses=68175, Ties=49162
p_tie: 0.234; rho (rank ICC): 0.0853

 Estimand Estimate    SE p-value              CI
    logWR    0.305 0.175  0.0804 (-0.037, 0.647)
\end{verbatim}
\end{spacing}

We next illustrate a single ordinal endpoint using the dataset \texttt{orData}. Here we therefore set create a single tier (\texttt{tier = 1} for all individuals).
\begin{spacing}{0.9}
\begin{verbatim}
library(WinCRT)
data(orData)

orData$tier <- 1L           # single endpoint -> one tier

fit_ord <- wincrt(
  data    = df,
  id      = "id",
  trt     = "z",
  cluster = "cluster",
  outcome = "y",
  tier    = "tier"
)

summary(fit_ord, test = "z", estimand = "logWR", alpha = 0.05)
\end{verbatim}
\end{spacing}
Example output:
\begin{spacing}{0.9}
\begin{verbatim}
Win Ratio Summary (Z-test, alpha=0.05, alternative=two.sided)
Clusters: M1=15, M0=15, M=30
Subjects: n1=423, n0=512
Totals (between arms): Wins=96826, Losses=73384, Ties=46366
p_tie: 0.214; rho (rank ICC): 0.105

 Estimand Estimate    SE p-value              CI
    logWR    0.277 0.196   0.156 (-0.106, 0.661)
\end{verbatim}
\end{spacing}

\end{document}